\documentclass[12pt,preprint]{aastex}
\usepackage{psfig}
\begin{document}
\title{The primordial abundance of $^4$He: a self-consistent empirical
analysis of systematic effects in a large sample of 
low-metallicity H {\sc ii} regions}
\author{Yuri I. Izotov}
\affil{Main Astronomical Observatory, Ukrainian National Academy of Sciences,
27 Zabolotnoho str., Kyiv 03680, Ukraine}
\email{izotov@mao.kiev.ua}
\author{Trinh X. Thuan}
\affil{Astronomy Department, University of Virginia, Charlottesville,
VA 22903}
\email{txt@virginia.edu}
\and
\author{Gra\.zyna Stasi\'nska}
\affil{LUTH, Observatoire de Meudon, F-92195 Meudon Cedex, France}
\email{grazyna.stasinska@obspm.fr}

\begin{abstract}
We determine the primordial
helium mass fraction $Y_p$ using 93 
spectra of 86 low-metallicity extragalactic H {\sc ii} regions. 
This sample constitutes the largest and most 
homogeneous high-quality data sets in existence for the determination of $Y_p$.
For comparison and to improve the statistics in our investigation of 
systematic effects affecting the $Y_p$ determination, 
we have also considered a  
sample of 271 low-metallicity 
H {\sc ii} regions selected from the Data Release 5 of the Sloan Digital 
Sky Survey. Although this larger sample 
shows more scatter, 
it gives results that are consistent at the 2$\sigma$ level with our original
sample. 
We have considered known systematic effects which may affect  
the $^4$He abundance determination. They include different sets of He {\sc i}
line emissivities and reddening laws, collisional and fluorescent
enhancements of He {\sc i} recombination lines, underlying He {\sc i} 
stellar absorption lines, collisional excitation of hydrogen lines,
temperature and ionization structure of the H {\sc ii} region, and 
deviation of He {\sc i} and H emission line intensities from case B. 
However,  the most likely value of $Y_p$ depends on the adopted set of 
He {\sc i} line emissivities.
Using Monte Carlo methods to solve simultaneously the above 
systematic effects we find a primordial helium mass fraction 
$Y_p$ = 0.2472 $\pm$ 0.0012 
when using the He {\sc i} emissivities from \citet{B99,B02} 
and 0.2516 $\pm$ 0.0011 when using those from \citet{P05}.
The first value agrees well with the value given by Standard Big Bang 
Nucleosynthesis (SBBN) theory, while the value obtained with likely more 
accurate emissivities of  \citet{P05} is higher at the 2$\sigma$ level. This 
latter value, if confirmed, would imply slight deviations from SBBN. 
\end{abstract}


\section{INTRODUCTION}\label{intro}

In the standard theory of big bang nucleosynthesis (SBBN), given the 
number of light neutrino species, the abundances of these light elements 
depend only on one cosmological parameter, the   
baryon-to-photon number ratio $\eta$, which in turn is directly related 
to the baryon density parameter $\Omega_b$, the present ratio of the 
baryon mass density to the critical density of the Universe.
This means that accurate measurements of the primordial abundances of each of 
the four light elements can provide, in principle,
a direct measurement of the baryonic density.  
As $\eta$ is a very small quantity, it is convenient to define 
the parameter $\eta_{10}$ = 10$^{10}$ $\eta$. Then $\eta_{10}$ is related 
to $\Omega_b$ by the expression $\eta_{10}$ = 274 $\Omega_bh^2$ ,
where $h$ = $H_0$/100 km s$^{-1}$ Mpc$^{-1}$ and 
$H_0$ is the present value of the Hubble parameter \citep{St05}. 

Because of the strong dependence of its abundance on $\eta$, deuterium has
become the baryometer of choice ever since accurate measurement of D/H 
in high-redshift low-metallicity QSO Ly$\alpha$ absorption
systems have become possible. Although the data is still scarce -- there  
are only six absorption systems for which such a D/H measurement has 
been carried out \citep{BT98a,BT98b,O01,O06,PB01,K03,C04} --
and the scatter remains large, 
 the measurements appear to converge to a mean primordial value 
D/H $\sim$ 2.4 $\pm$ 0.4 $\times$ 10$^{-5}$,
which corresponds to $\Omega_bh^2$ $\sim$ 0.023 $\pm$ 0.002.
This estimate of $\Omega_bh^2$
is in excellent agreement with the values of 0.022 -- 0.023 obtained 
from recent studies of
the fluctuations of the cosmic microwave background (CMB) by 
DASI \citep{Pr02}, BOOMERANG \citep{N02} and WMAP \citep{S03,S06}. 
It is also in good agreement with the value of $\Omega_bh^2$ derived 
from large scale structure \citep{T04}.

While a single good baryometer like D is sufficient to 
derive the baryonic mass density from BBN, accurate measurements of the 
primordial abundances of at least two different relic elements are required 
to check the consistency of SBBN. Among the remaining relic elements,
$^3$He can only be observed in the solar system and in the Galaxy, both of  
which have undergone significant chemical evolution, making it 
difficult to derive its primordial abundance \citep{BRB02}.
The derivation of the primordial 
$^7$Li abundance in metal-poor halo stars in the Galaxy is also beset 
by difficulties such as   
the uncertain stellar temperature scale and the temperature structures 
of the atmospheres of these very cool stars \citep{CP05}. 
The primordial abundance of the 
remaining relic element, $^4$He (hereafter He), can in principle be derived 
accurately from observations of the helium and 
hydrogen emission lines from low-metallicity blue compact dwarf (BCD) galaxies
which have undergone little chemical evolution. 
Several groups have used this technique to derive the primordial  
He mass fraction $Y_p$, with somewhat different results.
Our group \citep{ITL97, IT98b, IT04} has obtained 
$Y_p$ = 0.244 $\pm$ 0.002. This value is consistent at 
the two-sigma level with the prediction of SBBN, $Y_p$ = 0.2482 $\pm$ 0.0007 
\citep{W91,St05}, adopting the value of $\Omega_bh^2$ = 0.0223 $\pm$ 0.0008 
found by WMAP \citep{S06}.   
On the other hand, the $Y_p$ predicted by SBBN 
is not consistent with the lower values of 0.234 $\pm$ 0.002 
derived by \citet{O97}, of 0.2384 $\pm$ 0.0025
by \citet{P02} and of 0.2391 $\pm$ 0.0020 by \citet{L03}.
These lower estimates of $Y_p$ have led some authors to argue that the
discrepancy between the $\Omega_b$s derived from $Y_p$ and from D/H 
is a sign that SBBN is not valid 
\citep{C01,K01,B03a,B03b,C05} and that new physics are required.
Although He is not a sensitive baryometer ($Y_p$ depends only 
logarithmically on the baryon density), its primordial abundance depends much 
more sensitively on the expansion rate of the Universe and 
on a possible asymmetry between the numbers of neutrinos and anti-neutrinos in 
the early universe than D, $^3$He or $^7$Li. 
This, because of two reasons: 1) a faster expansion would leave less time for 
neutrons to convert into 
protons, and the resulting higher neutron abundance would result in a 
higher $Y_p$, and 2) $Y_p$ depends sensitively on the neutron to proton
ratio, which depends in turn on the numbers of electron neutrinos and 
anti-neutrinos. In that sense, He is both a chronometer 
and/or leptometer which is very sensitive to any small deviation from SBBN, 
and hence to new physics, much more 
so than the other three primordial light elements \citep{St06}. 

However, to detect small deviations from SBBN and 
make cosmological inferences, $Y_p$ has to be determined 
to a level of accuracy
of less than one percent. This is not an easy task. While it is relatively 
straightforward to derive the helium abundance
in an H {\sc ii} region with an accuracy of 10
percent if the spectrum is adequate,  gaining one order of magnitude
in the precision requires many conditions to be met.
First, the observational data has to be of excellent quality. 
This has been the concern of our group \citep{ITL94,ITL97,IT98b,IT04}. 
We have spent the last decade obtaining high signal-to-noise
spectroscopic data of low-metallicity extragalactic H {\sc ii} regions,
and our sample includes now a
total of 86 H {\sc ii} regions in 77 galaxies 
 \citep[see ][]{IT04}. 
This constitutes by far the
largest sample of high-quality data reduced in a homogeneous way 
to investigate the problem of the 
primordial helium abundance. To put things in perspective, 
the sample used in the pioneering work of 
\citet{PTP74,PTP76} comprised only 5 objects, with considerably larger 
observational errors. Later, the sample of \citet{P92}
included 36 objects and that of \citet{OS95} 49 objects. 

With such a large observational sample at hand, it is now the general 
consensus that the accuracy of the determination of the primordial He
abundance is limited presently, not so much 
by statistical uncertainties, but by our ability to account for systematic 
errors and biases.   
There are many known effects we need to correct for  
to transform the observed He {\sc i} line intensities into a He abundance. 
Neglecting or misestimating them may lead 
to systematic errors in the $Y_p$ determination that are larger than the 
statistical errors.
These effects are: (1) reddening, (2) underlying stellar
absorption in the He {\sc i} lines, (3) collisional 
excitation of the He {\sc i} lines which make their intensities 
 deviate from their recombination values, 
(4) fluorescence of the He {\sc i} lines which also make their intensities 
deviate from their 
recombination values, (5) collisional excitation of the hydrogen lines 
(hydrogen enters because the helium abundance is calculated relative to 
that of hydrogen), 
(6) possible departures from case B in the emissivities of H and He {\sc i} 
lines, (7) the temperature 
structure of the H {\sc ii} region and (8) its ionization structure. All
these corrections are at a level of a few percent 
except for effect (3) that can be much higher, exceeding 10\% in the case
of the He {\sc i} $\lambda$5876 emission line in hot 
and dense H {\sc ii} regions. All effects, 
apart from (8),
influence each other in a complicated way. At the present time, 
we are far from a
situation where all these corrections can be determined unambiguously
from observational constraints and theory. Given the
complexity of H {\sc ii} regions -- they have very perturbed morphologies,
possess complex velocity fields and a dust distribution difficult to quantify
-- the situation
is not likely to be improved considerably in the near future. 
Because of this, even if one were able to solve in an exact
manner the radiation transfer in the case of a simple geometry, 
there is still a long way between a simple model and a real
H {\sc ii} region. 

Effects (1), (3), (4) are relatively
easy to account for empirically from the available spectroscopic
information and have been already considered in the pioneering study of
\citet{PTP74}. Corrections (2) and (5) have been
discussed more recently \citep{R82,KS83,DK85}.
Effect (7) has been the subject of many papers by Peimbert and his 
coworkers. Elemental abundance determinations are affected by biases
in the presence of temperature inhomogeneities in 
the H {\sc ii} regions.
\citet{P67} and \citet{P02} have devised an empirical
method to correct for this bias. However, this method relies on
several assumptions that are not necessarily valid \citep[e.g. supposing that
the temperature fluctuation parameter is the same in the high and low
ionization regions, ][]{P02}. 
Furthermore, the source of temperature inhomogeneities at the
level invoked by Peimbert and colleagues remains unidentified
\citep{SS99,BL00,M95}, unless they are supposed to result 
from abundance inhomogeneities 
\citep[][ Stasi\'nska et al., in preparation]{TP05}.
This is also why the computation of temperature corrections from
photoionization models to derive He/H \citep[as proposed by ][]{SJ02},
is not necessarily relevant to real objects.
To check for possible temperature fluctuations, 
\citet{G06,G07} have compared the temperature in the 
O$^{2+}$ zone derived with the [O {\sc iii}] $\lambda$4363 line, with that in 
the H$^+$ zone derived with the Balmer and Paschen jumps in a large sample of 
hot low-metallicity H {\sc ii} regions used in the determination of $Y_p$.
They found that the two temperatures do not differ statistically, so that 
any temperature difference must be less than $\sim$ 5\%.   
Effect (8) has
first been corrected for using an empirical method \citep{PTP74}, and 
has been later the subject of many papers based on photoionization
modeling \citep{S80,Pe86,V00,B00,SJ02,G02}. Effect (6) has been discussed 
for H {\sc ii} regions by \citet{CF88} and \citet{HS92}, but 
has never been considered in the
various estimates of $Y_p$ thus far, case B having been assumed by all authors.

In the present determination of the primordial He abundance, 
we take into account all eight effects discussed  
above. We first use an empirical method to self-consistently
account for effects 1, 2, 3, 4, 5 and 7 in the derivation
of the abundance of He$^+$.
This method has been detailed in \citet{IT04}, where it has been applied to 
a small sample of 7
objects. It was shown that taking all these effects into
account in a systematic way can lead to helium
abundances that are significantly higher than those obtained by \citet{P02}
for the same objects (by up to 3\%). 
Here, we apply the same method, not only to our basic original sample of 
93 H {\sc ii} regions \citep{IT04} that we have 
assembled and observed ourselves (hereafter the HeBCD sample), but
also to a different and larger sample of 271 extragalactic  
H {\sc ii} regions selected from the Data Release 5 (DR5) of the 
Sloan Digital Sky Survey (hereafter the SDSS sample). 
We then considered effects 6 and 8 on the derivation of the abundance of 
He in each H {\sc ii} region and of the primordial He abundance.

We describe the two HeBCD and SDSS samples in \S\ref{sample}.
In \S\ref{method} we discuss the method used to derive He abundances in 
individual objects and the primordial He abundance from each sample.
In \S\ref{system}, we discuss the systematic effects considered.
Our new best values for $Y_p$ and the linear regression slope d$Y$/d$Z$ are
presented in \S\ref{primo}.
The cosmological implications of our new 
results are discussed in \S\ref{cosmo}.
We summarize our conclusions in \S\ref{summary}.

\section{THE SAMPLE}\label{sample}

\subsection{The HeBCD and SDSS subsamples}\label{sample1}

      We determine $Y_p$ and d$Y$/d$Z$ independently for  
two different samples. 
The HeBCD sample, is composed of
93 different observations of 86 H {\sc ii} regions in 77 galaxies.
The majority of these galaxies are low-metallicity BCD galaxies. This sample
is essentially the same as the one described in \citet{IT04}, except that 
two close pairs of H {\sc ii} regions in the spiral galaxy M101\footnote{
The coordinates of the two H {\sc ii} region pairs are: 
for the pair M101 No.1 + No.2, R.A.(J2000.0) = 14$^{\rm h}$04$^{\rm m}$29\fs5,
Dec (J2000.0) = +54\arcdeg23\arcmin47\arcsec\ and for the pair 
M101 No.3 + No.4,  
R.A.(J2000.0) = 14$^{\rm h}$03$^{\rm m}$01\fs2,
Dec (J2000.0) = +54\arcdeg14\arcmin29\arcsec.} have been added.
These four H {\sc ii} regions were observed with the 6.5m
MMT on the night of 2004 February 20. 
Observations were made with the Blue Channel of the MMT spectrograph. We used
a 2$''$$\times$300$''$ slit and a 300 grooves/mm grating in first order.
To remove the second-order contamination we use the L-38 blocking filter.
The above instrumental set-up gave a spatial scale along the slit of 0\farcs6
pixel$^{-1}$, a scale perpendicular to the slit of 1.96\AA\ pixel$^{-1}$,
a spectral range 3600 -- 7500\AA\ and a spectral resolution of 6\AA\ (FWHM).
The Kitt Peak IRS spectroscopic standard star HZ 44 was observed for flux
calibration. Spectra of He-Ar comparison arcs were obtained after each 
observation to calibrate the wavelength scale. 
The details of data reduction are described in \citet{TI05} and they are the
same as those for other H {\sc ii} regions from the HeBCD sample. 
The two-dimensional spectra were bias subtracted and flat-field corrected
using IRAF\footnote{IRAF is distributed by National Optical Astronomical 
Observatory, which is operated by the Association of Universities for 
Research in Astronomy, Inc., under cooperative agreement with the National 
Science Foundation.}. We then use the IRAF
software routines IDENTIFY, REIDENTIFY, FITCOORD, TRANSFORM to 
perform wavelength
calibration and correct for distortion and tilt for each frame. 
 Night sky subtraction was performed using the routine BACKGROUND. The level of
night sky emission was determined from the closest regions to the galaxy 
that are free of galaxian stellar and nebular line emission,
 as well as of emission from foreground and background sources.
One-dimensional spectra were then extracted from each two-dimensional 
frame using the APALL routine. Before extraction, distinct two-dimensional 
spectra of the same H {\sc ii} region
were carefully aligned using the spatial locations of the brightest part in
each spectrum, so that spectra were extracted at the same positions in all
subexposures. For all objects, we extracted the 
brightest part of the H {\sc ii} region. 
We have summed the individual spectra 
from each subexposure after manual removal of the cosmic rays hits. 
The spectra obtained from each subexposure
were also checked for cosmic rays hits at the location of strong 
emission lines, but none was found.

Particular attention was paid to the derivation of the sensitivity curve. 
It was obtained by 
fitting with a high-order polynomial the observed spectral energy 
distribution of the standard star HZ 44. 
Because the spectrum of HZ 44 has only a small number of a 
relatively weak absorption features, its spectral energy distribution is 
known with good accuracy of $\la$ 1\% \citep{O90}. 
Moreover, the response function of the CCD detector is smooth, so we could
derive a sensitivity curve with an accuracy better than 1\% over the
whole optical range. 

We show in Tables \ref{tab5} and \ref{tab6} (available only in electronic 
form), the emission line fluxes and the equivalent widths for
these H {\sc ii} regions, along with those of the remaining H {\sc ii} regions
in the HeBCD sample. They were measured using the IRAF SPLOT routine.
The line flux errors listed include statistical errors derived with
SPLOT from non-flux calibrated spectra, in addition to errors of 1\% 
of the line fluxes introduced in the standard star absolute flux calibration.
The line flux errors will be later propagated into the 
calculation of abundance errors.

The number of data points (93) is larger than the number of H {\sc ii}
regions (86) because several H {\sc ii} regions have independent observations
from different telescopes. We treat these independent observations as separate
data points in our least-square fitting.
In assembling the above sample, we have taken care not to include 
objects that, for one reason or another, are not appropriate 
for He abundance determination, as described in \citet{ITL94,ITL97}
and \citet{IT98a,IT98b}.
 In particular, the NW component of I Zw 18 has not been included 
because its He {\sc i} $\lambda$5876 emission line is strongly affected by the
Galactic interstellar Na {\sc i} $\lambda$5889,5895 absorption line.

The SDSS sample, which we will use as a comparison sample, 
is composed of 271 low-metallicity 
H {\sc ii} regions selected from the SDSS DR5. 
The SDSS \citep{Y00} offers a gigantic
data base of galaxies with well-defined selection criteria and observed in a
homogeneous way. In addition, the spectral resolution is much better than that
of most previous data bases on emission-line galaxies including all the 
spectra in the HeBCD sample.
It is possible to extract from the SDSS data base a sample of emission-line 
galaxies with well-defined criteria. Despite the lower quality of its data, 
this large comparison sample, observed and reduced 
in a different way, will allow us to
check for possible systematic shifts in the linear regressions 
introduced by different data sets. Its large size will also allow to check
for other systematic effects.

The SDSS DR5 provides spectra of some 800\,000 galaxies, quasars and stars 
selected over 
a sky area of 4783 square degrees, in the wavelength 
range $\sim$ 3800 -- 9200\AA, along 
with tables of measured parameters from these data.
From this data base accessible from the SDSS web page 
(http://www.sdss.org/dr5),
we have extracted flux-calibrated spectra of a total of  
271 H {\sc ii} regions which satisfy the
following selection criteria: 
1) the [O {\sc iii}] $\lambda$4363 is detected at the level of
$>$ 2$\sigma$ above the noise, allowing a direct heavy element abundance 
determination by the $T_e$ method, as was done for the 
H {\sc ii} regions in the 
HeBCD sample; 2) the equivalent width of the H$\beta$ emission line 
EW(H$\beta$) is $\geq$ 50\AA. 
This ensures that the effect of underlying He {\sc i} stellar absorption on 
the derived
He abundances is small; 3) the observed flux of the H$\beta$ emission line 
$F$(H$\beta$) is $\geq$ 10$^{-14}$ erg s$^{-1}$ cm$^{-2}$. 
This condition ensures that the SDSS sample contains only 
bright H {\sc ii} regions with strong emission lines, the fluxes of which 
can be measured accurately.  
Since the emission line [O {\sc ii}] $\lambda$3727 in
the SDSS spectra of galaxies with $z$ $\la$ 0.02 is out of the observed 
spectral range, we have estimated its flux from the
[O {\sc ii}] $\lambda$7320, 7330 emission line fluxes.
The measurements of the emission line fluxes and the determination of the
element abundances were done following the same procedures 
 as for our basic HeBCD sample.

The observed emission line fluxes $F(\lambda)$/$F$(H$\beta$) and their 
equivalent widths EW($\lambda$) for all the objects in the two samples 
are shown 
respectively in Tables \ref{tab5} and \ref{tab6}. Because of their large 
sizes, these tables are available only in electronic form.

\subsection{Statistical errors}\label{sample2}

Emission line fluxes of all spectra, in both the HeBCD and SDSS 
samples, were measured using Gaussian fitting
with the IRAF routine SPLOT.
The statistical errors of emission lines in the spectra of the HeBCD
sample are calculated using the photon statistics of the lines in the 
non-flux calibrated spectra. As for the SDSS sample, each spectrum in the
SDSS data base is supplemented by a file with the flux error in each pixel
which can be used to calculate the errors of the emission lines in 
each spectra.
The mean statistical errors are $\sim$ 4\% for the brightest 
He {\sc i} $\lambda$5876 emission line, $\sim$ 10\% for the He {\sc i} 
$\lambda$4471 and $\lambda$6678 emission lines, and $\sim$ 13\% for the 
He {\sc i} $\lambda$ 7065 emission line, for the combined HeBCD + SDSS sample. 
As the mean statistical error of the
weighted mean He abundance is primarily 
determined by the statistical error of the intensity of the brightest 
He {\sc i} $\lambda$5876 emission line, we expect it to be $\sim$ 3-4\%, 
which is the case (see Table \ref{tab2}).

\begin{figure*}
\figurenum{1}
\epsscale{1.1}
\plotfiddle{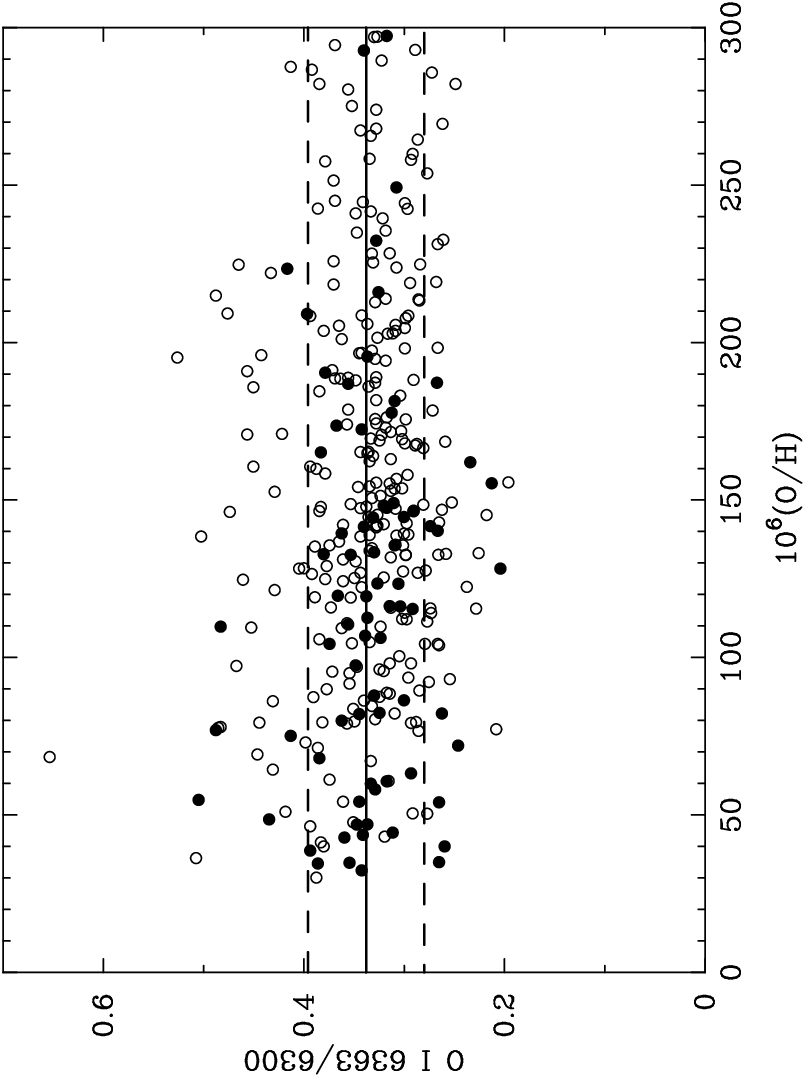}{1pt}{-90}{320}{420}{0.}{0.}
\figcaption{ [O {\sc i}] $\lambda$6363/$\lambda$6300 emission line flux
ratio vs oxygen abundance. The HeBCD and SDSS galaxies are shown respectively
by filled and open circles. The solid line shows the mean value of the
[O {\sc i}] $\lambda$6363/$\lambda$6300 emission line flux ratio and the dashed
lines 1$\sigma$ deviations from the mean.
\label{fig1}}
\end{figure*}

To check the reliability of error determinations for the 
emission lines in the spectra of our sample, we have considered
the flux ratios of the two [O {\sc i}] $\lambda$6363 and 
[O {\sc i}] $\lambda$6300 lines.
These flux ratios are plotted against oxygen abundance in Fig. \ref{fig1} 
where objects in the
HeBCD sample are shown by filled circles, and those in the SDSS sample
by open circles. The solid line shows
the mean value of the flux ratio for the combined HeBCD + SDSS sample, and the 
dashed lines represent 
1$\sigma$ dispersions, $\sim$ 17\% of the mean value. Since
the theoretical value of the [O {\sc i}] $\lambda$6363/$\lambda$6300 ratio is
constant (its value is $\sim$ 1/3)
 and independent of physical conditions in the H {\sc ii} region,
the dispersion of the points around the mean in Fig. \ref{fig1} 
gives a good representation of the observational uncertainties.
Typically, the flux of [O {\sc i}] $\lambda$6300 is comparable to the flux
of He {\sc i} $\lambda$4471 and $\lambda$6678 emission lines, 
with statistical errors of $\sim$ 10\%. The 
[O {\sc i}] $\lambda$6363 is $\sim$ 3 times weaker than 
the [O {\sc i}] $\lambda$6300 line, so its error is $\sim$ 17\%. 
 This is just the 1$\sigma$ scatter 
found in Fig. \ref{fig1} for the [O {\sc i}] line flux ratio.
We conclude that our estimates of the  
statistical errors for weak emission lines are reliable.

The statistical errors in the emission lines intensities are then 
propagated in the calculation of errors in   
electron temperatures, electron number densities and 
element abundances.

\section{THE METHOD}\label{method}

\subsection{Linear regressions}\label{method1}

As in 
our previous work \citep[see ][ and references therein]{IT04},
we determine the primordial He mass fraction
$Y_p$ by fitting the data points in the $Y$ -- O/H
and $Y$ -- N/H planes with linear regression lines of the
form \citep{PTP74,PTP76,P92}
\begin{equation}
Y = Y_p + \frac{{\rm d}Y}{{\rm d}({\rm O/H})} ({\rm O/H}),               \label{eq:YvsO}
\end{equation}
\begin{equation}
Y = Y_p + \frac{{\rm d}Y}{{\rm d}({\rm N/H})} ({\rm N/H}),               \label{eq:YvsN}
\end{equation}
where
\begin{equation}
Y=\frac{4y(1-Z)}{1+4y} \label{eq:Y}
\end{equation}
is the He mass fraction, $Z$ is the heavy element mass fraction,
$y$ = ($y^+$ + $y^{2+}$)$\times$$ICF$(He$^+$+He$^{2+}$) is the He 
abundance,
$y^+$ $\equiv$ He$^+$/H$^+$ and $y^{2+}$ $\equiv$ He$^{2+}$/H$^+$ are 
respectively the abundances of singly and doubly ionized He, and 
$ICF$(He$^+$+He$^{2+}$) is the ionization correction factor for He.
We have assumed $Z$ = 18.2(O/H) in Eq. \ref{eq:Y} which holds for a
metallicity $Z$=0.001 \citep{M92}.

The assumption of a linear dependence of $Y$ on O/H and N/H appears to be
reasonable as there are no evident non-linear trends
in the distributions of the data points in the $Y$ vs O/H and $Y$ vs N/H
diagrams \citep[e.g., ][]{IT04}. 
The linear regressions (Eqs. \ref{eq:YvsO} and \ref{eq:YvsN})
imply that the initial mass function (IMF) averaged stellar yields for 
different elements do not
depend on metallicity. This is the case if the stellar IMF is independent 
of metallicity. It has been suggested in the past \citep[e.g., ][]{B83} that,
at low metallicities,
the IMF may be top-heavy, i.e. there are relatively more massive stars 
as compared to lower mass stars than at high metallicities. 
If this is the case, then 
the IMF-averaged 
yields would be significantly different for low-metallicity 
stars as compared to those of more 
metal-enriched stars, resulting in a non-linear relationship between $Y$ 
and O/H or N/H 
\citep{SF03}. However, until now, there has not been persuasive 
evidence for a metallicity dependence of the IMF. Furthermore, 
the properties of
extremely metal-deficient stars remain poorly known, 
excluding quantitative estimates
of possible non-linear effects in the $Y$ -- O/H and $Y$ -- N/H
relations. Therefore, in the following analysis, we will continue to 
use linear
regressions (Eqs. \ref{eq:YvsO} and \ref{eq:YvsN}) to fit the data.

The slopes of the $Y$ -- O/H and $Y$ -- N/H linear regressions can be
written as:
\begin{equation}
\frac{{\rm d}Y}{{\rm d}({\rm O/H})} = 12\frac{{\rm d}Y}{{\rm d}{\rm O}} =
18.2\frac{{\rm d}Y}{{\rm d}Z}, \label{eq:dO}
\end{equation}
\begin{equation}
\frac{{\rm d}Y}{{\rm d}({\rm N/H})} = 10.5\frac{{\rm d}Y}{{\rm d}{\rm N}} =
564\frac{{\rm d}Y}{{\rm d}Z}, \label{eq:dN}
\end{equation}
where O, N and $Z$ are respectively the mass fractions of oxygen, nitrogen
and heavy elements. We have assumed that O = 0.66$Z$ \citep{M92} 
which holds for a
metallicity $Z$=0.001, an IMF slope $x$=1.35 (where $x$ is defined by
${\rm d}N/{\rm d}(\log M)$ $\propto$ $M^x$) and 
$\log$(N/H) -- $\log$(O/H) = --1.55 \citep{TIL95}.

To derive the parameters of the linear regressions,
 we use the maximum-likelihood method \citep{Pr92}
 which takes into account the errors in 
$Y$, O/H and N/H for each object.

\subsection{He$^+$ emissivities}\label{method2}

The derived He$^+$ abundance $y^+$ depends on the adopted He {\sc i} line
emissivities. We consider two sets of He {\sc i} emissivities: the old ones by 
\citet{B99,B02} 
which were used by \citet{IT04} [\citet{B02} take into account both 
collisional and fluorescent 
enhancements] and the new ones by \citet{P05} and \citet{B05}, 
which have been computed using improved radiative and collisional data. 
Following \citet{ITL94,ITL97} and \citet{IT98b}, we use the five 
strongest He {\sc i} $\lambda$3889, $\lambda$4471, 
$\lambda$5876, $\lambda$6678 and $\lambda$7065 emission lines to derive 
$N_e$(He$^+$) and $\tau$($\lambda$3889). \citet{B05} have estimated the
accuracy of new emissivities in the low-density limit
and found that accuracy is better than 1\% 
for He {\sc i} $\lambda$4471, $\lambda$5876, $\lambda$6678 and $\lambda$7065 
emission lines, but it is not as good for the He {\sc i} $\lambda$3889
emission line.
The He {\sc i} $\lambda$3889 and 
$\lambda$7065 lines play an important role because they are particularly 
sensitive to both quantities. Since the
He {\sc i} $\lambda$3889 line is blended with the H8 $\lambda$3889 line, 
we have subtracted the latter, assuming its intensity to be equal to 0.107 
$I$(H$\beta$) \citep{A84}. 
In our spectra, other He {\sc i} emission lines are seen, most often
He {\sc i} $\lambda$3820, $\lambda$4387, $\lambda$4026, $\lambda$4921,
$\lambda$7281. However, we do not attempt to use these lines for He 
abundance determination because they are much weaker as compared to the five 
brightest lines, and hence have larger uncertainties. 

We have used the simple and convenient fits 
provided by  \citet{B02} to
calculate $y^+$s with the \citet{B99} emissivities.
For the \citet{P05} emissivities, 
we have assumed the functional dependence of these emissivities on 
$N_e$ and $\tau$(He {\sc i} $\lambda$3889), 
which results from the collisional and fluorescence enhancements,
to be the same as the one for the \citet{B02} emissivities.
Eqs. \ref{y3889} -- \ref{y7065} give 
the linear fits we have adopted for the ratios of the two sets of
emissivities in the temperature range $T_e$= 10$^4$ -- 2$\times$10$^4$
and an electron number density 
$N_e$ = 100 cm$^{-3}$, 
for each of the five lines:
\begin{equation}
y^+_P(\lambda 3889) = y^+_B(\lambda 3889)/(1.079-0.052\times t),
\label{y3889}
\end{equation}
\begin{equation}
y^+_P(\lambda 4471) = y^+_B(\lambda 4471)/(1.020-0.026\times t),
\label{y4471}
\end{equation}
\begin{equation}
y^+_P(\lambda 5876) = y^+_B(\lambda 5876)/(0.956+0.011\times t),
\label{y5876}
\end{equation}
\begin{equation}
y^+_P(\lambda 6678) = y^+_B(\lambda 6678)/(0.938+0.028\times t),
\label{y6678}
\end{equation}
\begin{equation}
y^+_P(\lambda 7065) = y^+_B(\lambda 7065)/(1.051-0.040\times t),
\label{y7065}
\end{equation}
where $y^+_B$ and $y^+_P$ are respectively 
He$^+$ abundances calculated with \citet{B99} (B)
and \citet{P05} (P) emissivities and $t$ = 10$^{-4}$ $T_e$. 
We found that, with \citet{P05} emissivities, 
He$^+$ abundances are higher than those derived
with \citet{B99} emissivities by about 0 -- 2 percent for the $\lambda$4471
line, and by about 5 -- 6 percent for the $\lambda$5876 line, in the range
$t$ = 1 -- 2.

\subsection{A Monte Carlo algorithm for determining 
the best value of $y^+$}\label{method3}

In addition to the emissivities,
 the derived $y^+$ abundances depend also on a number of  
other parameters: the fraction $\Delta$$I$(H$\alpha$)/$I$(H$\alpha$) of the
H$\alpha$ emission line flux due to collisional excitation, the electron
number density $N_e$(He$^+$), the electron temperature $T_e$(He$^+$), 
the equivalent
widths EW$_{abs}$($\lambda$3889), EW$_{abs}$($\lambda$4471), 
EW$_{abs}$($\lambda$5876), 
EW$_{abs}$($\lambda$6678) and EW$_{abs}$($\lambda$7065) of He {\sc i} stellar 
absorption lines, and the optical depth 
$\tau$($\lambda$3889) of the He {\sc i} $\lambda$3889 emission line. 
To determine the best value of $y^+_{wm}$ (defined in Eq. \ref{eq2}), 
we use the Monte Carlo procedure described in \citet{IT04}, 
randomly varying each of the above parameters within a specified range. 
First, we take into account collisional excitation effects for hydrogen. 
The value of the fraction of the H$\alpha$ flux due to collisional  
excitation is randomly generated 100 times within an 
adopted range. The fraction of
the H$\beta$ emission line flux due to the collisional excitation
is adopted to be three times less than that of the H$\alpha$ flux. 
For each generated fraction, the fluxes of the H$\alpha$ and H$\beta$ lines
due to the collisional excitation are subtracted from the total observed
fluxes and then all emission line fluxes are corrected for interstellar
extinction and element abundances are calculated.

To calculate $y^+$ we vary simultaneously and randomly $N_e$(He$^+$),
$T_e$(He$^+$) and $\tau$($\lambda$3889) within their respective 
adopted ranges. We make a 
total of 10$^5$ such realizations for every H {\sc ii} region, for
a given fraction of the H$\alpha$ emission line flux due to 
collisional excitation.
Thus, the total
number of Monte Carlo realizations we have performed 
for each H {\sc ii} region is 
100 $\times$ 10$^5$ = 10$^7$.
As for the He {\sc i} underlying stellar absorption, we assume 
fixed values for EW$_{abs}$($\lambda$4471), chosen to be 
between 0 and 0.5\AA, and for the  
EW$_{abs}$($\lambda$3889)/EW$_{abs}$($\lambda$4471),
EW$_{abs}$($\lambda$5876)/EW$_{abs}$($\lambda$4471),
EW$_{abs}$($\lambda$6678)/EW$_{abs}$($\lambda$4471),
and EW$_{abs}$($\lambda$7065)/EW$_{abs}$($\lambda$4471) ratios.

For each H {\sc ii} region, we find the best solution for $y^+_{wm}$
 in 
the multi-parameter space defined above by 
minimizing the quantity
\begin{equation}
\chi^2=\sum_i^n\frac{(y^+_i-y^+_{wm})^2}{\sigma^2(y^+_i)}\label{eq1},
\end{equation}
where $y^+_i$ is the He$^+$ abundance derived from the flux of the He {\sc i}
emission line labeled $i$, and $\sigma(y^+_i)$ is the statistical error
of $y^+_i$. The quantity $y^+_{wm}$ is the weighted 
mean of the He$^+$
abundance as given by the equation
\begin{equation}
y^+_{wm}=\frac{\sum_i^k{y^+_i/\sigma^2(y^+_i)}}
{\sum_i^k{1/\sigma^2(y^+_i)}}\label{eq2}.
\end{equation}
We use all five He {\sc i} emission lines to calculate $\chi^2$ (i.e.,
$n$ = 5), but only three lines, He {\sc i} $\lambda$4471,
$\lambda$5876 and $\lambda$6678 to compute
$y^+_{wm}$ ($k$ = 3). This is because the fluxes of 
He {\sc i} $\lambda$3889 and $\lambda$7065 emission lines are
more uncertain as compared to the other three He {\sc i} emission lines.

An estimate of the 1$\sigma_{\rm sys}$
systematic error of $y^+_{wm}$ can 
be obtained from $\Delta\chi^2$. This quantity depends on the number of 
degrees of freedom in the problem which is the difference 
between the number of observational constraints, equal to 5 
(the five He {\sc i} emission line fluxes), and the number of the 
free parameters, equal to 4 
[the fraction $\Delta$$I$(H$\alpha$)/$I$(H$\alpha$) due to collisional
excitation of the H$\alpha$ emission line, $T_e$(He$^+$), $N_e$(He$^+$)
and $\tau$($\lambda$3889)]. 
Since the number of degrees of freedom is 5 -- 4 = 1, 
then $\Delta\chi^2 = 1$ \citep{Pr92}. Thus the systematic error 
$\sigma_{\rm sys}$ is the 1$\sigma$ dispersion of the computed values of 
$y^+_{wm}$ 
for solutions with $\chi^2$ between $\chi^2_{\rm min}$ and
$\chi^2_{\rm min}$ + 1.
The total error for $y^+_{wm}$ is then 
given by 
$\sigma^2_{\rm tot}$ = $\sigma^2_{\rm stat}$ + $\sigma^2_{\rm sys}$.

Additionally, in those cases when the nebular He {\sc ii} $\lambda$4686 
emission line was detected, we have added the abundance of doubly ionized 
helium $y^{2+}$ $\equiv$ He$^{2+}$/H$^+$ to $y^+$. Although the He$^{2+}$ zone
is hotter than the He$^{+}$ zone, we 
have adopted $T_e$(He$^{2+}$) = $T_e$(He$^{+}$).
The last assumption has only a minor effect on the $y$ value, because
 $y$$^{2+}$ is small ($\leq$ 3\% of $y^+$) in all cases.

\subsection{A basic set of parameters for Monte Carlo calculations}\label{method4}

In order to study the dependence of $Y_p$ on the various 
parameters characterizing 
different systematic effects, we define a reference set of parameters 
which we will call the ``basic set''. We wish to see how, by 
changing different parameters, the $y$ values change, as compared to those 
corresponding to the basic set. The basic set of parameters is defined in the 
following way : 1) the electron temperature
of the He$^+$ zone is varied in the range $T_e$(He$^+$) =
(0.95 -- 1.0)$\times$$T_e$(O {\sc iii}); 2) oxygen and nitrogen abundances 
are calculated adopting an electron temperature equal to $T_e$(O {\sc iii}); 
3) $N_e$(He$^+$) and $\tau$($\lambda$3889) vary 
respectively in the ranges 10 -- 450 cm$^{-3}$ and 0 -- 5; 4) 
the fraction of H$\alpha$ emission due to collisional excitation is
varied in the range 0\% -- 5\%; 5) the equivalent width of the 
He {\sc i} $\lambda$4471 absorption line is fixed to 
EW$_{abs}$($\lambda$4471) = 0.4\AA; 6) the equivalent 
widths of the other absorption lines are fixed according to the ratios  
EW$_{abs}$($\lambda$3889)/EW$_{abs}$($\lambda$4471) = 1.0,
EW$_{abs}$($\lambda$5876)/EW$_{abs}$($\lambda$4471) = 0.3,
EW$_{abs}$($\lambda$6678)/EW$_{abs}$($\lambda$4471) = 0.1 and 
EW$_{abs}$($\lambda$7065)/EW$_{abs}$($\lambda$4471) = 0.1. 
The justification of the values of these ratios 
is given in \citet{IT04} and in \S\ref{system4}.
The ionization correction factor $ICF$(He$^+$+He$^{2+}$) is 
discussed in \S\ref{system6}.

\section{SYSTEMATIC EFFECTS}\label{system}

We will consider successively the following systematic effects in the 
derivation of He abundances: 
1) He {\sc i} emissivities; 2) reddening;
3) the temperature structure of the H {\sc ii} region,
i.e. the temperature difference between $T_e$(He$^+$)
and $T_e$(O {\sc iii}); 4) underlying stellar He {\sc i} absorption; 
5) collisional excitation of
hydrogen lines;  6) the ionization structure of the H {\sc ii}
region; and 7) the deviation of hydrogen and He {\sc i} recombination 
line fluxes from case B. At the very end, we will 
also consider possible biases introduced by different samples 
not observed and reduced in the same way by deriving $Y_p$ independently 
for the HeBCD and SDSS samples and comparing the results. 
However, we will not discuss here collisional and fluorescent enhancements
of the He {\sc i} emission lines. These were discussed in detail by e.g.
\citet{B99,B02}. We only note that the importance of these two effects 
depends on the specific He {\sc i} emission line considered. In particular, 
the correction for collisional excitation of the He {\sc i} $\lambda$5876 
emission line can
decrease its flux by as much as $\sim$17\%, depending on
$T_e$(He$^+$) and $N_e$(He$^+$). In the following, we apply the corrections
for collisional and fluorescent enhancements, adopting the most recent
analytical fits of \citet{B99,B02}.

Effects 1 -- 6 have been considered in many previous papers on the 
determination of the 
primordial He abundance. However, 
previous analyses were typically based on small samples and/or 
not all systematic effects were considered. By contrast, the present study 
has two major advantages: 
1) it is based on the largest sample of H {\sc ii} regions ever 
assembled for the determination of $Y_p$; and 2) 
all important known systematic effects are taken into account 
using a Monte-Carlo approach.

\subsection{He {\sc i} emissivities}\label{system1}

First, we consider the difference in the primordial He mass fraction $Y_p$
caused by using two different sets of He {\sc i} line emissivities. 
We adopt the basic set of parameters from \S\ref{method4} except for:  
a) the range of $T_e$(He$^+$) variations, which we adopt here to be 
(0.9 -- 1.0)$\times$$T_e$(O {\sc iii}), the same as that used by
\citet{IT04}, and b) the oxygen and nitrogen abundances which are calculated 
with the electron temperature set to $T_e$(He$^+$).
There are two differences between the procedures of \citet{IT04} 
and the ones used here: 
1) the average $y^+$ is calculated using only three He {\sc i} 
emission lines while the computation of 
$\chi^2$ takes into account all five lines. \citet{IT04} 
included all five lines for the calculation of both 
$y^+$ and $\chi^2$; 2) \citet{IT04}
adopted EW$_{abs}$(H7 + $\lambda$3889) = 3.0\AA\ instead
of the relation EW$_{abs}$($\lambda$3889) = 
EW$_{abs}$($\lambda$4471) adopted here.
However, we will show later in this section that the variations of the 
EW$_{abs}$($\lambda$3889) have little effect on the derived $Y_p$.

\begin{figure*}
\figurenum{2}
\epsscale{1.1}
\plottwo{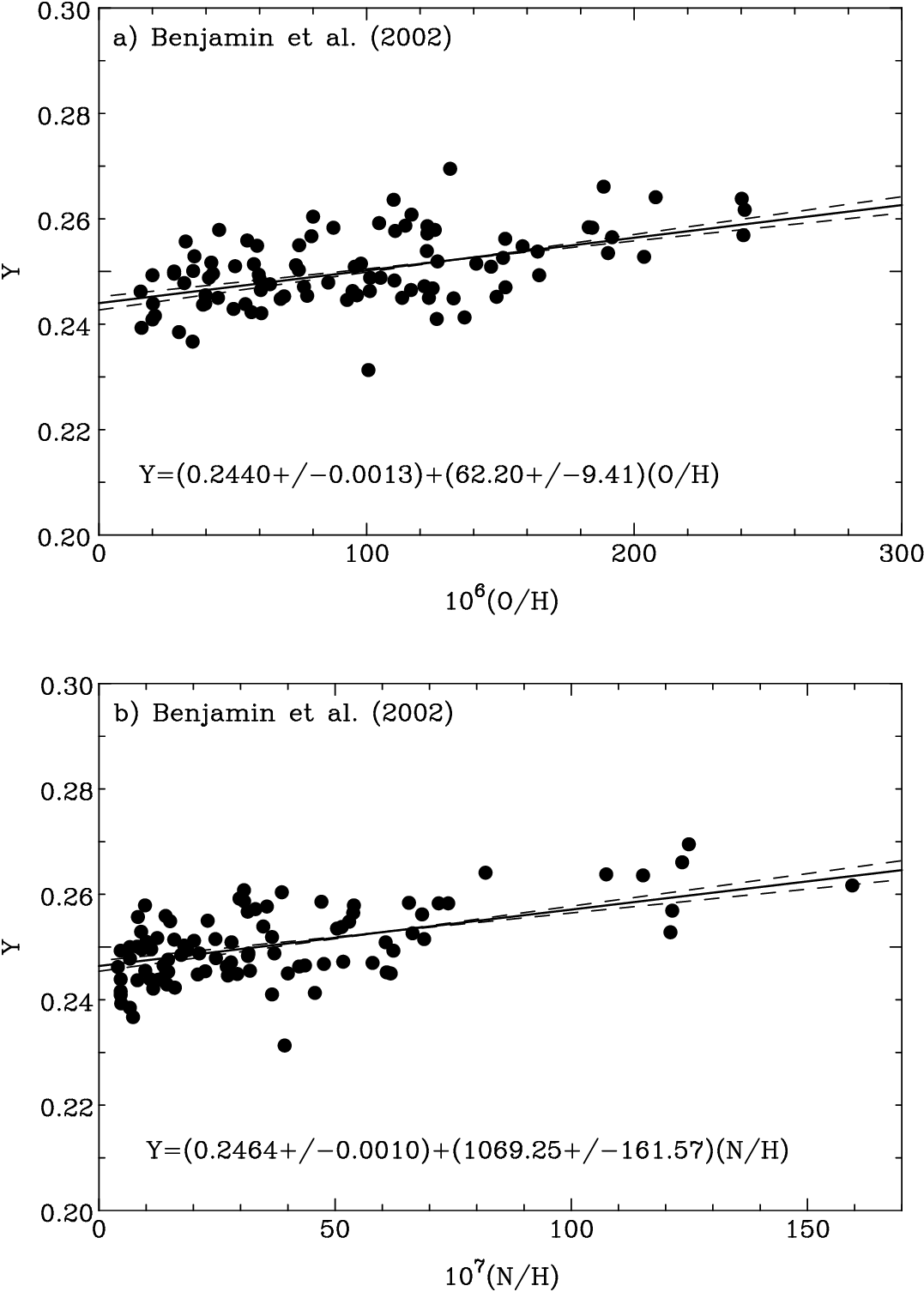}{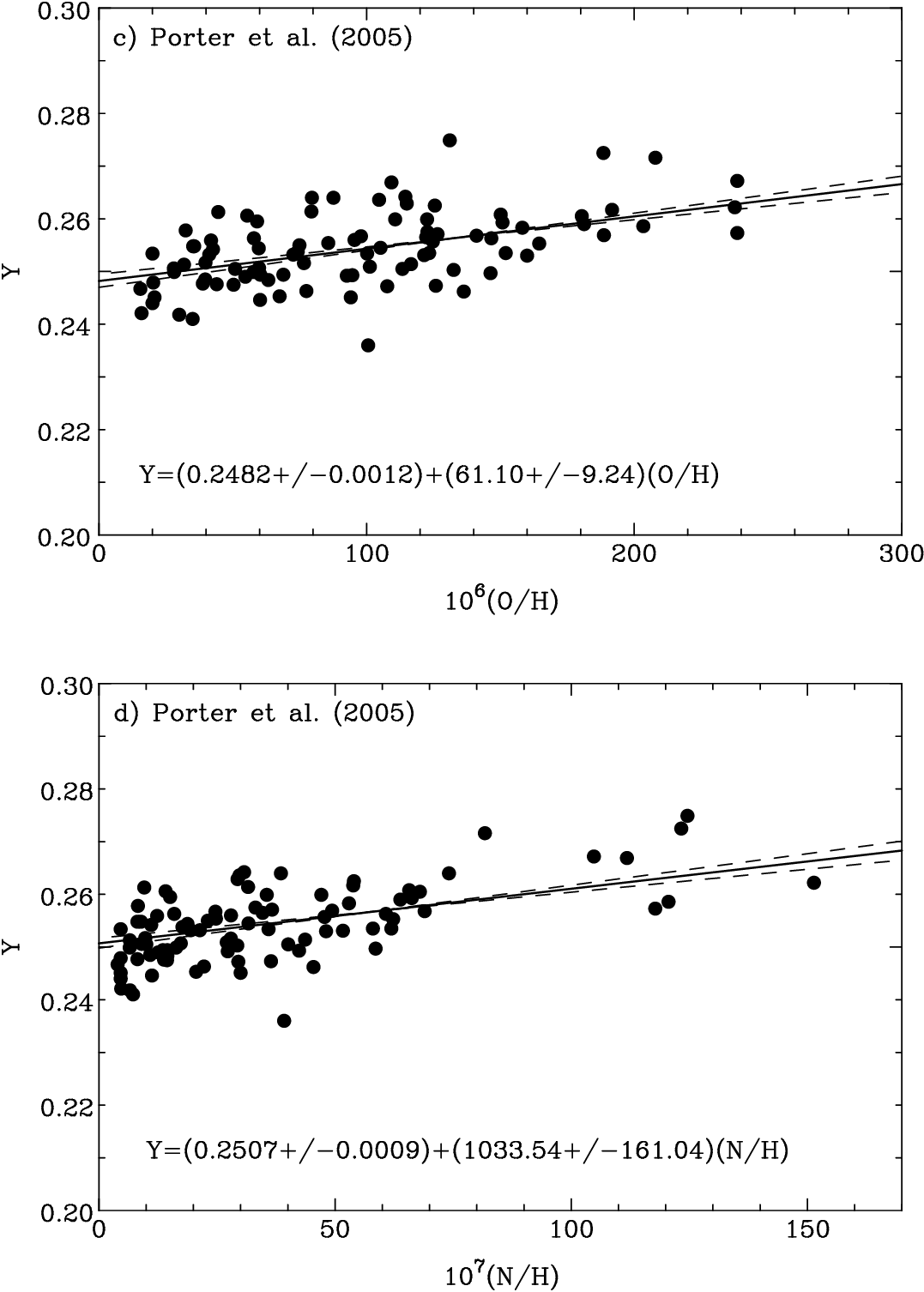}
\figcaption{  Linear regressions of the helium mass fraction $Y$ vs. oxygen 
and nitrogen abundances for H {\sc ii} regions in the HeBCD sample. 
The He {\sc i} emissivities
in (a) and (b) are from \citet{B99,B02} and in (c) and (d) from 
\citet{P05}.
In all panels, $Y$ was derived by minimizing $\chi^2$ and adopting
EW$_{abs}$($\lambda$4471) = 0.4 \AA, EW$_{abs}$($\lambda$3889) 
= EW$_{abs}$($\lambda$4471), EW$_{abs}$($\lambda$5876) = 
0.3$\times$EW$_{abs}$($\lambda$4471), EW$_{abs}$($\lambda$6678) = 
EW$_{abs}$($\lambda$7065) =0.1$\times$EW$_{abs}$($\lambda$4471).
The electron temperature 
$T_e$(He$^+$) is varied in the range (0.90 -- 1.00)$\times$$T_e$(O {\sc iii}).
The oxygen and nitrogen abundances are calculated setting the electron
temperature equal to $T_e$(O {\sc iii}).
\label{fig2}}
\end{figure*}

In Fig. \ref{fig2}a and \ref{fig2}b we show the
linear regressions $Y$ -- O/H and $Y$ -- N/H for the HeBCD sample where the 
values of $Y$ are calculated with the \citet{B02} He {\sc i} emissivities.
From these regressions
we derive $Y_p$ = 0.2440 $\pm$ 0.0013 and 0.2464 $\pm$ 0.0010.
If EW$_{abs}$(H7 + $\lambda$3889) is set to 3.0\AA\ instead
of adopting the relation 
EW($\lambda$3889) = EW($\lambda$4471),
then $Y_p$ = 0.2435 $\pm$ 0.0013 and 0.2462 $\pm$ 0.0010
respectively for the $Y$ -- O/H and $Y$ -- N/H regressions.
These values are in agreement with $Y_p$ = 0.2421 $\pm$ 0.0020 and 
0.2446 $\pm$ 0.0016 obtained by \citet{IT04} for their sample of 7 H {\sc ii}
regions. Note that the value of 
$Y_p$ derived from the $Y$ -- N/H regression is always
slightly greater than the one derived from the $Y$ -- O/H regression. This is
because the N/O abundance ratio tends to increase with increasing
oxygen abundance \citep[e.g. ][]{IT99,I06}. 

Figs. \ref{fig2}c and \ref{fig2}d show the linear regressions $Y$ -- O/H
and $Y$ -- N/H when the \citet{P05} He {\sc i} emissivities are used.
We obtain $Y_p$ = 0.2482 $\pm$ 0.0012 and 0.2507 $\pm$ 0.0009
from these regressions.
It is seen that the use of the new emissivities 
increases $Y_p$ by $\sim$ 1.7\%.

In the analysis of the other systematic effects,
 we will consider mainly helium abundances 
obtained with the \citet{P05} emissivities, with only occasional 
mention of helium abundances obtained
with the \citet{B02} emissivities for comparison.

\subsection{Reddening}\label{system2}

\citet{ITL94,ITL97} and \citet{IT98a,IT04} have used the \citet{W58} reddening
curve with $R_V$ = $A_V$/$E(B-V)$ = 3.2 to correct the line 
intensities for extinction. 
The derived value of $Y_p$ should not be
very sensitive to the choice of a particular reddening curve because 
of the way the extinction correction is applied: 
 the fluxes of the Balmer lines are corrected for extinction 
so that the resulting flux ratios correspond to 
the theoretical values, independently of the adopted 
reddening law.
Then, the corrected fluxes of the other emission lines are not
sensitive to the particular reddening law as well.

To check the sensitivity of $Y_p$ on the adopted reddening law, we have 
also considered the \citet{C89} 
reddening curves with $R_V$ = 2.5, 3.2 and 4.0. 
We find that the $Y_p$ derived
with the \citet{C89} reddening curve and $R_V$ = 3.2 is only $\sim$ 0.3\%
higher than the value derived with the \citet{W58} reddening curve. The $Y_p$
derived with $R_V$ = 2.5 is $\sim$ 0.4\% lower than the one with 
$R_V$ = 3.2, while the $Y_p$ derived with $R_V$ = 4.0 is $\sim$ 0.2\% higher.
The effect of various reddening laws on the derived $Y_p$ is indeed 
small. Therefore, for consistency with our previous work, 
we have adopted the \citet{W58} reddening law with
$R_V$ = 3.2, the same as the one used by \citet{ITL94,ITL97} and 
\citet{IT98a,IT04}.

\subsection{Temperature structure}\label{system3}


To derive the He abundances, \citet{ITL94,ITL97} and \citet{IT98b} 
have assumed 
that the temperatures $T_e$(He$^+$) and $T_e$(O {\sc iii}), averaged over
the whole H {\sc ii} region, are equal. $T_e$(O {\sc iii}) is 
determined from the observed 
[O {\sc iii}]$\lambda$4363/($\lambda$4959 + $\lambda$5007) emission line flux 
ratio. However, because of the high sensitivity of the flux of the auroral 
[O {\sc iii}]$\lambda$4363 emission line to temperature, 
$T_e$(O {\sc iii}) tends to be characteristic of the hotter zones in the 
H {\sc ii} region, and may be expected to be higher 
than $T_e$(He$^+$). 
On the other hand, $T_e$(H$^+$) and $T_e$(He$^+$) are  
derived from the recombination spectrum of ionized hydrogen and helium, 
which is not very sensitive to temperature. Therefore, the assumption 
$T_e$(He$^+$) = $T_e$(H$^+$), which we adopt here, seems 
reasonable, the H$^+$ and He$^+$ zones in high-excitation H {\sc ii} 
regions being nearly coincident.

To account for the difference between 
$T_e$(O {\sc iii}) and $T_e$(He$^+$),
\citet{P67} has  
developed a formalism based on the average temperature $T_0$ and 
the mean square temperature
variation $t^2$ in an H {\sc ii} region. Then 
$T_e$(He$^+$) and $T_e$(O {\sc iii})
can be expressed as functions
of $T_0$ and $t^2$, with $T_e$(O {\sc iii}) $\geq$ 
$T_e$(He$^+$). This approach has been applied 
by \citet{P02} for the determination of the He abundance in several  
low-metallicity dwarf galaxies. 
They find that the difference between
$T_e$(O {\sc iii}) and $T_e$(He$^+$) results in a reduction of the
He mass fraction by 2 -- 3 percent as compared to the case with 
$T_e$(O {\sc iii}) = $T_e$(He$^+$).

Until very recently, no direct measurement of
$T_e$(H$^+$) has been carried out for extremely 
metal-deficient H {\sc ii}
regions, those which carry the  
most weight in the determination of $Y_p$. All 
existing measurements of $T_e$(H$^+$) based on  
the Balmer jump or the Paschen jump,
have been done for relatively metal-rich H {\sc ii} regions. The 
lowest-metallicity H {\sc ii} region for which 
such a measurement was performed is Mrk 71 with  
12 + log O/H $\sim$ 7.9 \citep{G94}, $\sim$ 5 -- 6 times more 
metal-rich as compared to the lowest-metallicity H {\sc ii} regions in our
sample. Other measurements of $T_e$(H$^+$) have been done for 30 Dor
in the Large Magellanic Cloud \citep[12 + log O/H = 8.3, ][]{P03} and some
nearly solar-metallicity H {\sc ii} regions in our Galaxy 
\citep{P93,P00,E98,G04,G05}.

Using the Balmer and Paschen jumps, 
\citet{G06} have measured $T_e$(H$^+$) for an extensive sample
of 47 H {\sc ii} regions which includes 
the most metal-deficient BCDs known, such as
SBS 0335--052W, I Zw 18 and SBS 0335--052E. 
Contrary to the suggestions of 
\citet{P67} and \citet{P02}, they found 
no statistically significant difference between $T_e$(O {\sc iii}) and 
$T_e$(H$^+$), but could not exclude small differences of $\la$ 5\%.

To investigate the effect of these possible small differences between 
$T_e$(O {\sc iii}) and $T_e$(He$^+$) = $T_e$(H$^+$) on $Y_p$, we have derived 
$Y$ -- O/H regression fits for the two sets of He {\sc i} emissivities,
both for the case where $T_e$(He$^+$) is varied in the range 
(0.95 -- 1.0)$\times$ $T_e$(O {\sc iii}) 
(Fig. \ref{fig3}a), and for the one where 
$T_e$(He$^+$) = $T_e$(O {\sc iii}) (Fig. \ref{fig3}h). 
All other parameters are those from the basic set.
Comparison of Fig. \ref{fig3}a (the one corresponding 
to the basic set of parameters) and Fig. \ref{fig3}h shows that assuming
$T_e$(He$^+$) = $T_e$(O {\sc iii}) instead of 
varying it in the range (0.95 -- 1.0)$\times$ $T_e$(O {\sc iii}) 
gives a $Y_p$ larger by about  $\sim$ 1.0\%.  
 We note that if
$T_e$(He$^+$) is different from $T_e$(O {\sc iii}), it 
may also be used for the determination of O/H (Fig. \ref{fig3}i).
If we do so, then $Y_p$ is increased by only $\la$ 0.3\% compared to
the basic regression in Fig. \ref{fig3}a, but the slope
of the regression line 
is significantly shallower as compared to the case where
oxygen abundance is calculated with $T_e$(O {\sc iii}). This is because
the oxygen abundance derived with $T_e$(He$^+$) is larger than the one
derived with $T_e$(O {\sc iii}), and  
$\Delta$(O/H) = O/H[$T_e$(He$^+$)] -- O/H[$T_e$(O {\sc iii})] is 
 larger for high-metallicity H {\sc ii} regions than for 
low-metallicity ones.

We have also considered the changes on $Y_p$ if we adopt the 
temperature variations of $T_e$(He$^+$) proposed  
 by \citet{P02}. Their Fig. 1 shows that $T_e$(He$^+$) 
varies within a range
(0.97 -- 1)$\times$$T_e$(O {\sc iii}) for H {\sc ii} regions with
$T_e$(O {\sc iii}) = 20000K, with 
the mean square temperature variation $t^2$ ranging
between 0 and 0.04, but  
increases to (0.88 -- 1)$\times$$T_e$(O {\sc iii}) for H {\sc ii} regions with
$T_e$(O {\sc iii}) = 10000K, for the same range of $t^2$. 
By adopting the above temperature dependence of $T_e$(He$^+$),
we obtain $Y_p$ = 0.2506, essentially the same value as the one obtained
from the basic regression. 

\subsection{Underlying stellar He {\sc i} absorption}\label{system4} 

It has long been recognized 
\citep{R82,KS83,DS86,P92,O95} that absorption caused by hot stars in 
the He {\sc i} lines can induce an underestimation of 
the intensities of the nebular He {\sc i} lines. In particular, \citet{IT98a}
have shown that the neglect of He {\sc i} underlying stellar absorption has 
led to the derivation of a very low helium mass fraction in I Zw 18, the 
second most metal-deficient BCD known \citep{P92,O97}. Recently, 
\citet{IT04}, \citet{OS04} and \citet{F06} have taken into account  
this effect to derive $Y_p$, using a subsample of
the HeBCD sample of \citet{IT04}.

\citet{G99} have calculated synthetic spectra of H Balmer and He {\sc i}
absorption lines in starburst and poststarburst galaxies. They predict the 
equivalent width of the He {\sc i} $\lambda$4471 absorption line 
to be in the range $\sim$ 0.4 -- 0.5 \AA, or
$\la$ 10\% of the He {\sc i} $\lambda$4471 emission line equivalent
width for 
young starbursts with an age $\la$ 5 Myr, which is the case for 
the H {\sc ii} regions in our sample.

Unfortunately, those authors did not calculate absorption line equivalent 
widths for the other prominent He {\sc i} lines. 
We expect however that the effect of 
underlying absorption is smaller for the He {\sc i} $\lambda$5876,
$\lambda$6678 and $\lambda$7065 emission lines. This, 
for the following reason.
It is known that the equivalent widths are not the same for all 
hydrogen absorption lines. For a fixed 
age of the stellar population, EW$_{abs}$ is the largest for
H$\delta$, then decreases progressively for H$\gamma$, H$\beta$ and 
H$\alpha$ at longer wavelengths. 
A similar trend is likely to hold for He {\sc i} absorption lines: the
longer the wavelength of the line, the smaller its EW$_{abs}$.
It is not clear, however,
how strong is underlying absorption for the He {\sc i} $\lambda$3889 line, at 
the short wavelength end.
In any case, underlying stellar absorption must be taken into account for
all He {\sc i} lines if we are to achieve the desired high accuracy of
$\la$1\% in the primordial He abundance determination.

Among the five He {\sc i} emission lines used here,
the effect of the 
underlying stellar absorption is most important for the blue 
He {\sc i} $\lambda$4471 emission line because it has the lowest equivalent 
width. For the redder lines, 
especially for the He {\sc i} $\lambda$5876 emission line which carries 
the highest
weight in the $Y$ determination, the effect of underlying stellar absorption
is likely considerably 
lower. As for the blue He {\sc i} $\lambda$3889 emission line, its
flux is $\ga$ 2 times greater that of the 
He {\sc i} $\lambda$4471 emission line
and hence the effect of the underlying stellar absorption is also expected to 
be lower. However, the situation with this line is more complicated compared 
to other lines because of its blending with the hydrogen emission line H7.

We have presented our choice for the various ratios 
linking the absorption equivalent width of the He {\sc i} $\lambda$4471 line 
to those of the other He {\sc i} lines  
in our discussion of the basic set of parameters. While the 
choice of those ratios may appear arbitrary and is somewhat subjective since, 
with the exception of the $\lambda$4471 line, we do not have any  
guidance from observations or models concerning the equivalent widths 
of the He {\sc i} absorption lines,
we can estimate the effect of different adopted EW$_{abs}$ on $Y_p$.
Thus, assuming larger EW$_{abs}$ for the He {\sc i}
$\lambda$5876, $\lambda$6678 and $\lambda$7065 lines, in the range 0.5 -- 1.0 
that of the $\lambda$4471 line,
would result in a value of $Y_p$ 
that can be as much as $\sim$ 1\% larger than  
the basic value. However, our calculations show that for 
the majority of the H {\sc ii} regions, 
the $\chi^2_{min}$ values obtained with the larger EW$_{abs}$ for the 
$\lambda$5876, $\lambda$6678 and $\lambda$7065 lines
are $\sim$ 2 times larger than the $\chi^2_{min}$ value 
obtained with the basic parameter set. Additionally, the 
dispersion of the points in the $Y$ -- O/H and $Y$ -- N/H diagrams is higher. 
We also made calculations with the case 
EW$_{abs}$($\lambda$5876) =
0.1$\times$EW$_{abs}$($\lambda$4471). The resulting 
$Y_p$ is only $\la$0.3\% lower than the basic one. In summary, variations
of EW$_{abs}$s for He {\sc i} lines other than 
$\lambda$4471 do not affect much $Y_p$. Therefore, we will adopt 
the basic EW$_{abs}$ ratios in our $Y_p$ determination.

\begin{figure*}
\figurenum{3}
\epsscale{0.7}
\plotone{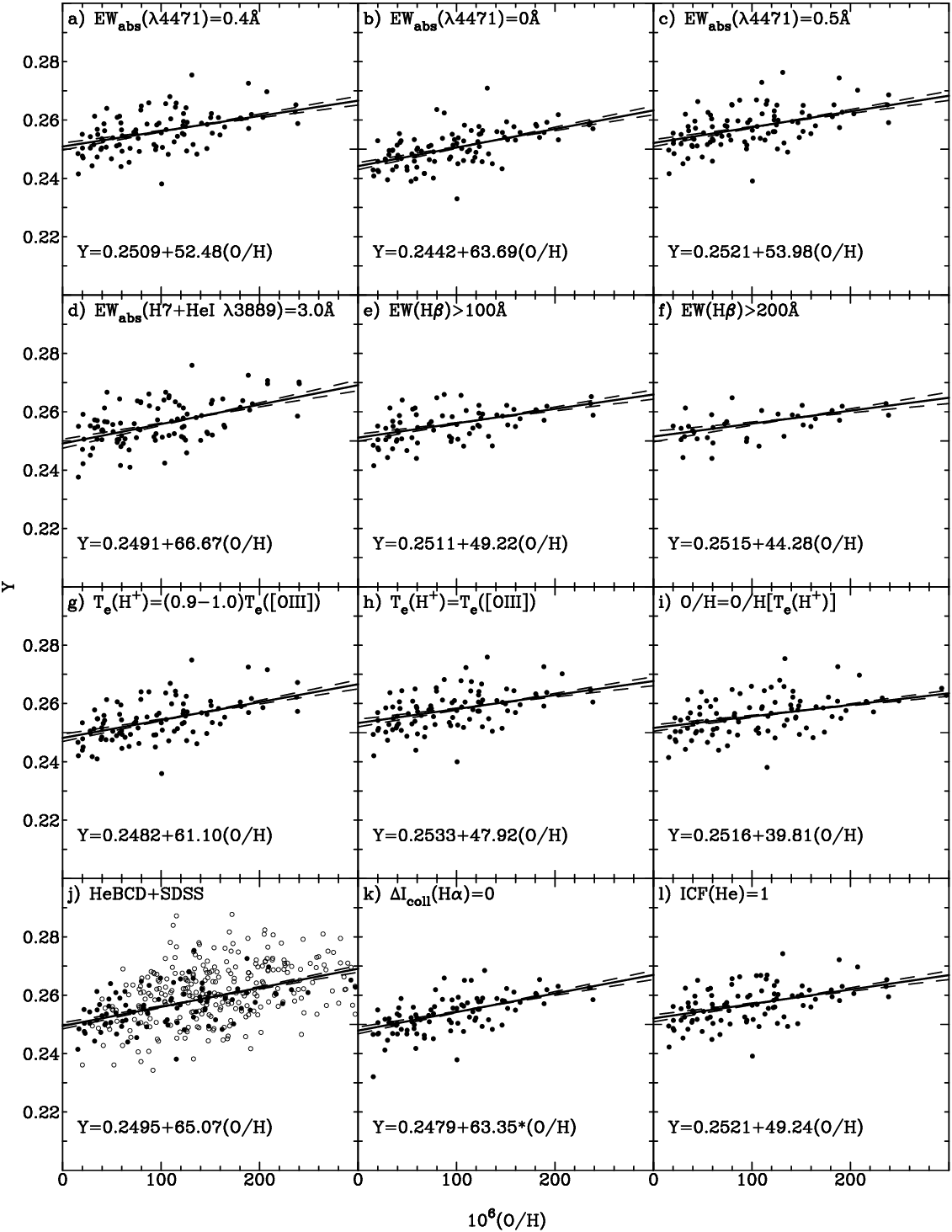}
\figcaption{
Linear regressions of the helium mass fraction $Y$  vs. oxygen 
abundances for different parameter sets. The He {\sc i} emissivities
are from \citet{P05}.
(a) $Y$s are derived for the HeBCD sample adopting
EW$_{abs}$($\lambda$4471) = 0.4 \AA. The electron temperature 
$T_e$(He$^+$) is varied in the range (0.95 -- 1.0)$\times$$T_e$(O {\sc iii})
(basic model). (b) same as in (a), but for EW$_{abs}$($\lambda$4471) 
= 0.0 \AA. (c) same as in (a), but for EW$_{abs}$($\lambda$4471) = 
0.5 \AA. (d) same as in (a), but for 
EW$_{abs}$(H7 + $\lambda$3889) = 3.0 \AA\ instead of
EW$_{abs}$($\lambda$3889) = EW$_{abs}$($\lambda$4471).
(e) same as in (a), but only objects with EW(H$\beta$) $\geq$
100\AA\ are shown. (f) same as in (a), but only objects with 
EW(H$\beta$) $\geq$ 200\AA\ are shown. (g) same as in (a), but 
$T_e$(He$^+$) is varied in the range (0.9 -- 1.0)$\times$$T_e$(O {\sc iii}).
(h) same as in (a), but with $T_e$(He$^+$) = $T_e$(O {\sc iii}).
(i) same as in (a), but oxygen abundance O/H is derived 
adopting $T_e$(He$^+$) instead of $T_e$(O {\sc iii}).
(j) same as in (a), but adding the H {\sc ii} regions
from the SDSS. (k) same as in (a), but for zero
collisional excitation of hydrogen lines. (l) same as in (a), but 
the ionization correction factor $ICF$(He$^+$ + He$^{2+}$) is set to 1.
\label{fig3}}
\end{figure*}

Once the EW$_{abs}$ ratios chosen, we need to fix the absorption 
equivalent width of the $\lambda$4471 line.   
To check the influence of the correction of He {\sc i} emission line fluxes 
for underlying stellar absorption on $Y_p$, we consider two cases:  
case a) where there is no underlying absorption, i.e. 
EW$_{abs}$($\lambda$4471) =0 and case b) 
where EW$_{abs}$($\lambda$4471) = 0.5\AA, the  
maximal equivalent width expected for underlying absorption. 
In Fig. \ref{fig3}b and \ref{fig3}c, we show 
the linear regressions respectively for cases (a) and (b).
It is seen that the difference in $Y_p$ between the two cases is $\sim$ 3\%.
In addition, we have also checked
 how variations of EW$_{abs}$($\lambda$3889)
affect $Y_p$.
Since the He {\sc i} $\lambda$3889 emission line is blended with the hydrogen 
H7 emission line, we have considered, in addition to the basic case 
shown in Fig. \ref{fig3}a where  
EW$_{abs}$($\lambda$3889) =
EW$_{abs}$($\lambda$4471), 
also the case where EW$_{abs}$(H7 + $\lambda$3889) = 3.0\AA, 
shown in Fig. \ref{fig3}d. 
The second case gives a $Y_p$ which is $\sim$1.0\% lower than the
value given by the basic model. However, it has a $\chi^2_{min}$ which 
is significantly larger than the one of the basic model. 

The effect of underlying stellar absorption is greater in H {\sc ii} regions
with lower equivalent widths of He {\sc i} emission lines. 
This allows us to check whether the value of 0.4\AA\ which we have adopted 
for EW$_{abs}$($\lambda$4471) in the basic set 
is reasonable or not. If the adopted value is not 
correct, then the values of $Y_p$ derived from H {\sc ii} 
regions with lower EW(H$\beta$)s and higher EW(H$\beta$)s will be different,
as the first ones will give underestimated He abundances as compared 
to the latter ones. 
In Figs. \ref{fig3}e and \ref{fig3}f, we show respectively 
the linear regressions for the 
subsamples of H {\sc ii} regions with EW(H$\beta$) $>$ 100\AA\ and of those 
with EW(H$\beta$) $>$ 200\AA.
It is seen that the $Y_p$ derived for the HeBCD subsample with 
EW(H$\beta$) $>$ 100\AA\ 
(Fig. \ref{fig3}e) is only slightly ($\la$ 0.2\%) lower than the value 
derived for
the HeBCD subsample with EW(H$\beta$) $>$ 200\AA\ (Fig. \ref{fig3}f).
This shows that  
EW$_{abs}$(He {\sc i} $\lambda$4471) = 0.4\AA\ and
the ratios of EW$_{abs}$ adopted in the basic model
characterize well the underlying He {\sc i} stellar absorption.

\begin{figure*}
\figurenum{4}
\epsscale{0.45}
\plotone{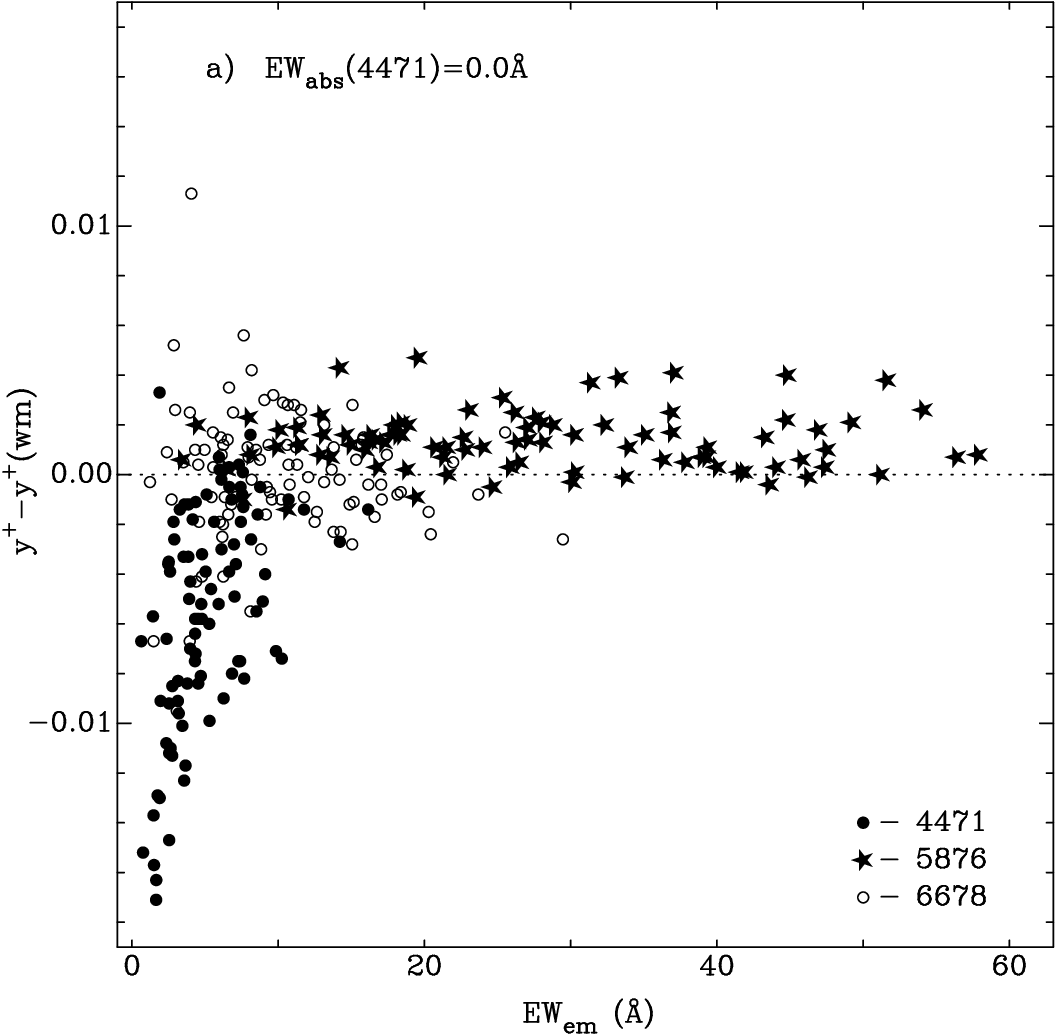}
\plotone{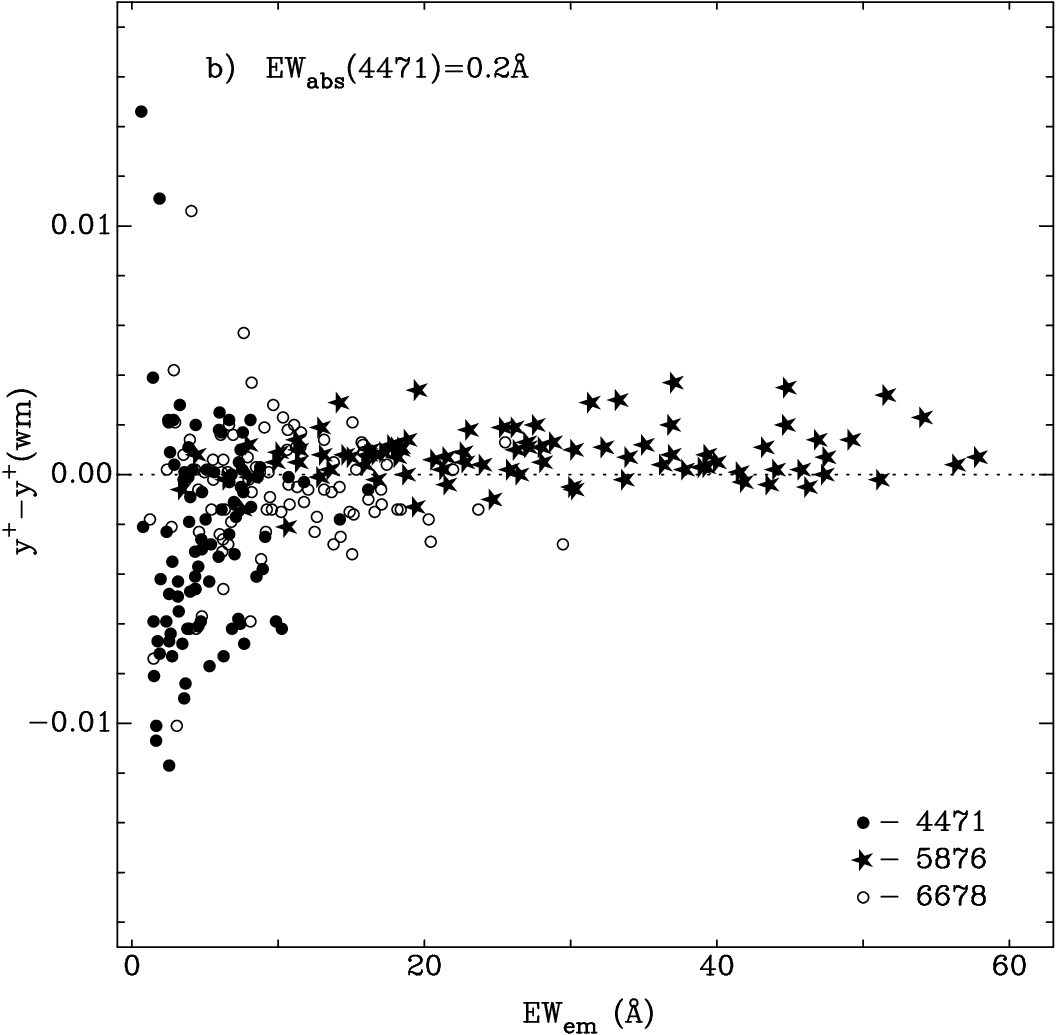}
\plotone{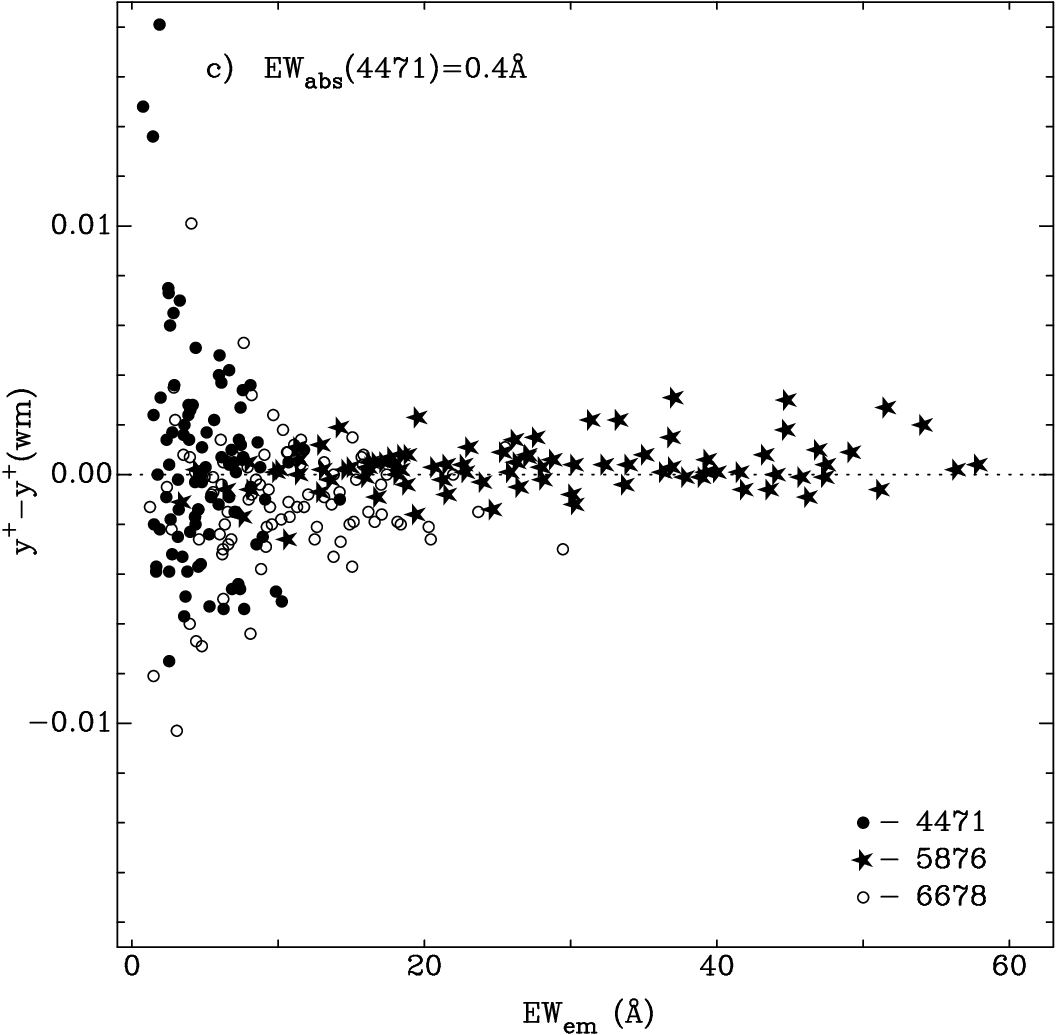}
\plotone{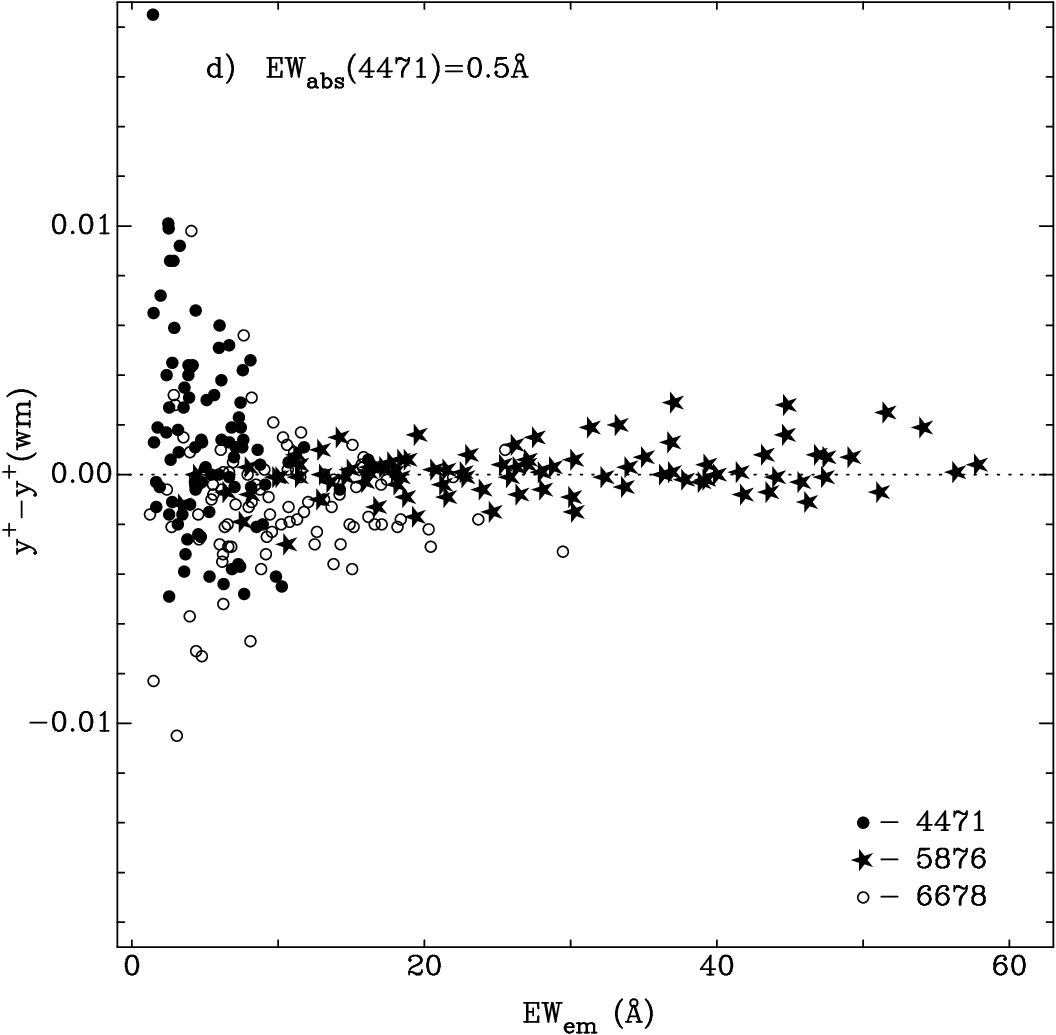}
\figcaption{($y^+$ -- $y^+_{wm}$) derived for each of the three He {\sc i}
emission lines $\lambda$4471 (filled circles), $\lambda$5876 (stars) and
$\lambda$6678 (open circles) vs the equivalent width EW$_{em}$ of the same 
lines. $y^+_{wm}$ is the weighted mean of the $y^+$ of each individual line.
Four different values of EW$_{abs}$($\lambda$4471) have been adopted: 
a) EW$_{abs}$($\lambda$4471) = 0, 
b) EW$_{abs}$($\lambda$4471) = 0.2\AA, c) EW$_{abs}$($\lambda$4471) = 0.4\AA\
and d) EW$_{abs}$($\lambda$4471) = 0.5\AA. Only data for the HeBCD sample are
shown. 
\label{fig4}}
\end{figure*}

\begin{figure*}
\figurenum{5}
\epsscale{0.45}
\plotone{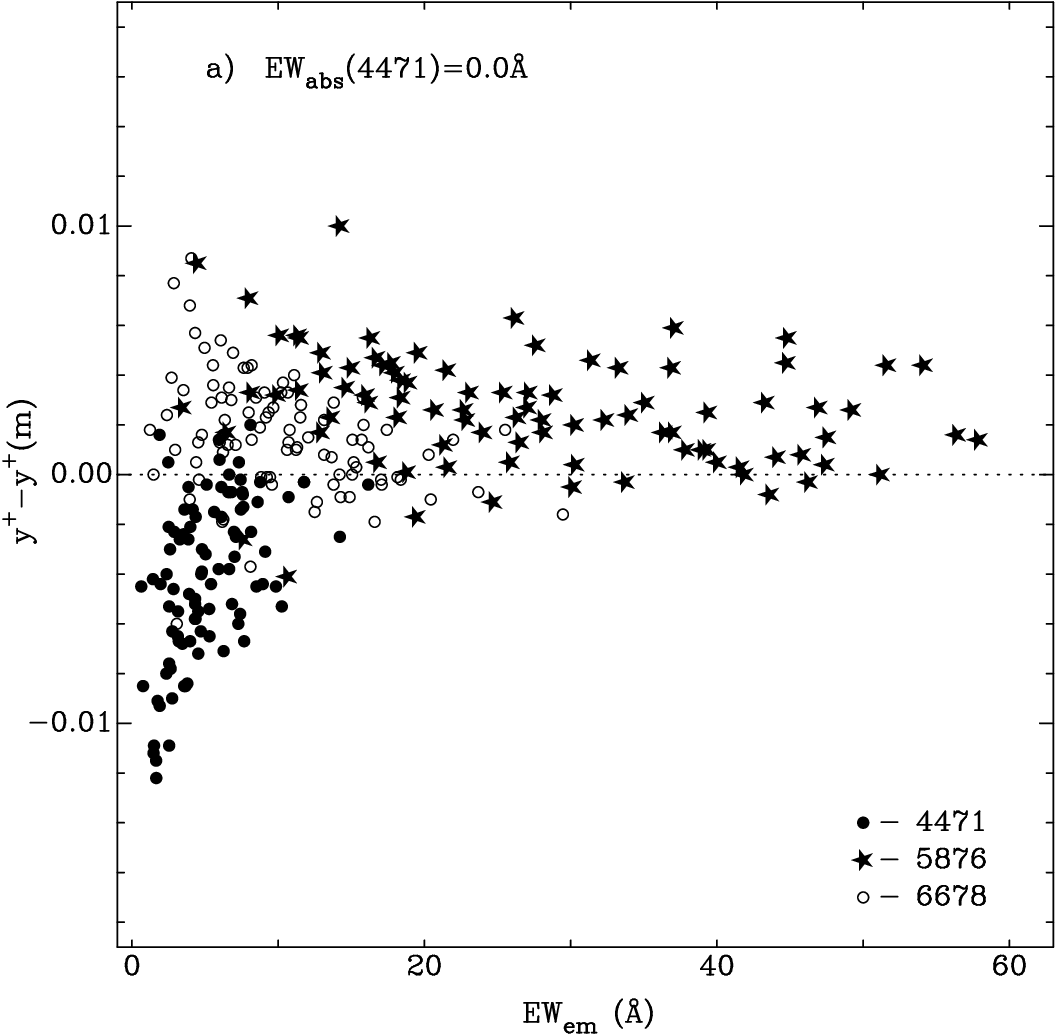}
\plotone{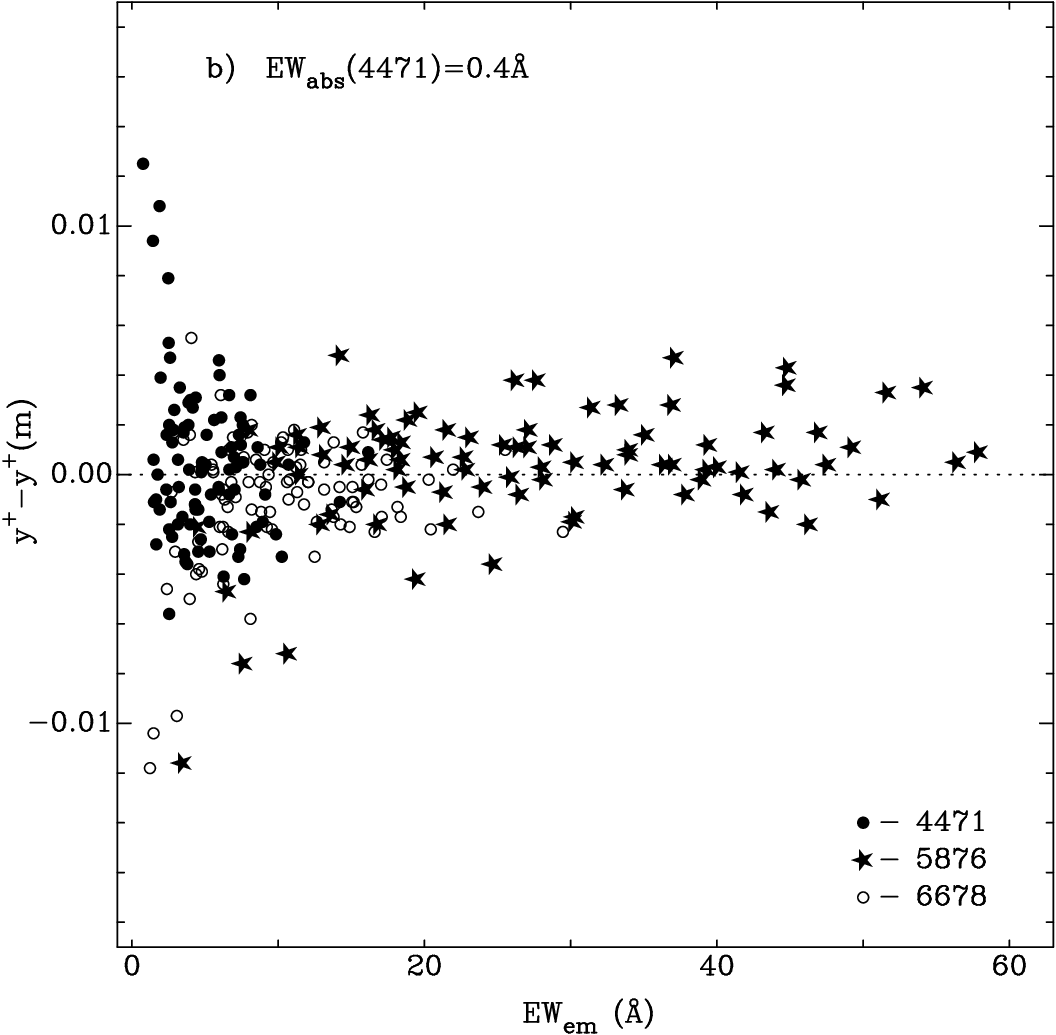}
\plotone{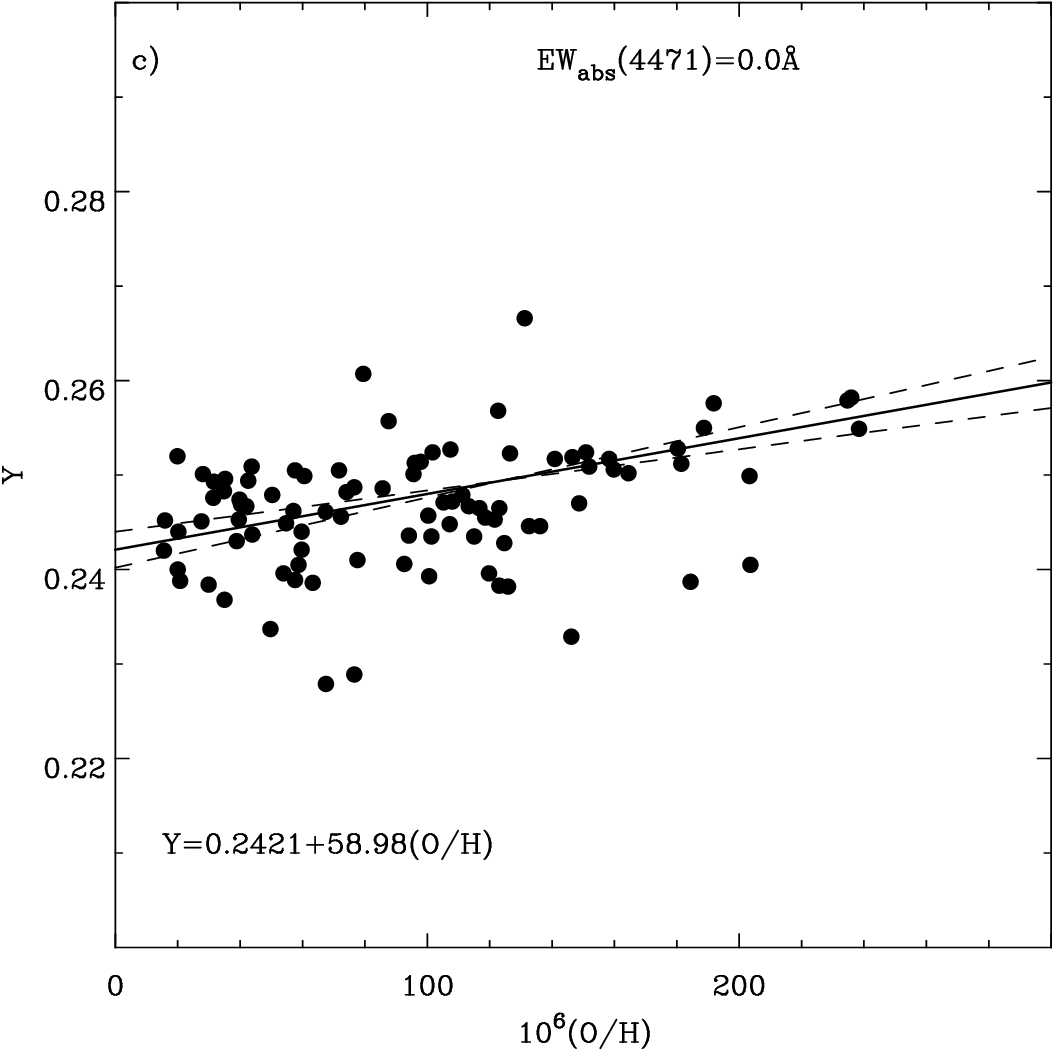}
\plotone{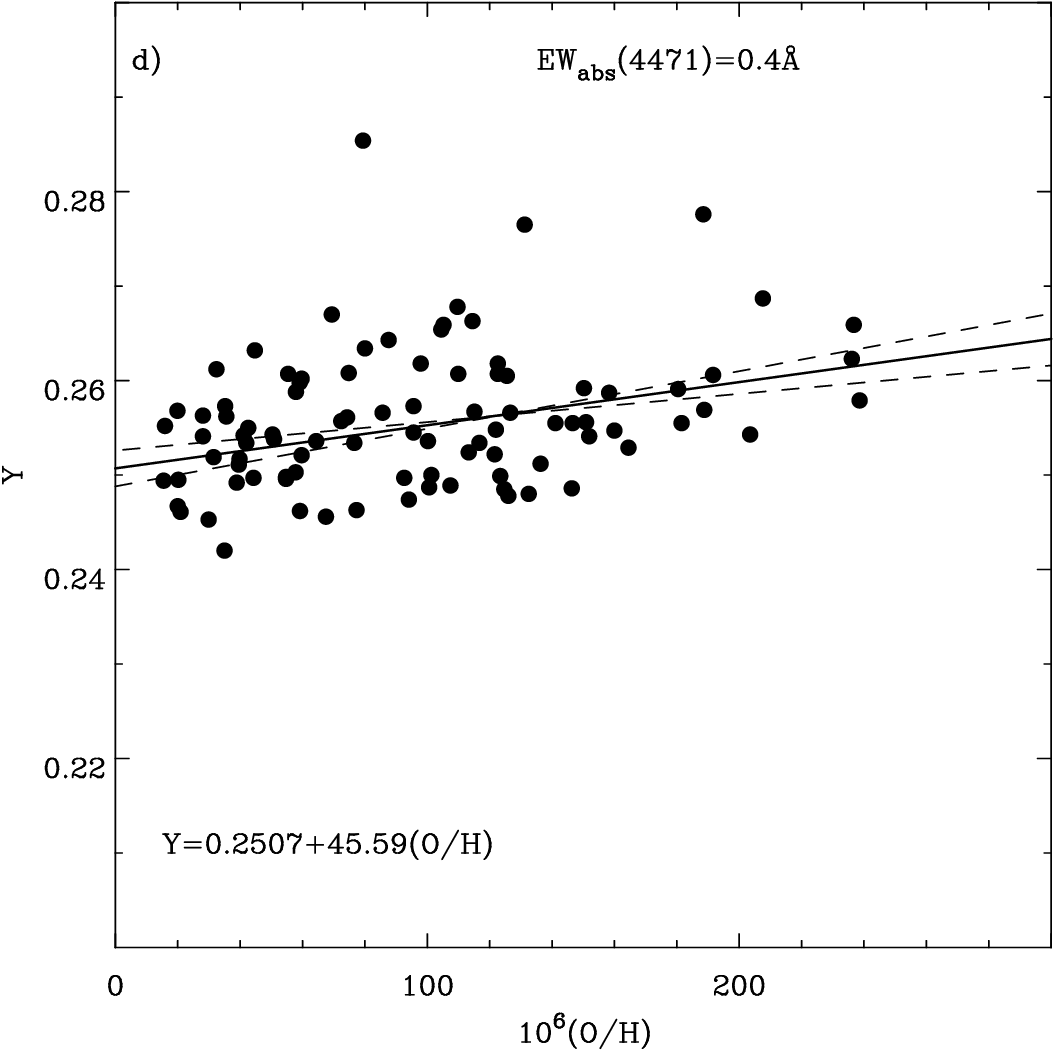}
\figcaption{a) Same as in Fig. \ref{fig4}a, except that $y^+_{m}$ is the 
simple mean instead of the weighted mean $y^+_{wm}$; 
b) Same as in Fig. \ref{fig4}c, except that $y^+_{m}$ is the 
simple mean instead of the weighted mean $y^+_{wm}$; 
c) Same as in Fig. \ref{fig3}c, except that $y^+_{m}$ is the 
simple mean instead of the weighted mean $y^+_{wm}$; 
d) Same as in Fig. \ref{fig3}a, except that $y^+_{m}$ is the 
simple mean instead of the weighted mean $y^+_{wm}$. 
\label{fig5}}
\end{figure*}

Probably, as proposed by \citet{P06}, the best check of the appropriateness of
the basic set of EW$_{abs}$s comes from comparing the abundance of singly
ionized He $y^+$, calculated separately for each of the three He {\sc i} 
$\lambda$4471, $\lambda$5876 and $\lambda$6678 emission lines, with their 
weighted mean $y^+_{wm}$. 
We show in Fig. \ref{fig4}a the dependence of ($y^+$ -- $y^+_{wm}$) on the 
He {\sc i} emission line equivalent width EW$_{em}$, for each of the three
emission lines, in the case where underlying 
absorption is not taken into account. It is seen that there are no significant
offsets between the $y^+$s derived from the He {\sc i} $\lambda$5876 and 
$\lambda$6678 emission lines (labeled respectively by stars and open circles), 
suggesting that the effect of underlying 
absorption for those lines is relatively small. On the other hand, the
$y^+$s derived from the He {\sc i} $\lambda$4471 emission line (filled circles)
are systematically lower than those derived from the other two emission lines, 
suggesting that underlying stellar absorption is significant.
Figs. \ref{fig4}b - \ref{fig4}d show
the same dependences as in Fig. \ref{fig4}a, but for
successively increasing values of EW$_{abs}$($\lambda$4471), varying from 
0.2\AA\ to 0.5\AA. For the other He {\sc i} lines we adopt the same ratios of 
EW$_{abs}$ to EW$_{abs}$($\lambda$4471) as those in the basic model.
It seen from Figs. \ref{fig4}c that the $y^+$s derived for 
He {\sc i} $\lambda$4471 with EW$_{abs}$($\lambda$4471) = 0.4\AA\ are in good
agreement with the ones for the other two lines. However, the
$y^+$($\lambda$4471) derived with lower and higher EW$_{abs}$($\lambda$4471)s 
are respectively lower and higher than the $y^+$ values derived for the
He {\sc i} $\lambda$5876 and $\lambda$6678 lines (compare Fig. \ref{fig4}b 
with Fig. \ref{fig4}d).
\citet{P06} have also suggested to use a simple mean $y^+_m$ instead of
the weighted mean $y^+_{wm}$ used in our analysis. In Figs. \ref{fig5}a and
\ref{fig5}b we show the dependence of ($y^+$ -- $y^+_{m}$) on EW$_{em}$ for
two choices of EW$_{abs}$($\lambda$4471): 0 (Fig. \ref{fig5}a) and 0.4\AA\
(Fig. \ref{fig5}b).
Similarly to Fig. \ref{fig4}a, there is a clear difference between the 
$y^+$ derived from the He {\sc i} $\lambda$4471 line and those derived from 
the other two He {\sc i} lines when underlying stellar absorption is not
taken into account (Fig. \ref{fig5}a).
On the other hand, there is no clear offset in the $y^+$ derived from the
different He {\sc i} lines 
in Fig. \ref{fig5}b, where EW$_{abs}$($\lambda$4471) = 0.4\AA.
The primordial He abundance $Y_p$ derived from the $Y$ -- O/H regression,
using a simple mean for $y^+$ and EW$_{abs}$($\lambda$4471) = 0 
(Fig. \ref{fig5}c) is $\sim$0.8\% lower than the $Y_p$ obtained using a 
weighted mean for $y^+$
(Fig. \ref{fig3}b). This is because the He {\sc i} $\lambda$4471 line
contributes more to the simple mean than to the weighted mean. 
On the other hand, 
no difference in $Y_p$ is found when EW$_{abs}$($\lambda$4471) is set 
to 0.4\AA, irrespective of whether $y^+$ is computed with a weighted mean
or a simple mean (compare Figs. \ref{fig5}d and \ref{fig3}a).

All the above considerations lead us to conclude that the basic set of 
EW$_{abs}$s adopted for the He {\sc i} lines is the most appropriate one to 
use.

\subsection{Collisional excitation of the H lines}\label{system5}

It has generally been assumed in abundance studies that deviations of the
observed H$\alpha$/H$\beta$ flux ratio from the theoretical recombination
value are entirely due to interstellar extinction. \citet{DK85} first
noted that in the hot and dense H {\sc ii} regions of BCDs, collisional 
excitation of hydrogen emission lines can be important and 
affect the derived He/H ratio. \citet{SI01} estimated
that this effect can result in an upward correction in the He abundances
of up to 5\%, assuming
that the excess of the H$\alpha$/H$\beta$ flux ratio above the
theoretical recombination value is due only to collisional excitation. 
However, new computations of the effective collision strength 
for excitation of hydrogen
\citep{A02} give a smaller enhancement of the Balmer lines due to 
collisional excitation. \citet{L03} have performed a very detailed analysis 
of 3 H {\sc ii} regions, including a tailored photoionization analysis and a 
discussion of reddening. They find that the collisional contribution to 
H$\alpha$ in those objects may reach 8 \% and that to H$\beta$ 
2 -- 2.5 \%. 
Thus, in our Monte Carlo approach, 
we have varied the fraction of the 
H$\alpha$ emission line flux due to collisional excitation, 
$\Delta$$I$(H$\alpha$)/$I$(H$\alpha$), 
in the range 0 -- 5\%. 
Its 
value is derived together with other
parameters by the minimization of $\chi^2$ as defined in Eq. \ref{eq1}. 
To check how the
inclusion of hydrogen collisional excitation influences $Y_p$, we have  
also calculated the linear regression for the case 
where that effect is not taken into account, i.e. 
$\Delta$$I$(H$\alpha$) = 0. The result
is shown in Fig. \ref{fig3}k. Comparison with the basic regression 
(Fig. \ref{fig3}a) shows that ignoring hydrogen collisional excitation
results in a decrease of $Y_p$ by $\sim$ 1.2\%.\footnote{In the above analysis,
we have adopted a ratio of H$\alpha$ to H$\beta$ emissivities due to 
collisional excitation equal to 3. The referee has pointed out that this 
ratio may be higher. We have estimated it to be 5 -- 7
in the temperature range $T_e$ = 10000K -- 20000K, consistent
with the \citet{A02} determination. The consequence of adopting a ratio of 
H$\alpha$ to H$\beta$ emissivities due to collisional excitation
equal to 5 would be to increase $Y_p$ by only 0.3\%, using the HeBCD sample and
in the case of the basic model.}

\subsection{Ionization structure}\label{system6}

    Another source of systematic uncertainty comes from the assumption
that the H$^+$ and He$^+$ zones in the H {\sc ii} region are 
spatially coincident. 
However, depending on the hardness of the ionizing radiation, the radius of 
the He$^+$ zone can be smaller than the radius of the H$^+$ zone in
the case of soft ionizing radiation, or larger in the case of hard
radiation. In the former case, a correction for unseen neutral helium
should be made, resulting in an ionization correction factor 
$ICF$(He$^+$ + He$^{2+}$) $>$ 1
and hence a higher helium abundance. In the latter case, the situation is 
opposite and $ICF$(He$^+$ + He$^{2+}$) $<$ 1. 
The ionization correction factor problem  has 
been discussed in several studies \citep{S80,P92,S97,O97,V00,P00,B00,SJ02}. 
It was concluded that the correction of the helium 
abundance can be as high as several percent in either downward or upward
directions, depending on the hardness of the radiation and the ionization
parameter $U$. \citet{SJ02} have calculated an
extensive grid of photoionized H {\sc ii} region models which give
correction factors as functions of hardness and $U$. Their conclusion was 
that a downward correction of $Y$ by as much as 6\% and 2\% is required 
respectively for ionization parameters log $U$ = --3.0 and --2.5.
However, if log $U$ $\ga$ --2.0, which is the case for the majority of our 
H {\sc ii} regions, the downward correction is $\la$ 1\%. 

\citet{B00} have suggested that $ICF$(He$^+$ + He$^{2+}$) 
can be estimated from the
[O {\sc iii}]$\lambda$5007/[O {\sc i}]$\lambda$6300 emission line flux ratio.
They have calculated an extensive grid of photoionized H {\sc ii} 
region models,
spanning a wide range of metallicity and excitation and with different
models for the ionizing stellar radiation. \citet{B00} have concluded that
the $ICF$ can be significantly lower than unity in hot H {\sc ii} regions.
However, if a H {\sc ii} region has a 
[O {\sc iii}] $\lambda$5007/[O {\sc i}]$\lambda$6300 ratio greater than 300,
then their models show that, regardless of its metallicity, it will have 
an $ICF$ very close to unity. 
Using the observations of \citet{IT98b}, \citet{B00} have
found that the scatter of the data points for H {\sc ii} regions with
[O {\sc iii}] $\lambda$5007/[O {\sc i}]$\lambda$6300 $>$ 300 in the $Y$ -- O/H
diagram is some 20\% smaller than that for the whole sample, and have derived
$Y_p$ = 0.2489 $\pm$ 0.0030 for the restricted sample. 
However, their analysis suffers from  
the small number of H {\sc ii} regions in the restricted sample. 
In particular, they found a negative
d$Y$/d(O/H) slope, which leads to 
a high value of $Y_p$. Such a negative slope
is unphysical and cannot be justified by any chemical evolution model, with
any star formation scenario. 

\begin{figure*}
\figurenum{6}
\epsscale{1.1}
\plotfiddle{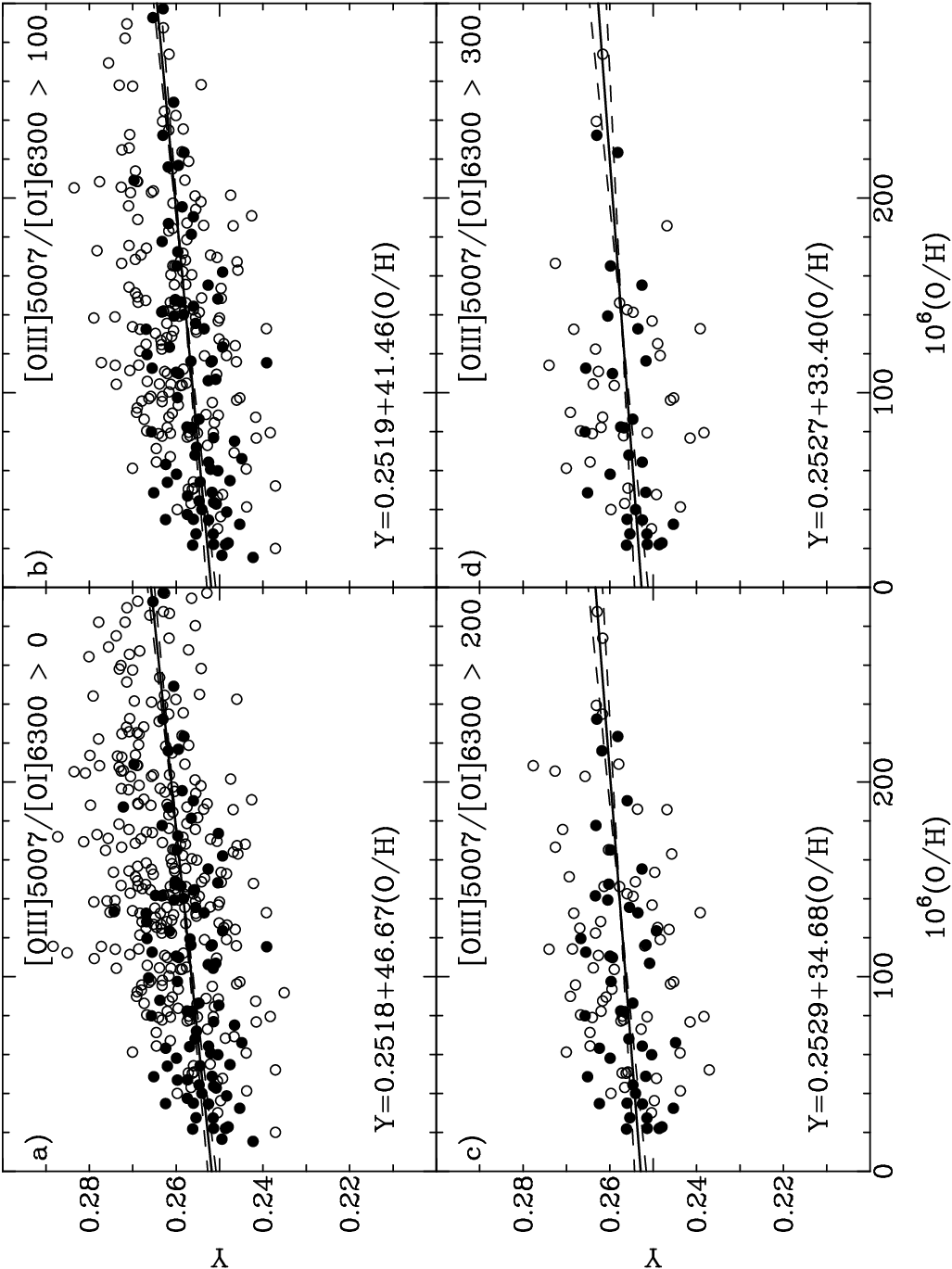}{1pt}{-90.}{420.}{520.}{-30.}{0.}
\figcaption{ Linear regressions of the helium mass fraction $Y$ vs. oxygen 
abundances for for H {\sc ii} regions with different values of
the [O {\sc iii}]$\lambda$5007/[O {\sc i}]$\lambda$6300 emission flux ratio. 
The He {\sc i} emissivities are from \citet{P05}. The ionization correction
factor $ICF$(He$^+$ + He$^{2+}$) is set to 1. In panel a) all H {\sc ii} regions are shown,
while in panels b), c) and d) are shown H {\sc ii} regions with 
the [O {\sc iii}]$\lambda$5007/[O {\sc i}]$\lambda$6300 emission flux ratio
respectively greater than 100, 200 and 300.
\label{fig6}}
\end{figure*}

To check how the $ICF$ is related to the
[O {\sc iii}] $\lambda$5007/[O {\sc i}]$\lambda$6300 ratio, we show in 
Fig. \ref{fig6} $Y$ -- O/H regression lines for the combined HeBCD + SDSS 
sample, with different
cutoffs of the [O {\sc iii}] $\lambda$5007/[O {\sc i}]$\lambda$6300 ratio. 
 The 
HeBCD H {\sc ii} regions are shown by filled circles and the SDSS H {\sc ii}
regions by open circles.  
 We adopt $ICF$(He$^+$ + He$^{2+}$) = 1 for all H {\sc ii} regions.
 All other parameters are those of the basic set (\S\ref{method4}).
It is seen that the scatter in the $Y$ -- O/H diagram of the HeBCD
H {\sc ii} regions with [O {\sc iii}] $\lambda$5007/[O {\sc i}]$\lambda$6300
$>$ 300 (Fig. \ref{fig6}d) is $\sim$ 20\% lower than that for the total 
HeBCD sample (Fig. \ref{fig6}a), confirming the finding by \citet{B00}. 
However,
for the SDSS sample which is larger and hence has 
better statistics, we do not find that the scatter decreases 
with increasing value of the cutoff. 
This suggests that the scatter of points of $\sim$ 3\% -- 4\%
around the regression line in Fig. \ref{fig6} is due mainly to
observational statistical uncertainties, 
and not so much to variations of the $ICF$. 
By contrast with 
\citet{B00}, we find only small changes in $Y_p$, by $\la$ 0.5\%,
when the cutoff is increased. These very small $Y_p$ changes 
can be due in part to the  
progressively smaller size of the H {\sc ii} region sample
 with increasing cutoff, 
and hence to larger uncertainties in the regression. 
The nearly constant value of $Y_p$
implies that $ICF$(He$^+$ + He$^{2+}$) is close to the adopted value of unity for 
our H {\sc ii} regions, regardless of the value of 
the 
[O {\sc iii}] $\lambda$5007/[O {\sc i}] $\lambda$6300 ratio which varies in
the range $\sim$ 30 -- 700.

\begin{figure*}
\figurenum{7}
\epsscale{1.1}
\plotfiddle{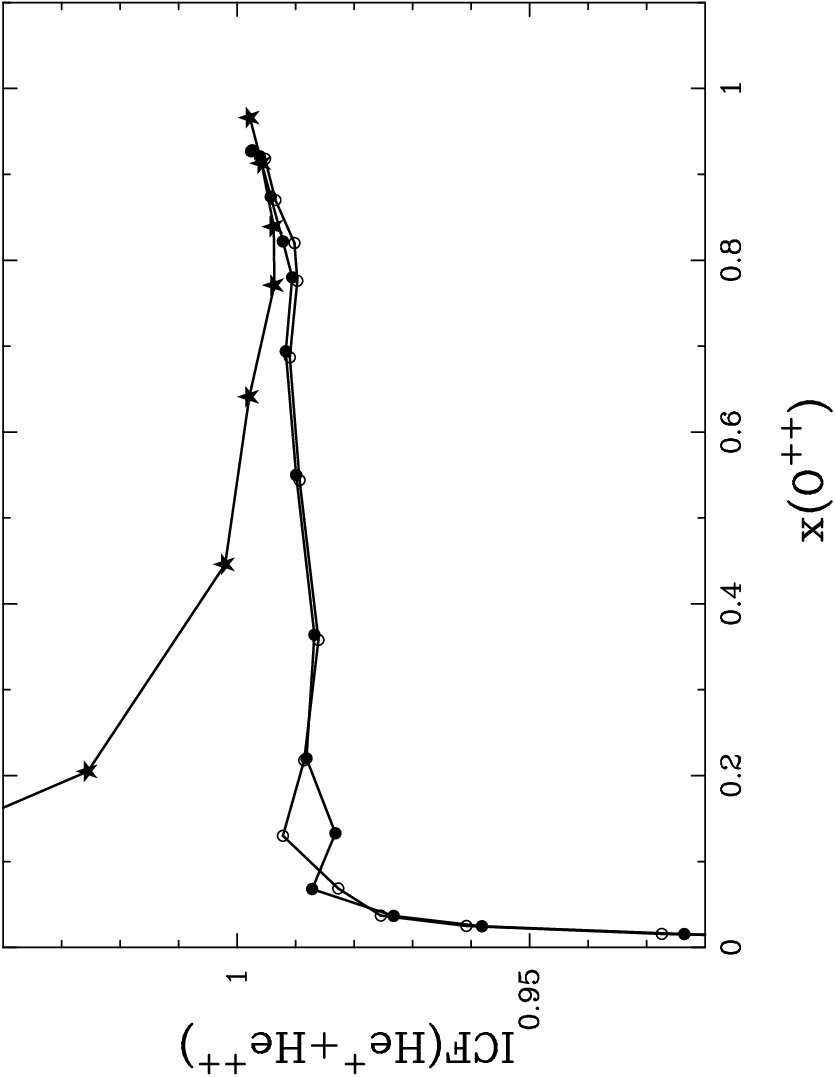}{1pt}{-90.}{320.}{420.}{0.}{0.}
\figcaption{ Ionization correction factor $ICF$(He$^+$ + He$^{2+}$) as a function
of the doubly ionized oxygen abundance fraction O$^{2+}$/(O$^+$+O$^{2+}$) 
\label{fig7}}
\end{figure*}

To derive $ICF$(He$^+$ + He$^{2+}$), we use the photoionized H {\sc ii} region models by
\citet{SI03}, but with an input radiation field computed with Starburst 99 
\citep{L99} using the stellar model atmospheres described in \citet{S02}.
In Fig. \ref{fig7}, we show $ICF$(He$^+$ + He$^{2+}$) as a function of the 
O$^{2+}$/(O$^+$+O$^{2+}$) abundance ratio, the latter being a measure of
the H {\sc ii} region excitation. Filled circles represent 
low-metallicity H {\sc ii} regions with 12 + log O/H = 7.2,
open circles intermediate-metallicity H {\sc ii} regions with
12 + log O/H = 7.6 and stars high-metallicity H {\sc ii} regions 
with 12 + log O/H = 8.3. It is seen from Fig. \ref{fig7} that the $ICF$ in
low-excitation H {\sc ii} regions, those with low
O$^{2+}$/(O$^+$+O$^{2+}$) abundance ratio, can significantly deviate from
unity. However, our sample consists only of high-excitation H {\sc ii}
regions with O$^{2+}$/(O$^+$+O$^{2+}$) $>$ 0.5 and for these 
objects, $ICF$ deviates
from unity by not more than 1\%. In our calculations, we use the $ICF$s
derived by interpolating or extrapolating in 
metallicity the values
given by the three curves in Fig. \ref{fig7}, for a given $x$(O$^{2+}$) =
$N$(O$^{2+}$)/$N$(O$^+$+O$^{2+}$).
 We find that the effect of ionization structure
on the derived $Y_p$ is small. Neglecting the correction for ionization, i.e.
adopting $ICF$(He$^+$ + He$^{2+}$) = 1, we obtain a $Y_p$ which is only $\la$0.5\% larger  
than the value derived when the correction for ionization 
is taken into account (compare Fig. \ref{fig3}l and \ref{fig3}a).

\subsection{Deviations from case B recombination}\label{system7}

All previous work on the primordial He abundance 
determination has assumed 
that case B recombination holds in H {\sc ii} regions. 
Case B assumes that there exists a balance between the absorption
and emission of photons in the resonant Lyman series transitions of hydrogen 
and helium and that there are no other processes. However in real
H {\sc ii} regions, processes do exist that may disrupt this balance. 
The first of these processes is
the leakage of Lyman line photons from the H {\sc ii} region because of
a finite optical depth. The second process is the absorption of UV 
photons by dust inside the H {\sc ii} region. The third process is line 
pumping of stellar photons.

Thanks to our large data base, we are able to 
estimate the effect of the deviations from case B for He {\sc i} emission 
lines. 
This is because some singlet emission lines, such as the He {\sc i}
$\lambda$5015 and $\lambda$7281 lines, are sensitive 
to this effect, while other
singlet lines, such as the He {\sc i} $\lambda$4922 and $\lambda$6678 lines, 
are much less 
sensitive \citep{A84}. Therefore, singlet He {\sc i} emission line flux
ratios may be good indicators of deviations from case B.
The SDSS H {\sc ii} region sample is particularly important in this regard, 
because the relatively high spectral 
resolution of the SDSS spectra allows to separate 
the He {\sc i} $\lambda$5015 emission line 
from the strong [O {\sc iii}] $\lambda$5007 emission line.

\begin{figure*}
\figurenum{8}
\epsscale{0.8}
\plotone{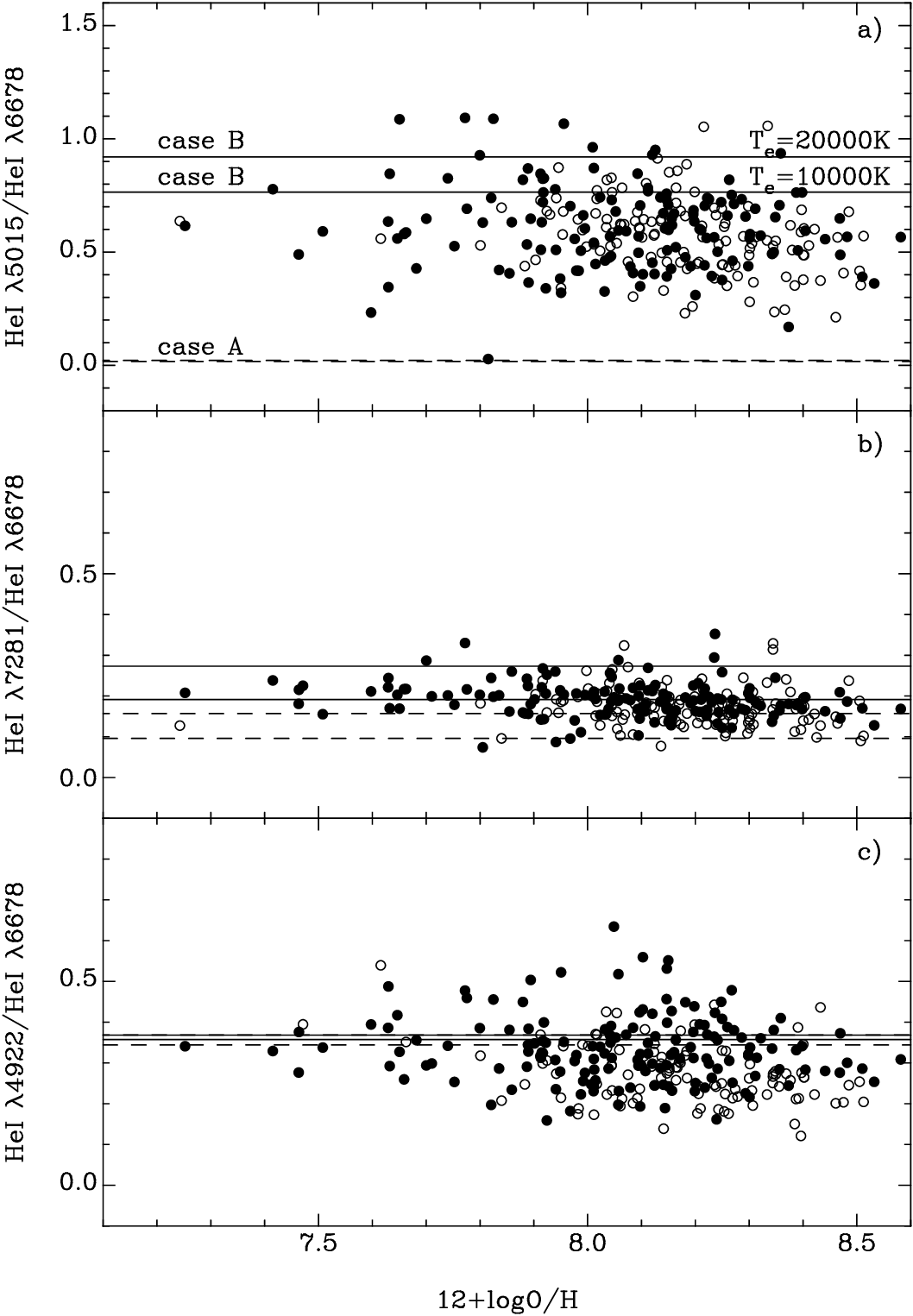}
\figcaption{Flux ratios of singlet He {\sc i} emission lines vs oxygen
abundance for the SDSS H {\sc ii} regions. Filled circles are H {\sc ii}
regions with EW(H$\beta$) $\geq$ 100\AA\ and open circles are H {\sc ii}
regions with EW(H$\beta$) $<$ 100\AA. The flux ratios for
cases A and B are shown respectively by dashed and solid horizontal lines.
\label{fig8}}
\end{figure*}

In Fig. \ref{fig8}, we show the dependence of the He {\sc i} 
$\lambda$5015/$\lambda$6678, $\lambda$7281/$\lambda$6678 and 
$\lambda$4922/$\lambda$6678 singlet flux ratios on oxygen abundance
12 + log O/H for the SDSS H {\sc ii} regions. 
We divide the SDSS objects into two groups:    
the filled circles represent   
H {\sc ii} regions with EW(H$\beta$) $\geq$ 100\AA\ and the open circles 
those with EW(H$\beta$) $<$ 100\AA. 
This division allows us to check 
whether or not deviations of the 
He {\sc i} flux ratios from case B are affected
by underlying stellar absorption. This effect is expected to be higher 
in H {\sc ii} regions
with low EW(H$\beta$). If it plays a role, then a systematic shift
between H {\sc ii} regions with high and low EW(H$\beta$) should be present
in Fig. \ref{fig8}.
The solid and dashed lines in Fig. \ref{fig8} show respectively 
the theoretical ratios for cases B and A, in the case of 
two electron temperatures, 
$T_e$ = 10000 K and $T_e$ = 20000 K. 
In panels a) and c), the two lines for case A are nearly indistinguishable.
The emission line which is most sensitive to deviations from case B is 
He {\sc i} $\lambda$5015. It is seen from Fig. \ref{fig8}a that
the He {\sc i} $\lambda$5015/$\lambda$6678 flux ratio is systematically lower
than the theoretical value for case B. The same is true, but to a 
lesser extent, for
the He {\sc i} $\lambda$7281/$\lambda$6678 flux ratio. It is also seen
that there is no systematic offset between H {\sc ii} regions with high and
low EW(H$\beta$). The situation is different for the
He {\sc i} $\lambda$4922/$\lambda$6678 flux ratio. There is no statistically 
significant
deviation for H {\sc ii} regions with high EW(H$\beta$), but the ratios for 
H {\sc ii} regions with low EW(H$\beta$) are systematically below the case
B theoretical value (Fig. \ref{fig8}c). Thus,
underlying absorption does play a role,
and its effect is most important for the He {\sc i} $\lambda$4922 
emission line. This
conclusion is supported by the synthetic spectra of 
\citet{G99} who predict rather high 
(0.3 -- 0.8 \AA) equivalent widths for the He {\sc i} $\lambda$4922 
absorption line.

Thus, the 
He {\sc i} $\lambda$5015/$\lambda$6678 and $\lambda$7281/$\lambda$6678
emission line flux ratios clearly show deviations
from case B. It is reasonable to suppose that deviations from case B are also
present for hydrogen. However, the overall effect of these H 
deviations on the derived element abundances is likely to be 
small. \citet{CF88} have considered the
effect of case B deviations on hydrogen line emissivities caused by 
a finite 
optical depth of the Lyman series lines, the presence of velocity 
gradients and dust. Applying their results to the low-metallicity H {\sc ii}
regions in our sample, we find that effect to be small. 
This, because of several reasons. First, the 
Ly$\alpha$ optical depth in our H {\sc ii} regions is likely to be very large,
more than $\sim$ 10$^5$, as evidenced by the 
very high H {\sc i} column
densities $\ga$ 10$^{21}$ cm$^{-2}$ measured 
in some of our objects \citep{TI97,K98}. Second,  
the ratio of the number of dust particles 
to the number of hydrogen atoms in our objects 
is more than one order of magnitude lower 
than the ratio in the interstellar medium of the Galaxy, as 
shown by \citet{I06} in their study of SDSS extragalactic H {\sc ii} regions.
For low-metallicity H {\sc ii} regions with these characteristics, 
we find from Figs. 3 and
4 of \citet{CF88} that the recombination rate for hydrogen is  
smaller than the rate 
for case B by not more than 1\%. Similar conclusions can be reached for
helium. 

As for case C which occurs when absorption of stellar photons at 
the wavelengths of Lyman lines is important \citep{F99}, our Cloudy 
(version C06.02) calculations show that this situation 
does not happen in
any significant degree in H {\sc ii} regions such as the ones in our sample.

For the objects considered here, 
the effects of deviations from case B 
work in the same direction
for both hydrogen and helium. Since the 
determination of the He abundance involves the ratio of $N$(He) to $N$(H),
these effects partially compensate each other. 
In the present paper, we have chosen not to take into account 
the deviations from case B in the calculation of He abundances. Accounting 
for them accurately would require a precise photoionization modeling 
for each individual object
which is not feasible, given the large size of our sample.
However, it would be interesting to  
attempt this modeling for a few cases in the future, for checking purposes.

\subsection{Effect of different samples}\label{system8}

In our previous papers \citep{ITL94,ITL97,IT98b,IT04}, we have only considered
H {\sc ii} regions from the HeBCD sample, data which we acquired and 
reduced ourselves. 
In this section, we consider the effect on the $Y_p$ value
when the H {\sc ii} regions from the SDSS sample, 
observed and reduced independently by the SDSS team,  
are added. 
In Fig. \ref{fig3}j, we show the linear regression $Y$ -- O/H
obtained with the basic parameter set for the combined HeBCD + SDSS sample.
It is seen that SDSS data (open circles) have a larger scatter
as compared to the HeBCD data (filled circles) which is expected
because of the lower signal-to-noise ratio of the SDSS spectra.
However, there is no visible offset between the two data sets. The derived 
$Y_p$ value for the combined sample is  
close to the one obtained for the basic case (compare Fig. \ref{fig3}a and 
\ref{fig3}j). The difference in $Y_p$ is $\sim$ 0.5\%.

We have also derived the regression line for the SDSS subsample only 
with the basic 
parameter set. We obtain $Y$ = (0.2459$\pm$0.0021) + (90.60$\pm$13.5) O/H,
as compared to $Y$ = (0.2509$\pm$0.0012) + (52.48$\pm$9.0) O/H for 
the HeBCD subsample only, with the same parameters (see Fig. \ref{fig3}a). 
The $Y_p$s for the 
two subsamples are consistent with each other at the 2$\sigma$ level.
The SDSS subsample was not designed specifically for the $Y_p$ problem.  
The derived lower $Y_p$ and the steeper slope are due to both 
the lack of very low-metallicity H {\sc ii} regions and the lower quality 
spectra in the SDSS sample. 

\subsection{Budget of systematic errors}\label{system9}

Table \ref{tab1} gives a summary budget of the various systematic errors 
discussed above. After the important effect of collisional enhancement of the
He {\sc i} lines (not listed in Table \ref{tab1}) which can be, 
depending on the particular line, larger than
10\%, the second most important systematic effect
is that of underlying He I stellar absorption, as already emphasized 
by \citet{IT98a}. Its effect is $\la$ 3\%. 
The remaining 
systematic effects -- temperature fluctuations, collisional excitation 
of H lines and ionization correction factors -- are all $\la$ 1\%, some 
increasing $Y_p$, and some decreasing it. If we adopt the \cite{B02} 
emissivities as in our previous work, then all systematic effects 
listed in Table \ref{tab1} add up to a net increase
of $\la$ 2\%. Thus, the value $Y_p$ = 0.243 obtained by 
\citet{IT04} for the HeBCD sample without considering the  
systematic effects listed in Table \ref{tab1} should be increased to 
$\sim$ 0.248. This is indeed the case.
Examination of the maximum likelihood linear regressions in Table \ref{tab4} 
for the entries with \cite{B02} emissivities and $N$ = 93 show that   
$Y_p$ = 0.247$\pm$0.001, whether O/H is calculated with $T_e$(O {\sc iii}) or 
$T_e$(He$^+$). Using the \citet{P05} emissivities increases $Y_p$ by 
$\la$1.7\%, so that $Y_p$ becomes equal to $\sim$ 0.251.

\section{The primordial He mass fraction $Y_p$ and the slope d$Y$/d$Z$}\label{primo}

To account for the various systematic errors, including those caused
by collisional and fluorescent enhancements of He {\sc i} lines and those 
summarized in Table \ref{tab1}, we have used a Monte Carlo method as 
described in \S\ref{method3} and applied by \citet{IT04} to a selected sample 
of 7 H {\sc ii} regions. We need to fix some parameters and 
vary others within ``reasonable'' ranges. We discuss our choices below.   
 
One of the most
important sources of uncertainty concerns the He {\sc i} emissivities.
Changing from the \citet{B99} and
\citet{B02} emissivities to those of \citet{P05}
changes $Y_p$ by $\sim$ 1.7\%.  
Because the \citet{P05} emissivities have been calculated with updated 
atomic data, we will adopt them. But to compare with our previous work,
and to assess the change in $Y_p$ caused by the change in emissivities, 
we have also derived $Y_p$ with \citet{B02} emissivities. 
As for the equivalent widths of He {\sc i}
stellar absorption lines, we adopt the basic ones (\S\ref{method4}).
This set of equivalent widths appears to be the most reasonable for the
reasons discussed in \S\ref{system4} and also because it gives 
the
lowest $\chi^2_{min}$ for the derived $Y$s in individual H {\sc ii} regions.
For $T_e$(He$^+$), variations in the range 
(0.95 -- 1.0)$\times$$T_e$(O {\sc iii}) appear to be reasonable and 
are consistent 
with the measurements of \citet{G06,G07}.
We vary $N_e$(He$^+$) and $\tau$($\lambda$3889) in
the ranges 10 -- 450 cm$^{-3}$ and 0 -- 5.
The ionization correction factor $ICF$(He$^+$+He$^{2+}$) is interpolated 
or extrapolated from the curves in
Fig. \ref{fig7}. 

In Table \ref{tab2} (available only in electronic form), 
we show for the whole HeBCD+SDSS sample oxygen abundances 
O/H and nitrogen abundances N/H calculated for two different cases, one where 
the electron temperature $T_e$ is set to $T_e$(O {\sc iii}) and 
one where it is set to $T_e$(He$^+$), and
He mass fractions $Y$. 
Table \ref{tab3} (also in electronic form only) shows the values of
the parameters with which the abundances in Table \ref{tab2} are calculated. 

\begin{figure*}
\figurenum{9}
\epsscale{1.1}
\plottwo{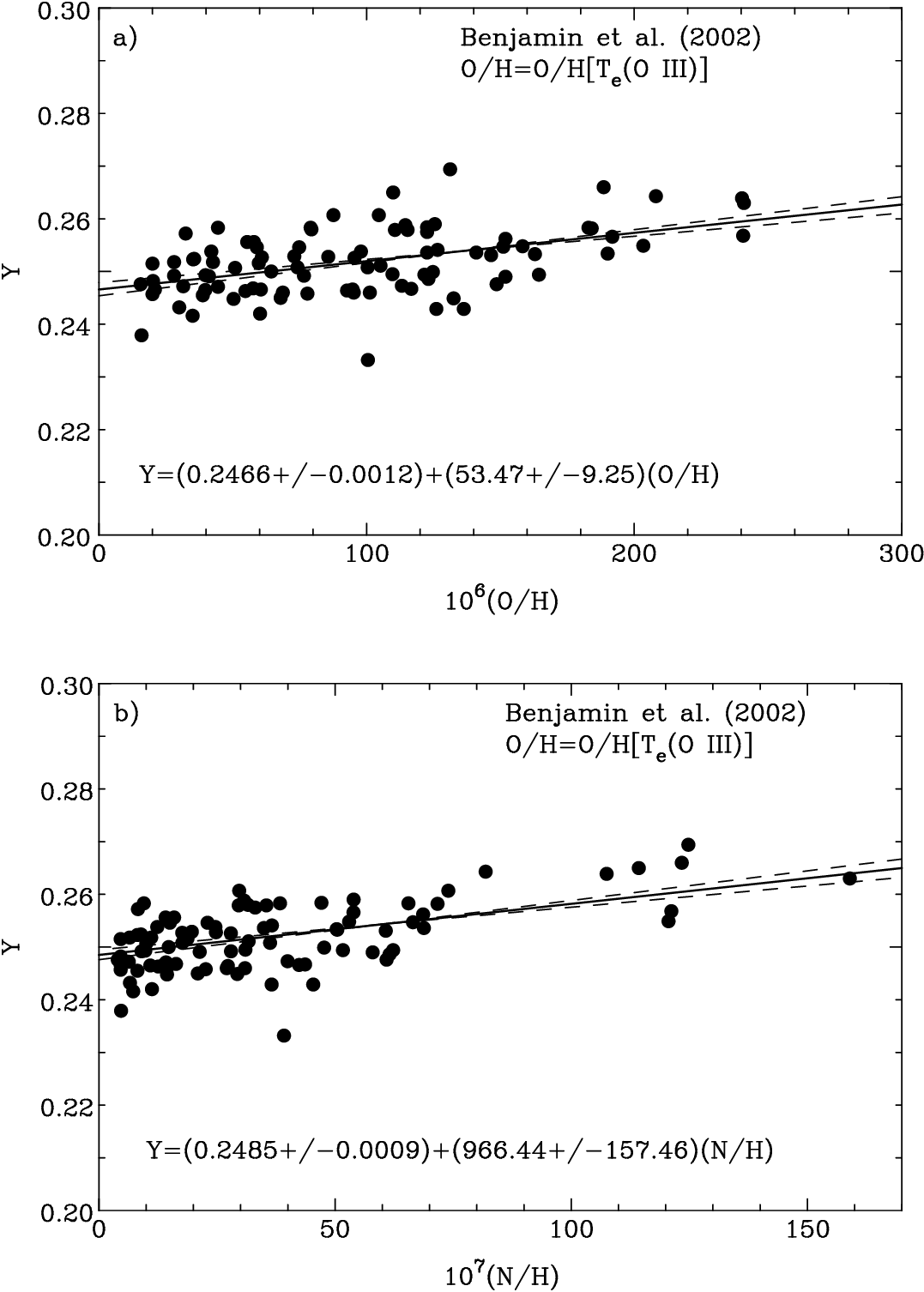}{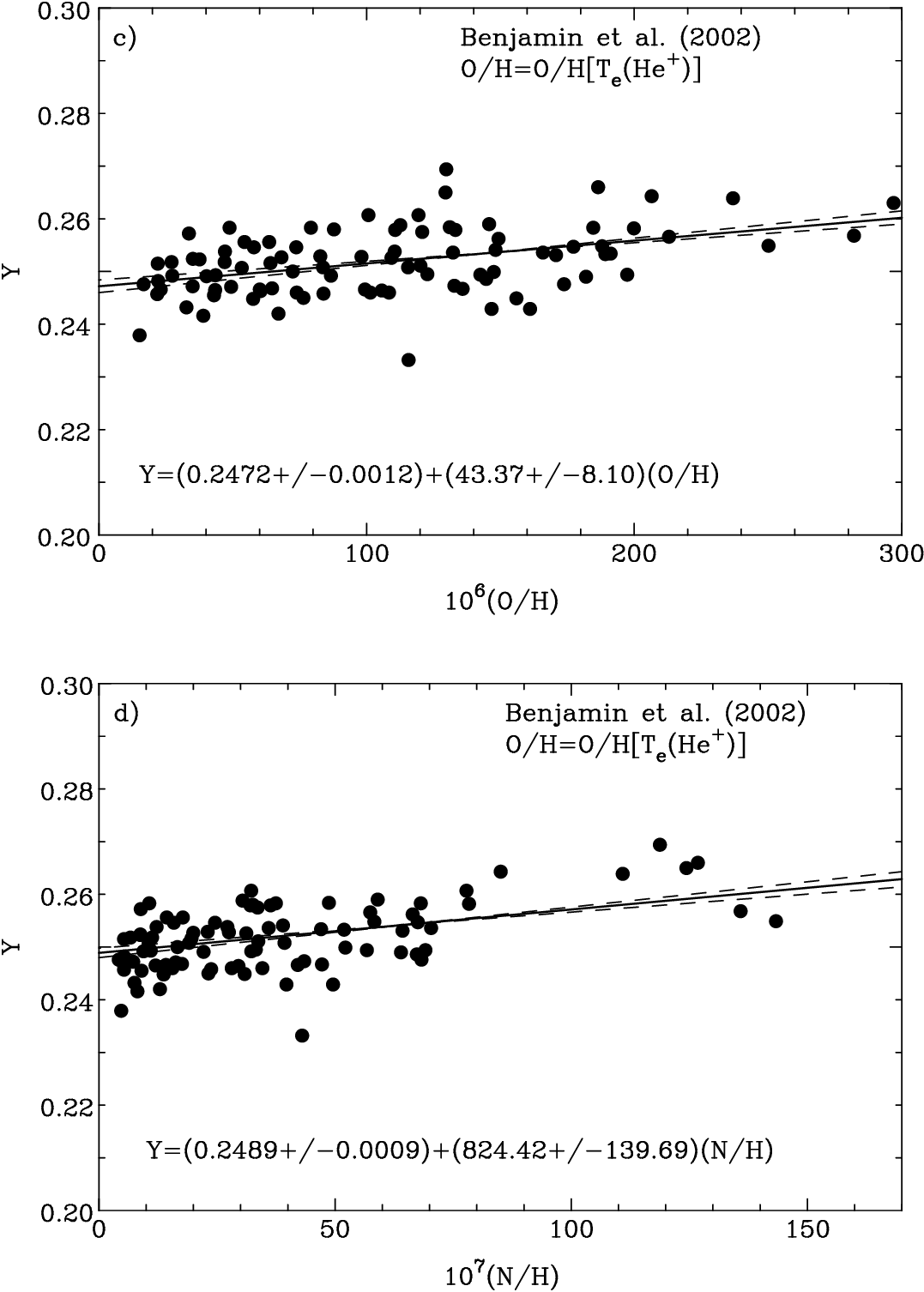}
\figcaption{  Linear regressions of the helium mass fraction $Y$ vs. oxygen 
and nitrogen abundances for H {\sc ii} regions from the HeBCD sample. 
The $Y$s are derived with the He {\sc i} emissivities
from \citet{B99,B02}. The electron temperature 
$T_e$(He$^+$) is varied in the range (0.95 -- 1)$\times$$T_e$(O {\sc iii}).
The oxygen abundance is derived adopting $T_e$(O {\sc iii}) in a) and b)
and $T_e$(He$^+$) in c) and d).
\label{fig9}}
\end{figure*}

\begin{figure*}
\figurenum{10}
\epsscale{1.1}
\plottwo{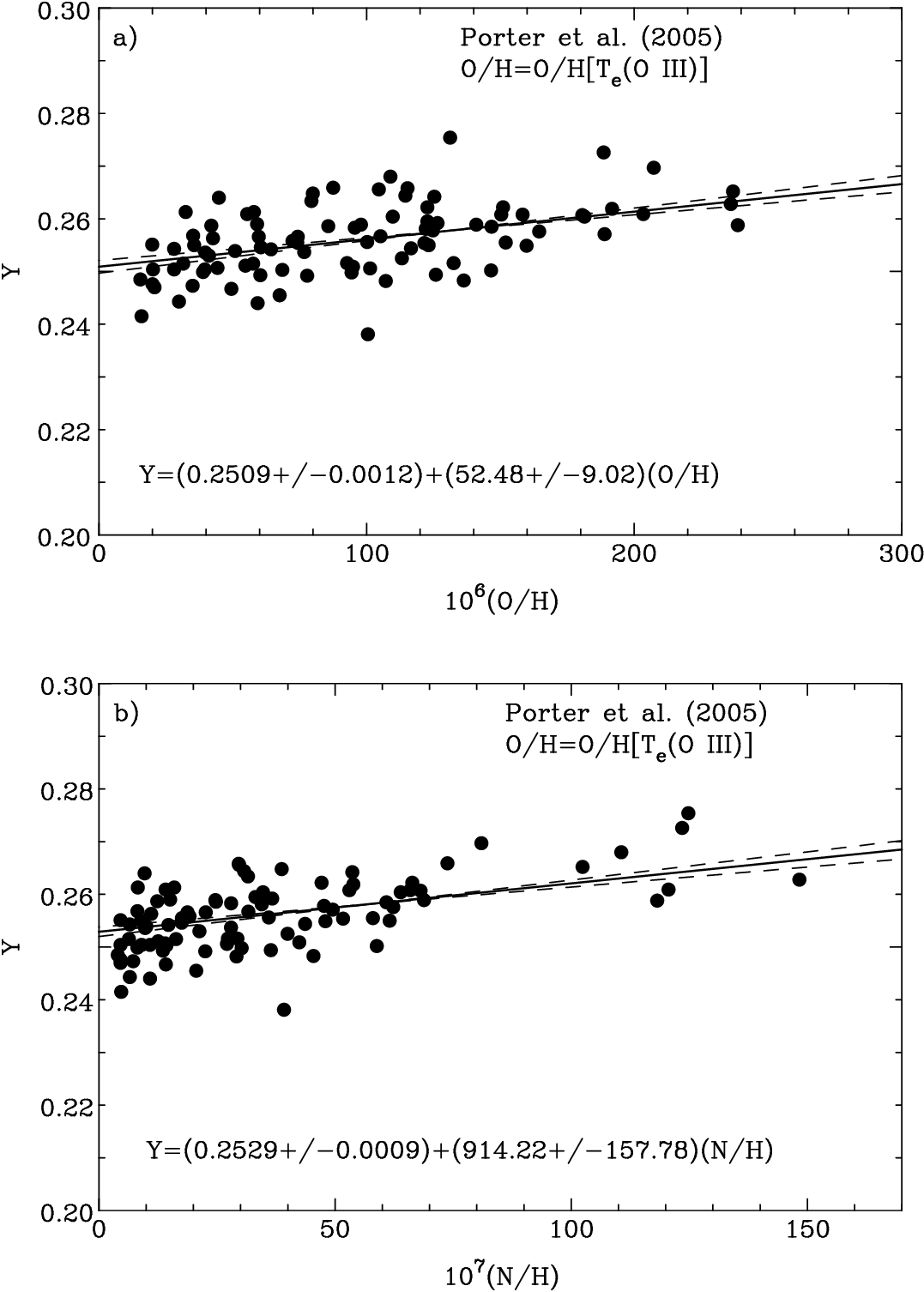}{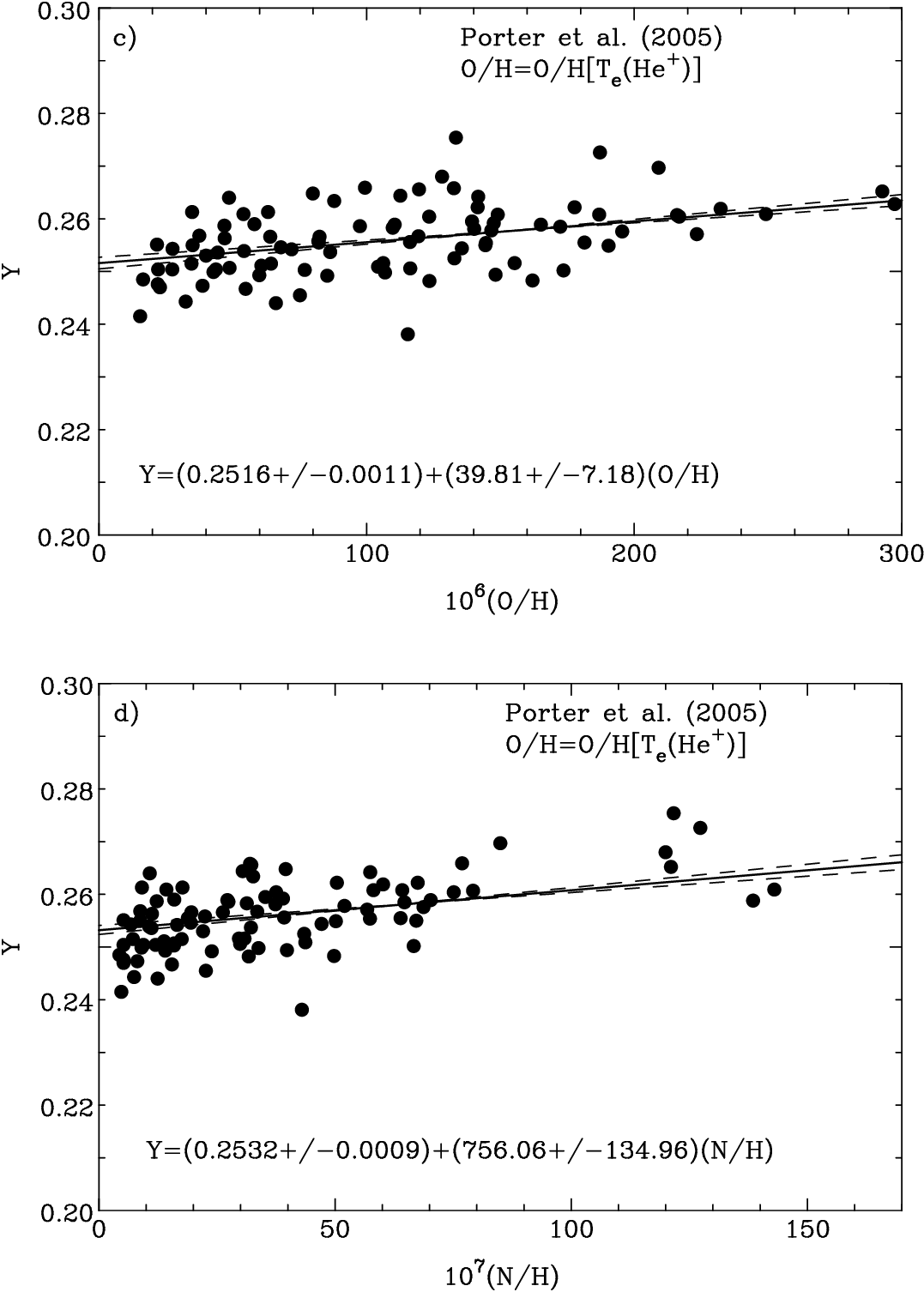}
\figcaption{  Same as in Fig. \ref{fig9} but $Y$s are derived
with the \citet{P05} He {\sc i} emissivities.
\label{fig10}}
\end{figure*}

Linear regressions $Y$ -- O/H and $Y$ -- N/H for the HeBCD sample calculated 
with the \citet{B02} He {\sc i} emissivities are shown in Fig. \ref{fig9}, 
in the cases where O/H = O/H[$T_e$(O {\sc iii})] (Fig. \ref{fig9}a,b) 
and O/H = O/H[$T_e$(He$^+$)] (Fig. \ref{fig9}c,d). 
The corresponding regression lines calculated with the \citet{P05} He {\sc i} 
emissivities  are shown in Fig. \ref{fig10}. 
Comparison of Figs. \ref{fig9} and \ref{fig10} shows that $Y_p$ 
does not depend sensitively on whether 
O/H is computed with an electron temperature equal to 
$T_e$(O {\sc iii}) or $T_e$(He$^+$), while the slope
d$Y$/d(O/H) is quite sensitive to the adopted $T_e$.
This is because $Y_p$ is determined mainly by low-metallicity H {\sc ii}
regions for which changes in O/H caused by different $T_e$ are small 
compared to the whole O/H range for the sample. 
On the other hand, d$Y$/d(O/H) is mainly 
determined by high-metallicity H {\sc ii} for which changes in O/H 
caused by 
different temperatures are comparable to the whole O/H range.
The same is true for the $Y$ -- N/H linear regressions.

From the $Y$ -- O/H regressions for the HeBCD sample with the \citet{B02} and
\citet{P05} He {\sc i} emissivities and O/H = O/H[$T_e$(He$^+$)],
we obtain respectively $Y_p$ = 0.2472 $\pm$ 0.0012 and
0.2516 $\pm$ 0.0011 (Fig. \ref{fig9}c, \ref{fig10}c).  
The parameters of the regression lines are also given in Table \ref{tab4}.
The values of $Y_p$ obtained from the $Y$ -- N/H linear regressions 
(Fig. \ref{fig9}d, \ref{fig10}d, Table \ref{tab4}) are higher, being 
respectively  $Y_p$ = 0.2489 $\pm$ 0.0009
and $Y_p$ = 0.2532 $\pm$ 0.0009 for the \citet{B02} and \citet{P05}
He {\sc i} emissivities. 
As noted before \citep[e.g. ][ and references therein]{IT04},
the systematically higher $Y_p$ from the $Y$ -- N/H regression is due to the
nonlinear dependence of the nitrogen abundance on oxygen abundance. 
This makes the $Y$ -- N/H linear regression less reliable 
for the determination of $Y_p$. 

\begin{figure*}
\figurenum{11}
\epsscale{1.1}
\plottwo{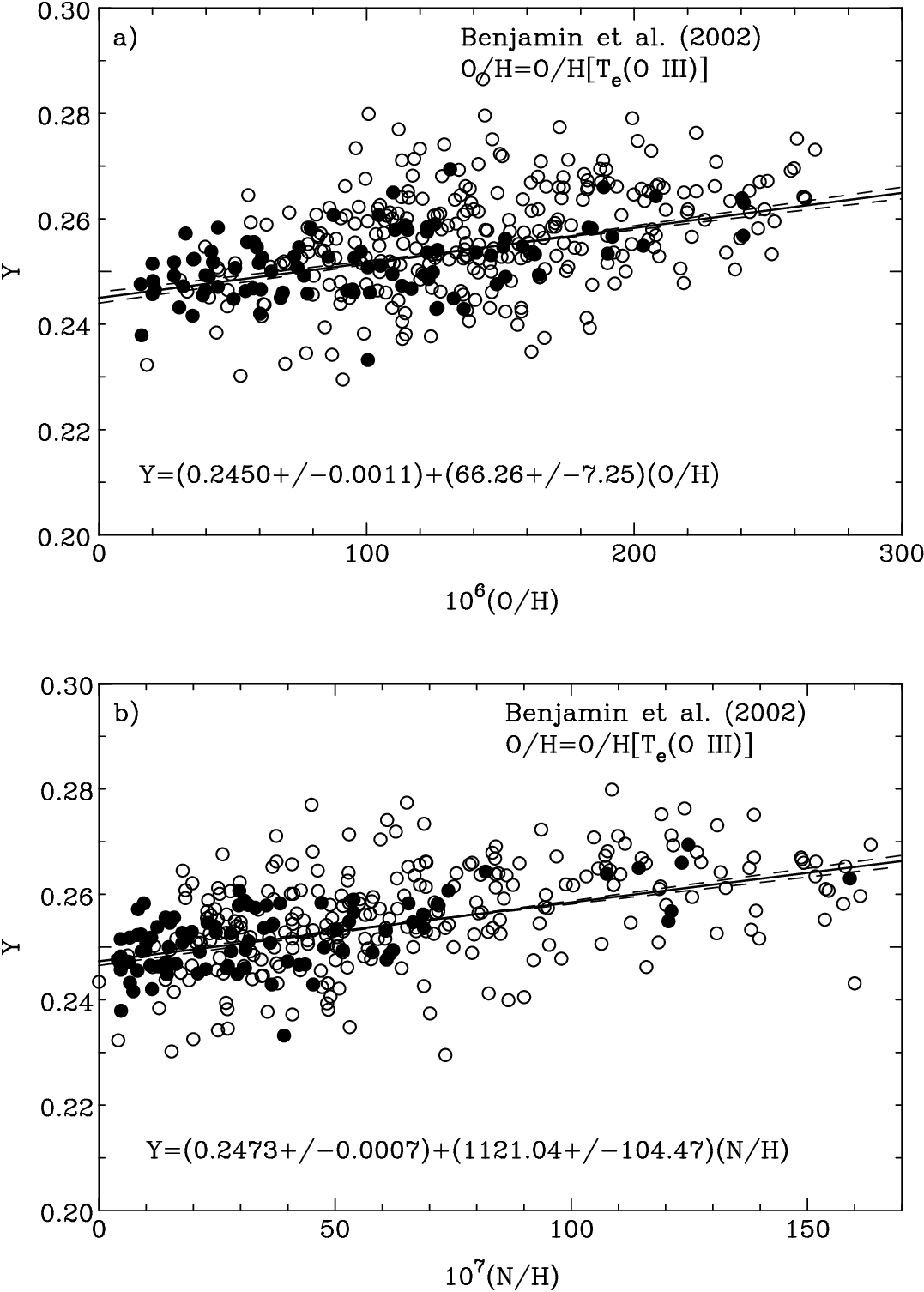}{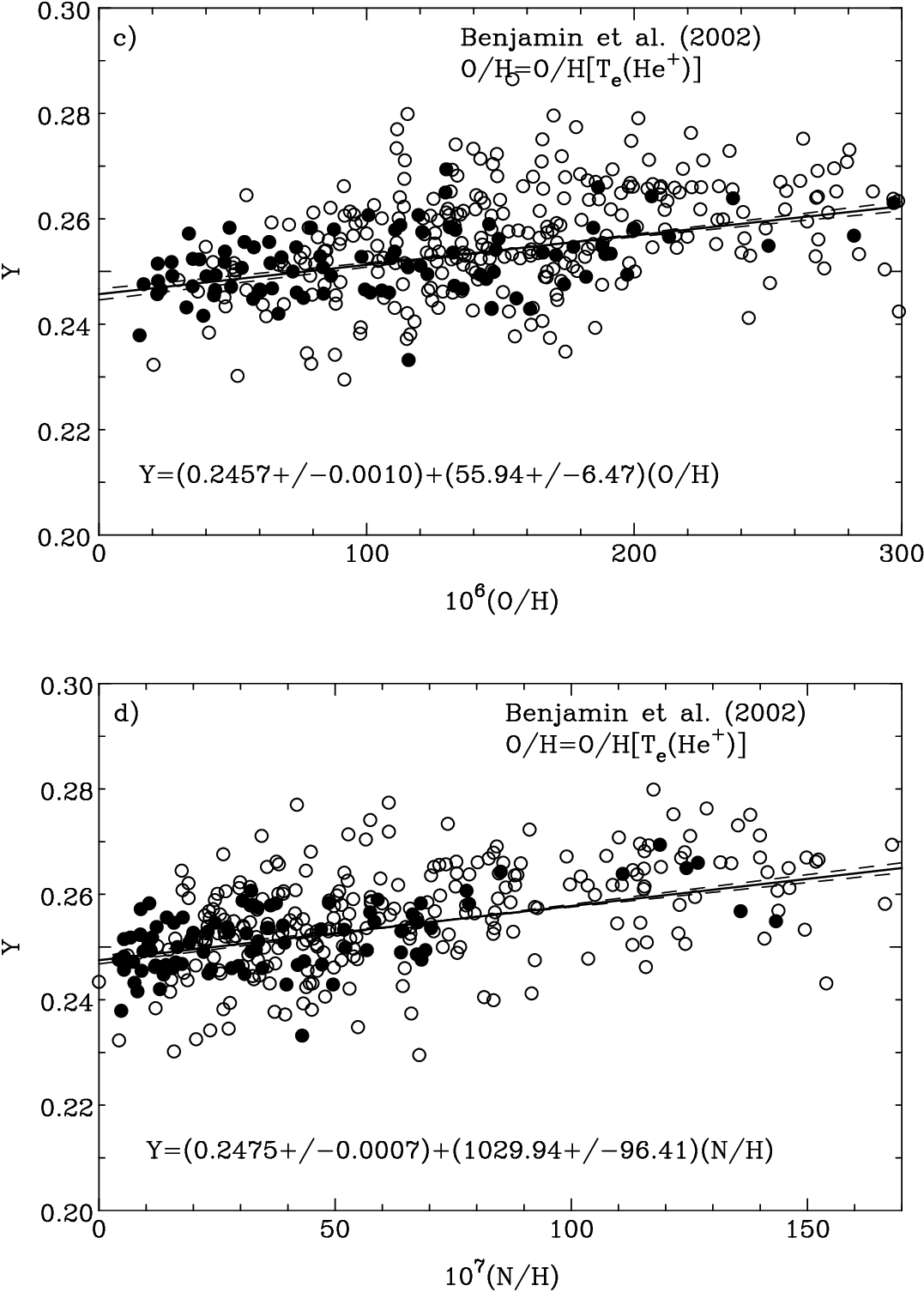}
\figcaption{ Same as in Fig. \ref{fig9}, but for the
HeBCD + SDSS sample. HeBCD H {\sc ii} regions are shown by filled
circles and SDSS H {\sc ii} regions by open circles.
\label{fig11}}
\end{figure*}

\begin{figure*}
\figurenum{12}
\epsscale{1.1}
\plottwo{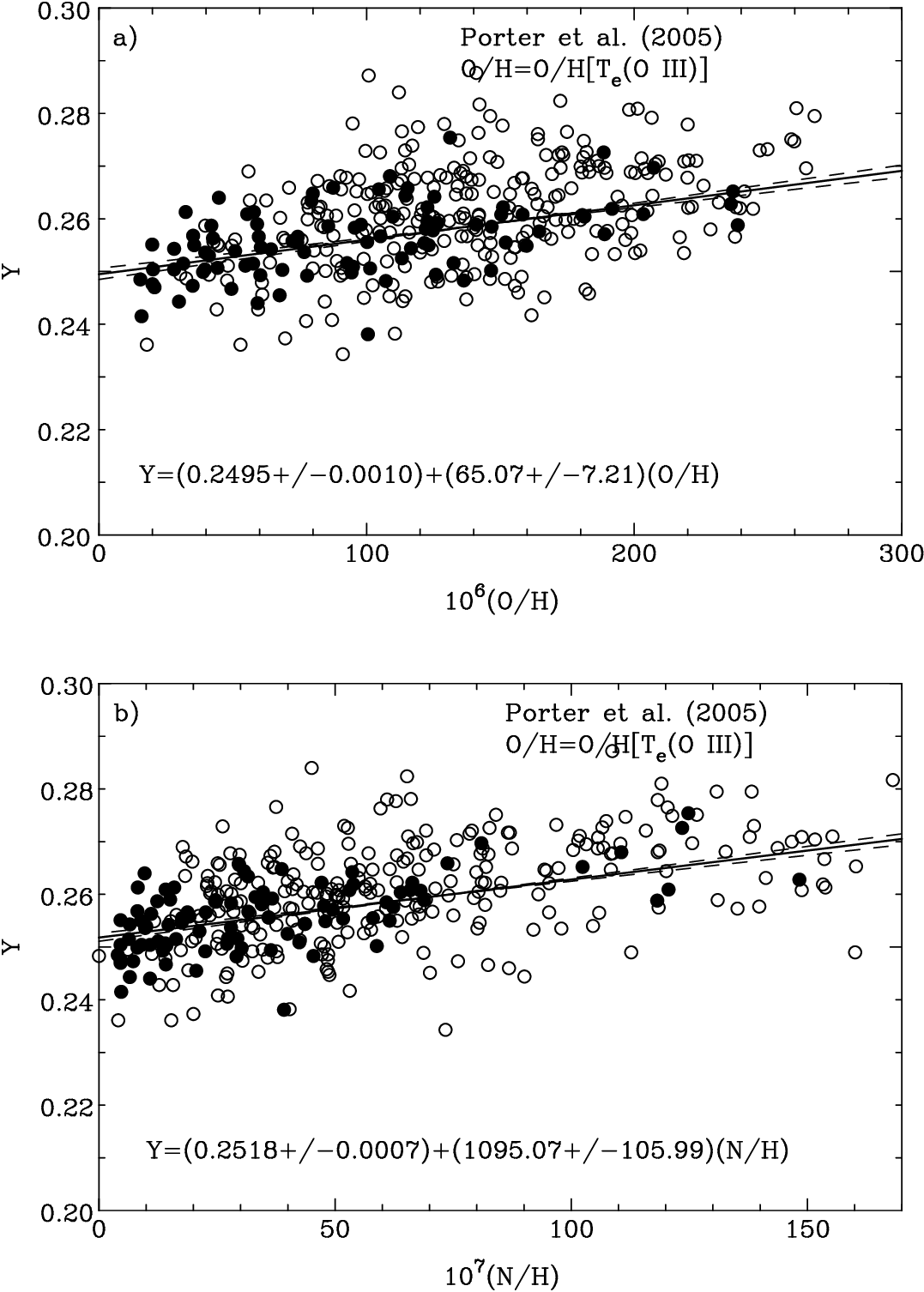}{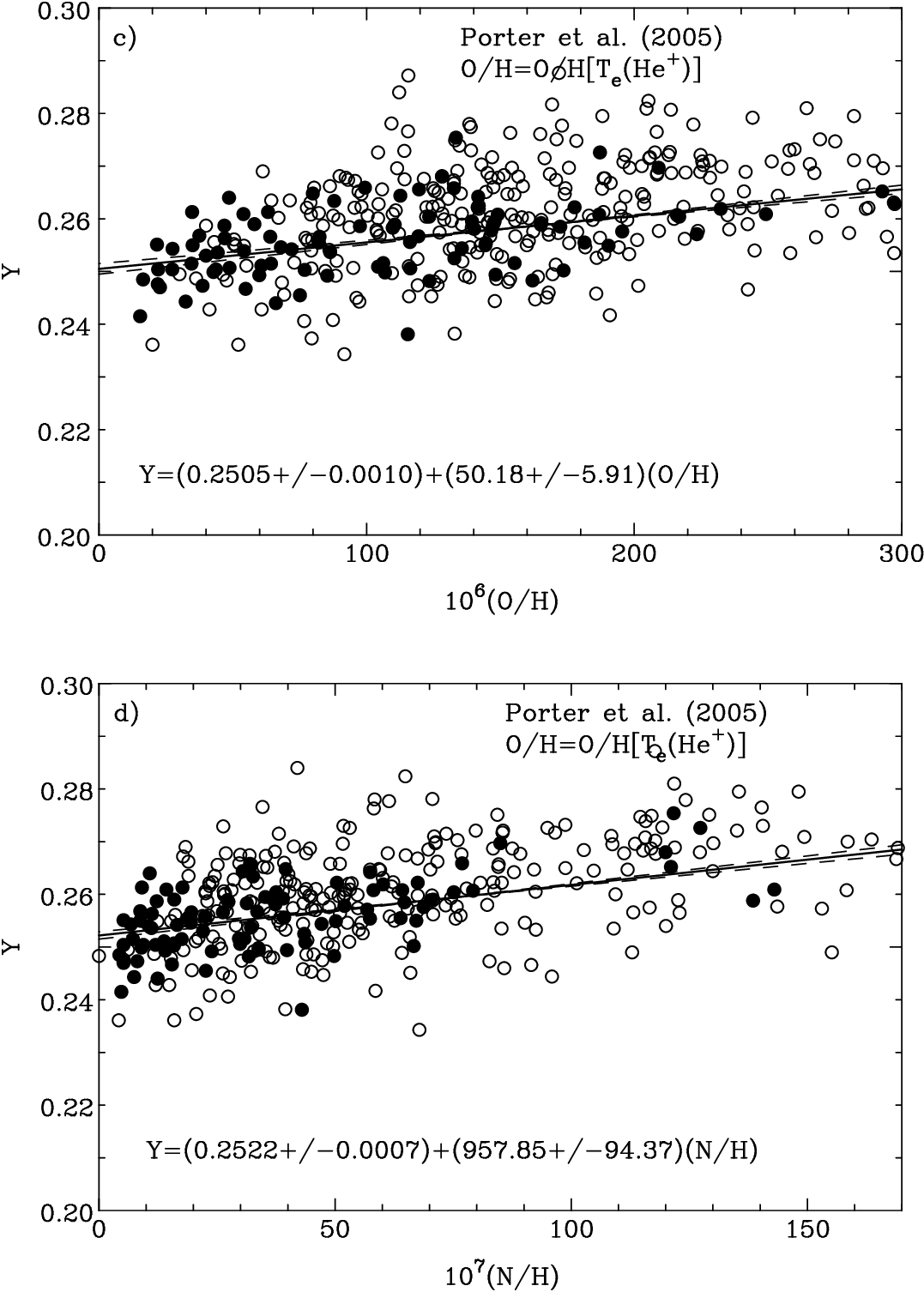}
\figcaption{  Same as in Fig. \ref{fig10}, but for the
HeBCD + SDSS sample. HeBCD H {\sc ii} regions are shown by filled
circles and SDSS H {\sc ii} regions by open circles.
\label{fig12}}
\end{figure*}

For comparison, the linear regressions for the total HeBCD + SDSS sample 
with the \citet{B02} and \citet{P05} He {\sc i} emissivities
are shown respectively in Figs. \ref{fig11} and 
\ref{fig12}.
The HeBCD H {\sc ii} regions are represented by filled circles and the SDSS 
H {\sc ii} regions by open circles. Using the $Y$ -- O/H linear 
regressions, we obtain
$Y_p$ = 0.2457 $\pm$ 0.0010 (Fig. \ref{fig11}c) when  
\citet{B02} emissivities are used and $Y_p$ = 0.2505 $\pm$ 0.0010
(Fig. \ref{fig12}c) when \citet{P05} emissivities are used.
The $Y_p$s derived for the HeBCD+SDSS
sample are consistent within the errors with the ones derived 
from the HeBCD sample alone (compare the regression parameters in Table 
\ref{tab4}, where the total sample is characterized by $N$ = 364 H {\sc ii} 
regions, and the HeBCD sample by $N$= 93 H {\sc ii} regions).
As can be seen from Figures \ref{fig11} and \ref{fig12}, the scatter about the 
regression lines is considerably larger for the SDSS sample than for 
the HeBCD sample.
Thus since the data points entering in the 
determination of the regression lines are 
weighted by their errors, most of the weight in the $Y_p$ determination still 
comes from the HeBCD sample, even in the total HeBCD+SDSS sample.
Going from 93 to 364 H {\sc ii} regions does not change 
appreciably the 1$\sigma$ error in the determination of $Y_p$: it 
decreases only by 0.0001.  
Thus, while we show the results for the total sample for comparison, 
we consider the $Y_p$ determined from the HeBCD sample, for which 
we understand well the selection effects, as being the best value.  

Using Eq. \ref{eq:dO}, we derive from the $Y$ -- O/H linear regressions
(Table \ref{tab4})  
the slopes d$Y$/d$Z$ = 2.38$\pm$0.45
and 2.19$\pm$0.39   
for the HeBCD sample, in the case where O/H is calculated 
with $T_e$ (He$^+$), and for the \citet{B02} and
\citet{P05} He {\sc i} emissivities respectively 
(Fig. \ref{fig11}c, \ref{fig12}c). 
These slopes are steeper than the ones predicted by closed-box
chemical evolution models of dwarf galaxies \citep[$\sim$ 1, ][]{L01}. 
It is reasonable to expect
that some fraction of chemical elements, especially oxygen, escapes
the galaxy due to e.g. supernova explosions.
The slopes d$Y$/d$Z$ derived from the $Y$ -- N/H linear regressions
(Fig. \ref{fig11}d,\ref{fig12}d) are not as steep: d$Y$/d$Z$ = 1.46$\pm$0.25
and 1.34$\pm$0.24 respectively for the \citet{B02} and \citet{P05} He {\sc i}
emissivities, and for the case  where O/H is calculated 
with $T_e$(He$^+$).
 Again, the difference is due to the 
nonlinear dependence of the N abundance
on O abundance, as discussed above. As for $Y_p$, this makes the 
determination of d$Y$/d$Z$ from the 
$Y$ -- N/H regression not as reliable as the one from the $Y$ -- N/H 
regression.
The d$Y$/d$Z$ slopes are steeper when 
$T_e$(O {\sc iii}) is used for the O/H determination. In the latter case, 
we obtain for the HeBCD sample, 
from the $Y$ -- O/H regressions, d$Y$/d$Z$ = 2.94$\pm$0.51 and 
2.88$\pm$0.50 respectively for the two sets of emissivities 
(Fig. \ref{fig11}a and \ref{fig12}a). The true electron temperature is 
probably somewhere between $T_e$(He$^+$) and $T_e$(O {\sc iii}), so we expect 
the true slopes to be in the range 2.2 -- 2.9 for both \citet{B02} and 
\citet{P05} emissivities.   
    
Our $Y_p$ range of $\sim$ 0.001 is much smaller than the one 
of $\sim$ 0.009 claimed by \citet{OS04}. This difference is due to 1) 
different methods used in this paper and in \citet{OS04} and 
2) much larger samples used in this paper 
(93 observations in the HeBCD sample and 364 observations in the total 
sample as compared to 7 objects in the \citet{OS04} sample). 
We emphasize that the use of large sample significantly reduces the
uncertainties in $Y_p$ to the level that is acceptable for 
cosmological implications.
    

\section{COSMOLOGICAL IMPLICATIONS}\label{cosmo}

\subsection{The baryonic mass density}\label{cosmo1}

\begin{figure*}
\figurenum{13}
\epsscale{1.1}
\plottwo{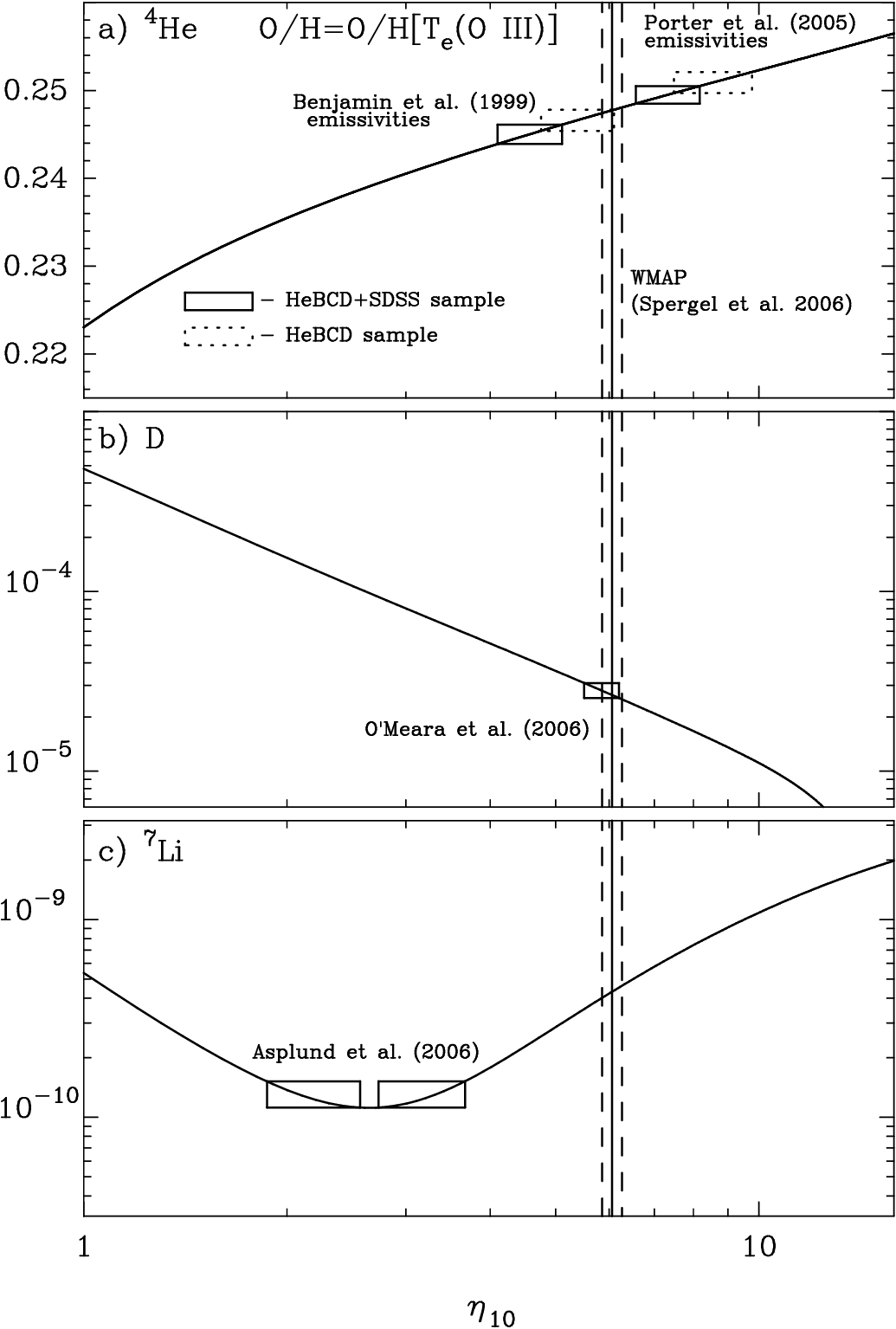}{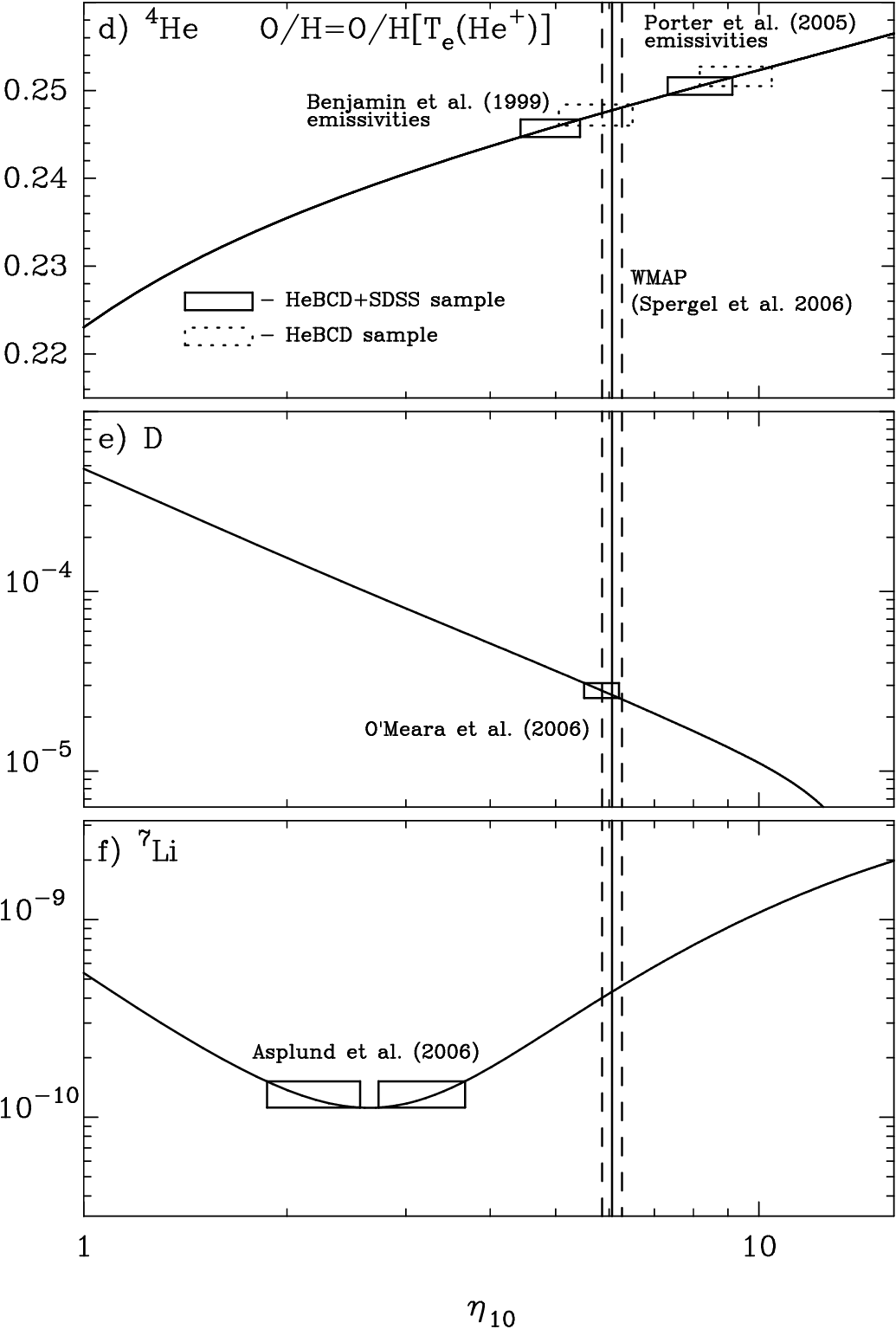}
\figcaption{ Dependence of the helium mass fraction (a),(d), deuterium (b),(e)
and $^7$Li (c),(f) abundances on the photon-to-baryon number ratio $\eta$.
Solid curves are predictions of the SBBN, solid and dashed vertical lines
indicate the $\eta$ and 1$\sigma$ deviations derived from WMAP \citep{S06}.
Boxes show the observed light element abundances 
along with their 1$\sigma$ deviations. The solid boxes correspond to the
combined HeBCD + SDSS sample and the dotted boxes to the HeBCD sample only.
Boxes in (a) are calculated adopting O/H = O/H[$T_e$(O {\sc iii})], and
boxes in (d) are calculated adopting O/H = O/H[$T_e$(He$^+$)].
\label{fig13}}
\end{figure*}

We now investigate whether our derived values of $Y_p$ are consistent 
with the predictions of SBBN and whether the baryonic mass density 
corresponding to $Y_p$ agrees with the one derived from measurements 
of the CMB.   
In Fig. \ref{fig13}, we show by solid curves 
the dependence of the abundances of the three  
light elements He, D and $^7$Li on the 
baryon-to-photon number ratio $\eta$ as predicted by SBBN 
\citep[e.g. ][]{St06}. We do not discuss $^3$He because the derivation 
of its primordial value from its presently observed value in the Galaxy 
is complicated by uncertain effects of chemical evolution.
The solid and dashed vertical lines show the value of 
$\eta$ and its 1$\sigma$
deviations as obtained by WMAP \citep{S06}. The presently best observational 
abundances of the three light elements with their
1$\sigma$ deviations are shown by boxes. Concerning He, we show 
in the left and right panels 
the $Y_p$ derived from regression fitting of the HeBCD sample and 
with O/H obtained respectively with 
the electron temperature set equal to  
$T_e$(O {\sc iii}) and $T_e$(He$^+$).
We shall discuss the results concerning our preferred values of $Y_p$, 
the ones 
corresponding to  $T_e$ =  $T_e$(He$^+$).
With $Y_p$ = 0.2472$\pm$0.0012 and 0.2516$\pm$0.0011
(solid boxes in Fig. \ref{fig13}d) 
for the two different sets of He {\sc i} emissivities, and with
an equivalent number of light neutrino species equal to 3, the SBBN model 
gives $\eta_{10}$ = 10$^{10}$$\eta$ =
5.5$^{+0.7}_{-0.6}$ and 8.7$^{+1.1}_{-1.0}$, respectively, where the
error bars denote 1$\sigma$ errors. These values correspond to a baryonic mass
fractions $\Omega_b$$h^2$ = 0.020$^{+0.003}_{-0.002}$ and 
0.032$^{+0.004}_{-0.004}$ respectively for \citet{B02} and \citet{P05} 
He {\sc i} emissivities. The value derived with the old \citet{B02} 
He {\sc i} emissivities is much higher than $\Omega_b$$h^2$ derived
from the primordial $^7$Li abundance \citep{BM97,As06,B06} but
is in excellent agreement with
$\Omega_b$$h^2$ = 0.021$\pm$0.002 derived from recent 
measurements of the deuterium abundance in damped Ly$\alpha$ systems 
\citep{K03,O06} and $\Omega_b$$h^2$ = 0.0223$\pm$0.0008 derived from 
measurements of the fluctuations of the microwave
background radiation by WMAP \citep{S06}. Another way of saying 
the same thing is that the observational value of 
$Y_p$ derived with the \citet{B02} 
He {\sc i} emissivities is in excellent agreement with the theoretical 
SBBN value of 0.248.    

On the other hand, the $Y_p$ derived with the new \citet{P05} He {\sc i} 
emissivities is larger than the predicted SBBN value and is only 
consistent with it at the 2$\sigma$ level. Consequently, the corresponding  
$\Omega_b$$h^2$ = 0.032$^{+0.004}_{-0.004}$ is also consistent only 
at the 2$\sigma$ level with the value inferred from the D abundance 
measurements and the WMAP data. In principle, 
the \citet{P05} He {\sc i} emissivities should be more reliable than
those from \citet{B02}, since they are based on 
new updated atomic data for He.

\citet{St06} has shown that, while He may not be as sensitive
a baryometer as D, it is an excellent chronometer (in the sense that 
it is sensitive to small deviations from the standard Hubble expansion 
rate) and/or leptometer (in the sense that it is sensitive to any 
asymmetry between the numbers of neutrinos and antineutrinos).
We investigate next possible small deviations from SBBN if we take at face 
value the relatively high value of $Y_p$ obtained with the 
\citet{P05} He {\sc i} emissivities.

\subsection{Deviations from SBBN}\label{cosmo2}  
 
Deviations from the standard rate of Hubble expansion in the early Universe 
can be caused by an extra contribution to the total energy density (for 
example by additional flavors of active or sterile neutrinos) which can 
conveniently be parameterized by an equivalent number of neutrino 
flavors $N_\nu$. 
Combining $\Omega_b$$h^2$ = 0.00223 $\pm$ 0.0008 obtained by WMAP \citep{S03}
with $Y_p$ = 0.2516 $\pm$ 0.0011, we obtain $N_\nu$ $\sim$ 3.2
\citep{W91}.

We use the statistical $\chi^2$ technique with the code described by
\citet{Fi98} and \citet{Li99} to analyze the constraints 
that the measured He, D and $^7$Li abundances put 
on $\eta$ and $N_\nu$. 
For the primordial D and $^7$Li abundances, we use respectively the 
values obtained by \citet{O06} and \citet{As06}. With  
$Y_p$ = 0.2472$\pm$ 0.0012 \citep[emissivities by ][]{B02}, 
the minimum $\chi^2_{min}$ = 9.036 is obtained
when $\eta_{10}$ = 5.79 and $N_\nu$ = 2.972. 
With $Y_p$ = 0.2516$\pm$0.0011 \citep[emissivities by ][]{P05},
the minimum $\chi^2_{min}$ = 9.334 is obtained  
when $\eta_{10}$ = 5.97 and $N_\nu$ = 3.280.
 
\begin{figure*}
\figurenum{14}
\epsscale{0.7}
\plotone{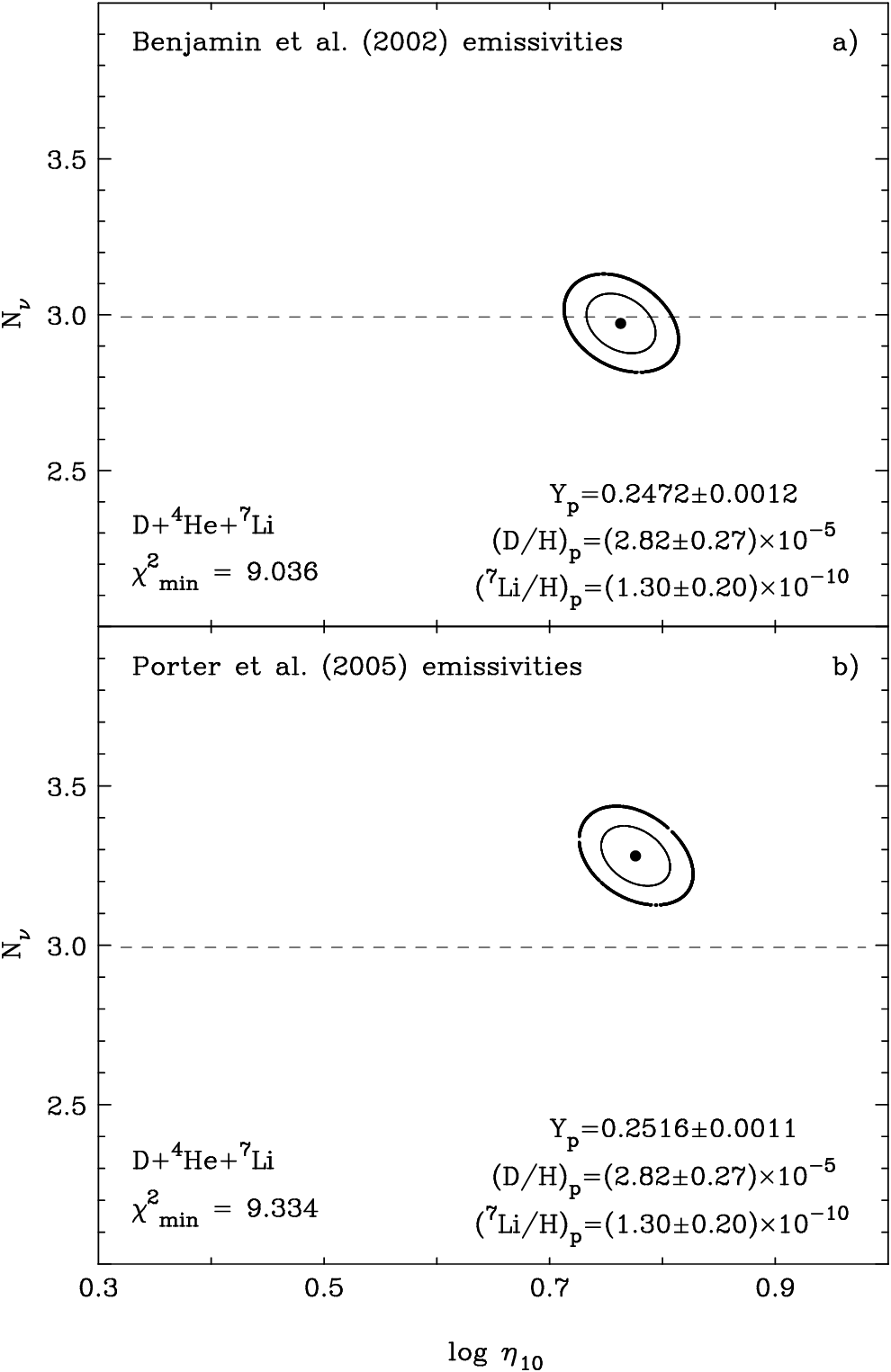}
\figcaption{ Joint fits to the baryon-to-photon number ratio, 
log $\eta_{10}$, and the equivalent number of light neutrino species $N_\nu$, 
using a $\chi^2$ analysis with the code developed by \citet{Fi98}
and \citet{Li99} (a) for the primordial abundance value $Y_p$ 
derived with the \citet{B02} He {\sc i} emissivities (this paper), 
(D/H)$_p$ from \citet{O06} and ($^7$Li/H)$_p$ from \citet{As06} 
and (b) for the primordial abundance value $Y_p$ 
derived with the \citet{P05} He {\sc i} emissivities (this paper),
and the same (D/H)$_p$ and ($^7$Li/H)$_p$ as in (a). 
Thin and thick solid lines are respectively 1$\sigma$ and 2$\sigma$ deviations.
The experimental value $N_\nu$ = 2.993 \citep{Ca98} is shown by the
dashed line.
\label{fig14}}
\end{figure*}

The joint fits of $\eta$ and $N_\nu$ with the $Y_p$s derived with
the two sets of emissivities are shown respectively in
Figure \ref{fig14}a and \ref{fig14}b. 
The 1$\sigma$ ($\chi^2$ -- $\chi^2_{min}$ = 1.0) and
2$\sigma$ ($\chi^2$ -- $\chi^2_{min}$ = 2.71) deviations are shown 
respectively by the thin and thick solid lines. 
We find the equivalent number of
light neutrino species to be respectively in the range
$N_\nu$ = 2.97$\pm$0.16 (2$\sigma$) (Fig. \ref{fig14}a), and 
$N_\nu$ = 3.28$\pm$0.16 (2$\sigma$) (Fig. \ref{fig14}b). The first value 
of $N_\nu$ obtained with $Y_p$ derived with
 the old set of He {\sc i} emissivities 
is entirely consistent with the experimental 
value of 2.993$\pm$0.011 \citep{Ca98} shown by the dashed line. 
On the other hand, the second value of $N_\nu$ obtained with $Y_p$  
derived with the new set of He {\sc i} emissivities is significantly higher 
than the experimental value, suggesting a slight deviation from 
SBBN. We note that the primordial helium abundance sets very tight
constraints on the effective number of neutrino species. These constraints
are much tighter than those derived using the CMB and galaxy clustering
power spectra. For example, using these two sets of data, \citet{Ic06}
derive 0.8 $<$ $N_\nu$ $<$ 7.6 at the 95\% confidence level.

Alternatively, deviations from the SBBN model can be checked by 
introducing two parameters, the expansion
rate parameter $S$, and the electron neutrino asymmetry parameter $\xi_e$,
as suggested by \citet{St05}. 
If deviations from the standard expansion are caused by an extra non-standard 
energy density component, then it is convenient to 
express that extra energy component in terms of an extra effective number 
of neutrino species $\Delta$$N_\nu$ defined as: $N_\nu$ = 3 + $\Delta$$N_\nu$.
Then $S$ is related to $\Delta$$N_\nu$ by: 
\begin{equation}
S=\left(1+\frac{7}{43}\Delta N_\nu\right)^{1/2}. \label{S}
\end{equation}
For the SBBN model, $\Delta$$N_\nu$ = 0, $N_\nu$ = 3 and $S$ = 1.

As for the parameter $\xi_e$, it is related to the difference between the 
number of electron neutrino species $n_{\nu_e}$ and the number of 
electron antineutrino species $n_{{\bar\nu}_e}$ by :
\begin{equation}
\xi_e \approx 1.33 \left(\frac{n_{\nu_e} - n_{{\bar\nu}_e}}{n_\gamma}\right),
\label{xi}
\end{equation}
where $n_\gamma$ is the number of photons. A  non-zero $\xi_e$ would 
result in a different n/p ratio during the BBN period, 
therefore changing the light element abundances.
In particular, if 
$\xi_e$ $>$ 0, there are more neutrinos than antineutrinos, and reactions
such as n + $\nu_e$ $\rightarrow$ p + $e^-$ drive down the n/p ratio 
\citep{St05}.

\begin{figure*}
\figurenum{15}
\epsscale{0.7}
\plotone{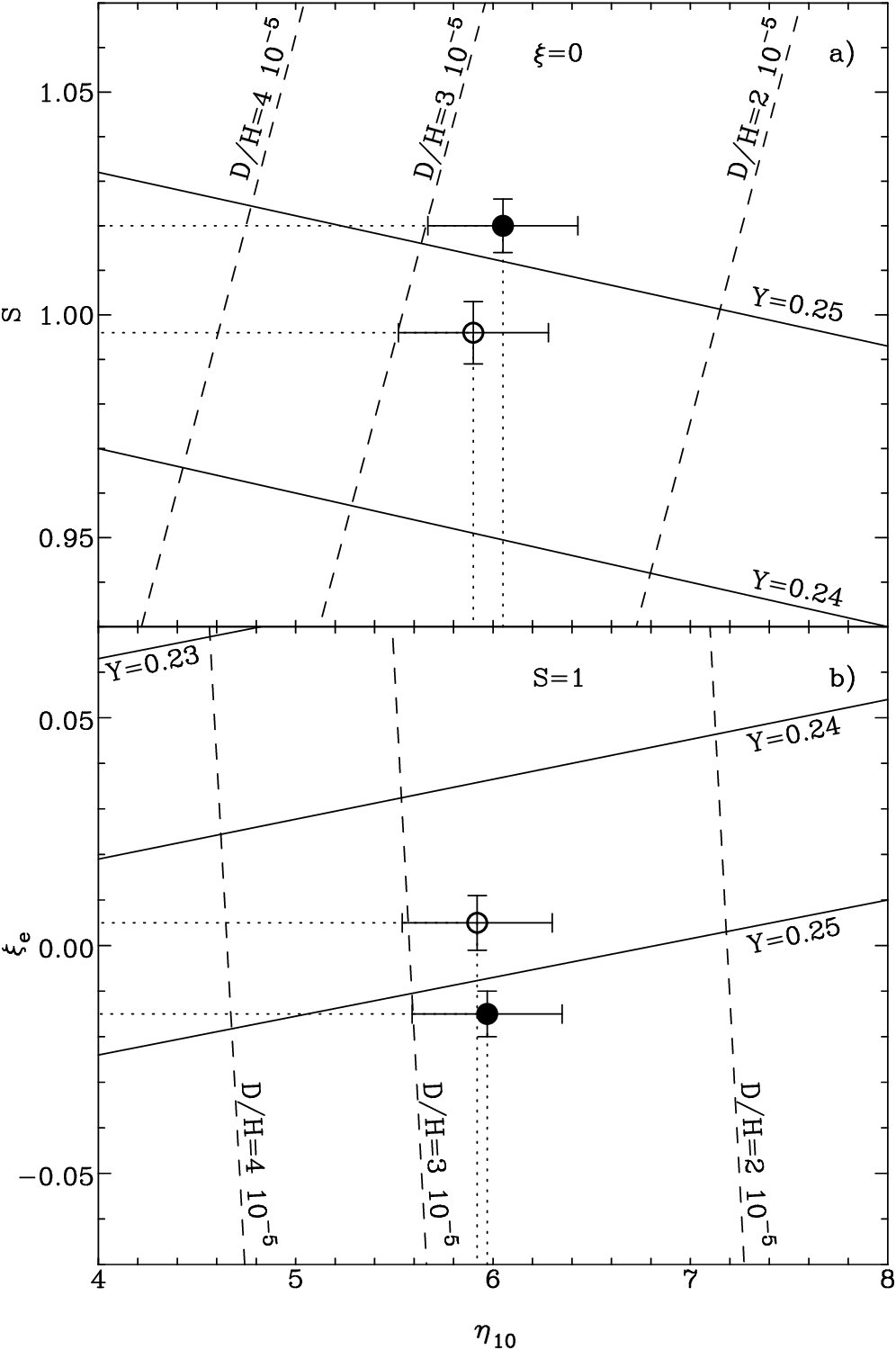}
\figcaption{(a) Diagram showing the expansion rate factor $S$ vs the 
baryon-to-photon number
ratio $\eta$. Isoabundance curves for D and He are shown respectively 
by dashed and solid lines and labeled by the abundance values. 
The filled circle
corresponds to the primordial D/H abundance from \citet{O06} and the primordial
He mass fraction derived in this paper with the \citet{P05} emissivities. 
The open circle corresponds
to the primordial D/H abundance from \citet{O06} and the primordial
He mass fraction derived in this paper with the \citet{B99} emissivities. 
Error bars are 1$\sigma$
deviations. (b) Same as in (a) except the diagram shows the
electron neutrino asymmetry
parameter $\xi_e$ vs the baryon-to-photon number ratio $\eta$.
\label{fig15}}
\end{figure*}

Following \citet{St05}, we show in Fig. \ref{fig15}a 
the $S$ -- $\eta$ diagram 
with the nearly orthogonal 
isoabundance curves of D (dashed lines) and He (solid lines), 
in the case where $\xi_e$ = 0. The open circle corresponds to the 
value of the primordial D/H
abundance ratio derived by \citet{O06} and that of 
the primordial He mass fraction 
derived in this paper, using the \citet{B99} He {\sc i} emissivities. 
The filled circle
corresponds to the same D/H abundance ratio, but to 
$Y_p$ derived with the \citet{P05}
He {\sc i} emissivities. From Fig. \ref{fig15}a, we obtain 
$S$ = 0.996 $\pm$ 0.007 (1$\sigma$) and 1.020 $\pm$ 0.006 (1$\sigma$) 
for $Y_p$ obtained respectively with the old and new He {\sc i} emissivities.
As before, we find that the first value is consistent 
within the errors with the SBBN value, $S$ = 1. On the other hand, 
the second value indicates a slight deviation from the SBBN and 
implies the presence of additional neutrino species.
The same conclusion can be reached with Fig. \ref{fig15}b. It shows 
the $\xi_e$ -- $\eta$ diagram with the isoabundances curves 
of D (dashed lines) and 
He (solid lines), in the case where $S$ = 1. Again, it 
can be seen that the $Y_p$ derived with the \citet{B99} He {\sc i} 
emissivities
is consistent with SBBN ($\xi_e$ = 0), while the $Y_p$ derived with the
\citet{P05} He {\sc i} emissivities gives a negative value of 
$\xi_e$, indicating deviation from SBBN.

\section{SUMMARY AND CONCLUSIONS}\label{summary}

     We present in this paper the determination of the primordial
helium mass fraction $Y_p$ by linear regressions of a sample of
93 spectra of 86 H {\sc ii} low-metallicity extragalactic regions 
(the HeBCD sample). 
This sample is one of the largest and most 
homogeneous data sets in existence for the determination of $Y_p$.
For comparison and to improve the statistics for investigating 
systematic effects in the determination of $Y_p$, we have also considered a 
sample of 271 low-metallicity 
H {\sc ii} regions selected from the Data Release 5 of the Sloan Digital 
Sky Survey (the SDSS sample).  

In the determination of $Y_p$, we have considered several   
known systematic effects. 
We have used Monte Carlo methods to take
into account the effects of collisional and fluorescent enhancements 
of He {\sc i} recombination lines, of
collisional excitation of hydrogen emission lines, of underlying stellar
He {\sc i} absorption, of the difference between the temperature
$T_e$(He$^+$) in the He$^+$ zone and the temperature $T_e$(O {\sc iii})
derived from the [O {\sc iii}]$\lambda$4363/($\lambda$4959+$\lambda$5007)
flux ratio, and of the ionization correction factor $ICF$(He$^+$ + He$^{2+}$). 
We have also considered the effects of different sets of He {\sc i}
line emissivities and of different reddening laws.
We discuss the effect of possible 
deviations of He {\sc i} and H emission line intensities from case B. 

We have obtained the following results:

1. After the effect of collisional enhancement of the He {\sc i} lines,
the second most important systematic effect comes from underlying He {\sc i}
stellar absorption ($\sim$ 3\%). Other effects such as variations in 
temperature, collisional excitation of hydrogen emission lines or ionization
corrections are smaller ($\sim$ 1\%). Because those systematic effects  
can work in either direction, either increasing or decreasing $Y_p$, they
tend to cancel each other, so that the net effect of all the 
considered systematic effects is an increase of $Y_p$ of $\sim$ 2\% as 
compared to the value derived by \citet{IT04}. With the old 
set of He {\sc i} emissivities by \citet{B02} used by \citet{IT04},
and with the electron temperature set, not to $T_e$(O {\sc iii}), but to 
$T_e$(He$^+$),  
we derive $Y_p$ = 0.2472$\pm$0.0012 which, according to SBBN, corresponds 
to a baryonic mass fraction $\Omega_b$$h^2$ = 0.020$^{+0.003}_{-0.002}$.
This value of $\Omega_b$$h^2$ is consistent with 
the ones derived from deuterium abundance observations 
and WMAP microwave background radiation fluctuation measurements, and 
all three measurements concur to support the validity of the SBBN model 
\citep[SBBN predicts $Y_p$ = 0.248 when $\Omega_b$$h^2$ is set to the value 
derived by WMAP, ][]{S06}. We have checked that 
the derived $Y_p$ does not depend on the particular sample of H {\sc ii}
regions used. We have also performed regression fits for the SDSS sample 
observed and reduced by the Sloan team, and found that the derived $Y_p$ 
is consistent at the 2$\sigma$ level with the value obtained from 
the HeBCD sample.  

2. On the other hand, if the new 
set of emissivities, based on updated He {\sc i} atomic data, 
by \citet{P05} is used, then we obtain  $Y_p$ = 0.2516$\pm$0.0011, 
corresponding to $\Omega_b$$h^2$ = 0.032$^{+0.004}_{-0.004}$, 
significanly larger (at the 2$\sigma$ level) than the $\Omega_b$$h^2$ 
values derived from the deuterium abundance
and microwave background radiation fluctuation measurements. 
If we take the higher value of $Y_p$ at its face value, then this would 
imply the existence of small deviations from SBBN. 
In order to bring the high value of 
$Y_p$ into agreement with the deuterium and WMAP measurements, 
we would need an equivalent number of neutrino flavors equal to 3.303 instead 
of the canonical 3. 

3. The $dY/dZ$ slopes derived from the $Y$ -- O/H linear
regressions for the HeBCD sample, using He {\sc i} emissivities 
by \citet{B02} and \citet{P05}, are respectively equal to 2.38 $\pm$ 0.45 and 
2.19 $\pm$ 0.39, consistent with previous determinations by \citet{ITL97},
\citet{IT98b,IT04} using BCDs, and by \citet{J03} from nearby K dwarf stars.

We have considered here, as best as we can, all known systematic 
uncertainties that may affect the determination of the primordial He 
abundance. 
However, the real situation may be more complicated. 
The most important issue that we do 
not yet fully understand appears to be 
the temperature structure of H {\sc ii} regions. For 
example, detailed photoionization models of H {\sc ii} regions often 
underpredict the electron temperature that is measured from the 
[O {\sc iii}] $\lambda$4363/($\lambda$4959 + $\lambda$5007) line ratio 
\citep[see e.g. references in][]{L03}. Also, there are indications that the 
matter inside H {\sc ii} regions is not chemically homogeneous 
\citep[][ Stasi\'nska et al, in preparation]{TP05}, so that the oxygen 
abundance obtained by traditional H {\sc ii} region abundance analysis may be 
biased. How much this affects the He/H abundance ratio, as well as the slope 
in the $Y$ vs O/H relation is not known yet. Finally, as suggested long ago 
by \citet{B83}, very massive primordial stars may produce a significant 
amount of helium without overproducing metals, in which case the primordial 
helium abundance could be smaller than 
obtained from a linear extrapolation of 
the $Y$ - O/H relation. Recent computations of Population III star yields 
\citep{M03} associated with measurements of the near-infrared background 
suggest that this may be the case \citep{SF03}.

\acknowledgements
Y.I.I. is grateful to the staff of the Astronomy Department at the
University of Virginia for warm hospitality. He thanks the support
of the Deutsche Forschung Geselschaft (DFG) grant No. 436 UKR 17/25/05.
Y.I.I. and T.X.T. thank the support of National Science Foundation
grant AST 02-05785. The research described in this publication was made
possible in part by Award No. UP1-2551-KV-03 of the US Civilian Research
\& Development Foundation for the Independent States of the Former Soviet
Union (CRDF). All the authors acknowledge the work of the Sloan Digital Sky
Survey (SDSS) team.
Funding for the SDSS has been provided by the
Alfred P. Sloan Foundation, the Participating Institutions, the National
Aeronautics and Space Administration, the National Science Foundation, the
U.S. Department of Energy, the Japanese Monbukagakusho, and the Max Planck
Society. The SDSS Web site is http://www.sdss.org/.
     The SDSS is managed by the Astrophysical Research Consortium (ARC) for
the Participating Institutions. The Participating Institutions are The
University of Chicago, Fermilab, the Institute for Advanced Study, the Japan
Participation Group, The Johns Hopkins University, the Korean Scientist Group,
Los Alamos National Laboratory, the Max-Planck-Institute for Astronomy (MPIA),
the Max-Planck-Institute for Astrophysics (MPA), New Mexico State University,
University of Pittsburgh, University of Portsmouth, Princeton University, the
United States Naval Observatory, and the University of Washington.



\clearpage

\clearpage

\clearpage

\clearpage

\clearpage

\clearpage

\clearpage

\clearpage

\clearpage

\clearpage

\clearpage

\clearpage

\clearpage

\clearpage

\clearpage

\begin{deluxetable}{lrrrrrrrr}
 \tabletypesize{\scriptsize}
 \tablenum{1}
 \tablecolumns{9}
 \tablewidth{0pc}
 \tablecaption{Emission line fluxes \label{tab5}}
 \tablehead{
 \colhead{Object\tablenotemark{a}} &

 \colhead{[O {\sc ii}] 3727} &
 \colhead{H12 3750} &
 \colhead{H11 3771} &
 \colhead{H10 3797} &
 \colhead{H9 3835} &
 \colhead{H8+He {\sc i} 3889} &
 \colhead{H$\delta$ 4102} &
 \colhead{H$\gamma$ 4340} \\
 \colhead{} &
 \colhead{[O {\sc iii}] 4363} &
 \colhead{He {\sc i} 4471} &
 \colhead{He {\sc ii} 4686} &
 \colhead{H$\beta$ 4861} &
 \colhead{[O {\sc iii}] 4959} &
 \colhead{[O {\sc iii}] 5007} &
 \colhead{He {\sc i} 5876} &
 \colhead{H$\alpha$ 6563} \\
 \colhead{} &
 \colhead{[N {\sc ii}] 6583} &
 \colhead{He {\sc i} 6678} &
 \colhead{[S {\sc ii}] 6717} &
 \colhead{[S {\sc ii}] 6731} &
 \colhead{He {\sc i} 7065} &
 \colhead{[O {\sc ii}] 7320} &
 \colhead{[O {\sc ii}] 7330} 
 } 
 \startdata
CGCG 007-025 (No. 1) &  88.7$\pm$   1.4&   2.8$\pm$   1.0&   3.5$\pm$   1.0&   4.3$\pm$   1.0&   6.1$\pm$   1.0&  17.6$\pm$   1.0&  23.7$\pm$   1.1&  47.0$\pm$   1.1\\    
                         &  11.9$\pm$   1.0&   3.5$\pm$   1.0&   1.3$\pm$   1.0& 100.0$\pm$   1.0& 185.8$\pm$   2.1& 564.2$\pm$   5.7&  12.2$\pm$   1.0& 349.9$\pm$   3.7\\
                         &   4.1$\pm$   1.0&   3.6$\pm$   1.0&   9.0$\pm$   1.0&   6.9$\pm$   1.0&   4.0$\pm$   1.0&   1.6$\pm$   1.0&   1.3$\pm$   1.0\\                  
CGCG 007-025 (No. 2) & 135.2$\pm$   2.1&   2.3$\pm$   1.2&   3.4$\pm$   1.2&   4.9$\pm$   1.2&   6.8$\pm$   1.2&  16.7$\pm$   1.2&  24.9$\pm$   1.1&  45.7$\pm$   1.2\\    
                         &  10.2$\pm$   1.1&   3.4$\pm$   1.1&   0.8$\pm$   1.1& 100.0$\pm$   1.1& 151.2$\pm$   1.9& 463.4$\pm$   4.8&  11.1$\pm$   1.1& 325.7$\pm$   3.5\\
                         &   5.3$\pm$   1.1&   3.3$\pm$   1.1&  13.3$\pm$   1.1&   9.2$\pm$   1.1&   3.0$\pm$   1.1&       \nodata~~ &       \nodata~~ \\                  
HS 0029+1748         &  72.0$\pm$   1.4&   1.4$\pm$   1.1&   1.6$\pm$   1.1&   3.0$\pm$   1.1&   3.9$\pm$   1.1&  12.7$\pm$   1.1&  18.7$\pm$   1.1&  41.2$\pm$   1.1\\    
                         &   2.5$\pm$   1.0&   3.9$\pm$   1.0&       \nodata~~ & 100.0$\pm$   1.0& 134.1$\pm$   1.7& 413.4$\pm$   4.3&  13.0$\pm$   1.0& 347.7$\pm$   3.6\\
                         &  23.7$\pm$   1.0&   4.1$\pm$   1.0&  18.7$\pm$   1.0&  15.1$\pm$   1.0&   3.5$\pm$   1.0&   3.2$\pm$   1.0&   2.8$\pm$   1.0\\                  
HS 0111+2115         & 240.9$\pm$   3.5&       \nodata~~ &       \nodata~~ &       \nodata~~ &       \nodata~~ &  11.0$\pm$   1.5&  18.1$\pm$   1.5&  38.6$\pm$   1.5\\    
                         &   7.1$\pm$   1.1&   3.5$\pm$   1.1&   0.8$\pm$   1.1& 100.0$\pm$   1.0& 200.3$\pm$   2.3& 609.6$\pm$   6.2&  13.0$\pm$   1.1& 381.8$\pm$   4.0\\
                         &   7.4$\pm$   1.1&   4.1$\pm$   1.1&  13.8$\pm$   1.1&  10.3$\pm$   1.1&   3.3$\pm$   1.1&   2.0$\pm$   1.1&   1.4$\pm$   1.1\\                  
HS 0122+0743         &  66.7$\pm$   1.4&   2.0$\pm$   1.1&   2.8$\pm$   1.1&   3.6$\pm$   1.1&   6.0$\pm$   1.1&  16.6$\pm$   1.1&  23.1$\pm$   1.1&  45.1$\pm$   1.2\\    
                         &   3.5$\pm$   1.4&   3.7$\pm$   1.5&       \nodata~~ & 100.0$\pm$   1.3& 153.3$\pm$   2.2& 476.4$\pm$   5.1&  15.1$\pm$   1.4& 413.5$\pm$   4.7\\
                         &  27.5$\pm$   1.4&   4.5$\pm$   1.4&  36.1$\pm$   1.5&  24.7$\pm$   1.5&   4.4$\pm$   1.5&       \nodata~~ &       \nodata~~ \\                  
\multicolumn{9}{c}{\bf (Abridged. Table with all H II regions will be published in the online edition of ApJ)} \\
 \enddata
\tablenotetext{a}{The first 93 entries are HeBCD H {\sc ii} regions with
names in alphabetical order. The remaining 271 entries are SDSS H {\sc ii}
regions. Their names are in the format xxxxx-yyyy-zzz, where xxxxx is the 
middle Julian date (MJD) of the
observation, yyyy is the plate number, and zzz is the fiber number.}
 \end{deluxetable}

 \begin{deluxetable}{lrrrrrrrr}
 \tabletypesize{\scriptsize}
 \tablenum{2}
 \tablecolumns{9}
 \tablewidth{0pc}
 \tablecaption{Emission line EWs \label{tab6}}
 \tablehead{
 \colhead{Object\tablenotemark{a}} &
 \colhead{[O {\sc ii}] 3727} &
 \colhead{H12 3750} &
 \colhead{H11 3771} &
 \colhead{H10 3797} &
 \colhead{H9 3835} &
 \colhead{H8+He {\sc i} 3889} &
 \colhead{H$\delta$ 4102} &
 \colhead{H$\gamma$ 4340} \\
 \colhead{} &
 \colhead{[O {\sc iii}] 4363} &
 \colhead{He {\sc i} 4471} &
 \colhead{He {\sc ii} 4686} &
 \colhead{H$\beta$ 4861} &
 \colhead{[O {\sc iii}] 4959} &
 \colhead{[O {\sc iii}] 5007} &
 \colhead{He {\sc i} 5876} &
 \colhead{H$\alpha$ 6563} \\
 \colhead{} &
 \colhead{[N {\sc ii}] 6583} &
 \colhead{He {\sc i} 6678} &
 \colhead{[S {\sc ii}] 6717} &
 \colhead{[S {\sc ii}] 6731} &
 \colhead{He {\sc i} 7065} &
 \colhead{[O {\sc ii}] 7320} &
 \colhead{[O {\sc ii}] 7330} 
 } 
 \startdata
CGCG 007-025 (No. 1) & 127.8$\pm$   0.6&   4.1$\pm$   0.3&   5.1$\pm$   0.3&   6.4$\pm$   0.3&   9.1$\pm$   0.2&  26.3$\pm$   0.3&  42.8$\pm$   0.2& 100.4$\pm$   0.3\\      
                         &  25.7$\pm$   0.2&   8.1$\pm$   0.2&   3.4$\pm$   0.2& 270.2$\pm$   0.5& 535.6$\pm$   0.6&1633.0$\pm$   1.1&  47.4$\pm$   0.3&1544.0$\pm$   1.3\\  
                         &  15.7$\pm$   0.2&  15.8$\pm$   0.3&  40.4$\pm$   0.3&  31.3$\pm$   0.3&  20.4$\pm$   0.4&   9.1$\pm$   0.4&   7.6$\pm$   0.4\\                    
CGCG 007-025 (No. 2) & 131.0$\pm$   1.1&   2.4$\pm$   0.6&   3.6$\pm$   0.5&   5.2$\pm$   0.5&   7.4$\pm$   0.4&  21.7$\pm$   0.8&  32.8$\pm$   0.4&  72.3$\pm$   0.5\\      
                         &  16.5$\pm$   0.4&   5.9$\pm$   0.4&   1.6$\pm$   0.3& 204.3$\pm$   0.8& 308.8$\pm$   0.9& 952.1$\pm$   1.5&  34.0$\pm$   0.6&1152.0$\pm$   2.4\\  
                         &  18.9$\pm$   0.6&  12.1$\pm$   0.6&  47.2$\pm$   0.8&  34.2$\pm$   0.8&  12.5$\pm$   0.8&       \nodata~~ &       \nodata~~ \\                    
HS 0029+1748         & 106.3$\pm$   0.9&   2.1$\pm$   0.6&   2.3$\pm$   0.5&   4.1$\pm$   0.4&   5.3$\pm$   0.3&  19.2$\pm$   0.3&  24.3$\pm$   0.3&  54.5$\pm$   0.3\\      
                         &   2.9$\pm$   0.1&   4.8$\pm$   0.1&       \nodata~~ & 153.0$\pm$   0.2& 215.9$\pm$   0.2& 681.9$\pm$   0.4&  30.3$\pm$   0.1& 919.1$\pm$   0.5\\  
                         &  62.4$\pm$   0.2&  11.6$\pm$   0.1&  51.6$\pm$   0.2&  41.7$\pm$   0.2&  10.6$\pm$   0.2&  10.2$\pm$   0.2&   8.7$\pm$   0.2\\                    
HS 0111+2115         & 117.8$\pm$   1.0&       \nodata~~ &       \nodata~~ &       \nodata~~ &       \nodata~~ &   5.1$\pm$   0.4&   8.8$\pm$   0.4&  21.4$\pm$   0.4\\      
                         &   9.3$\pm$   0.2&   4.7$\pm$   0.2&   1.0$\pm$   0.2& 149.4$\pm$   0.4& 298.6$\pm$   0.6& 912.1$\pm$   0.9&  22.8$\pm$   0.3& 793.9$\pm$   1.1\\  
                         &  15.5$\pm$   0.3&   8.7$\pm$   0.3&  29.5$\pm$   0.4&  22.0$\pm$   0.3&   7.8$\pm$   0.4&   5.1$\pm$   0.5&   3.4$\pm$   0.5\\                    
HS 0122+0743         & 103.1$\pm$   0.8&   3.1$\pm$   0.4&   4.3$\pm$   0.4&   5.5$\pm$   0.4&   9.1$\pm$   0.3&  24.2$\pm$   0.4&  38.7$\pm$   0.4&  83.6$\pm$   0.5\\      
                         &   1.8$\pm$   0.3&   2.0$\pm$   0.3&       \nodata~~ &  63.0$\pm$   0.5&  98.3$\pm$   0.6& 307.0$\pm$   0.8&  11.5$\pm$   0.4& 357.9$\pm$   1.5\\  
                         &  23.9$\pm$   0.6&   3.9$\pm$   0.5&  32.4$\pm$   0.7&  22.1$\pm$   0.6&   4.2$\pm$   0.6&       \nodata~~ &       \nodata~~ \\                    
\multicolumn{9}{c}{\bf (Abridged. Table with all H II regions will be published in the online edition of ApJ)} \\
 \enddata
\tablenotetext{a}{The first 93 entries are HeBCD H {\sc ii} regions with
names in alphabetical order. The remaining 271 entries are SDSS H {\sc ii}
regions. Their names are in the format xxxxx-yyyy-zzz, where xxxxx is the 
middle Julian date (MJD) of the
observation, yyyy is the plate number, and zzz is the fiber number.}
 \end{deluxetable}



   \begin{deluxetable}{lrrrrrr}
   \tablenum{3}
   \tablecolumns{7}
   \tablewidth{0pc}
   \tabletypesize{\scriptsize}
   \tablecaption{Oxygen, nitrogen and helium abundances for
the best solutions with the \citet{P05} He {\sc i}
emissivities \label{tab2}}
   \tablehead{
   \colhead{Object\tablenotemark{a}} &
   \colhead{O/H\tablenotemark{b,d}} &
   \colhead{O/H\tablenotemark{c,d}} &
   \colhead{N/H\tablenotemark{b,e}} &
   \colhead{N/H\tablenotemark{c,e}} &
   \colhead{$Y$} &
   \colhead{EW(H$\beta$)} 
   }
   \startdata
CGCG 007-025 (No. 1) &   6.0$\pm$   0.1&   6.0$\pm$   0.1&  13.5$\pm$   0.3&  14.0$\pm$   0.3& 0.2493$\pm$ 0.0060& 270 \\ 
CGCG 007-025 (No. 2) &   5.5$\pm$   0.2&   6.1$\pm$   0.2&  12.6$\pm$   0.5&  13.8$\pm$   0.5& 0.2511$\pm$ 0.0069& 204 \\ 
HS 0029+1748         &  11.3$\pm$   0.4&  13.3$\pm$   0.4&  40.0$\pm$   1.4&  43.5$\pm$   1.4& 0.2525$\pm$ 0.0043& 149 \\ 
HS 0111+2115         &  20.7$\pm$   2.8&  20.9$\pm$   2.8&  81.0$\pm$   8.7&  85.0$\pm$   8.7& 0.2697$\pm$ 0.0100&  63 \\ 
HS 0122+0743         &   4.0$\pm$   0.1&   4.4$\pm$   0.1&   9.8$\pm$   0.4&  11.1$\pm$   0.4& 0.2536$\pm$ 0.0055& 232 \\ 
\multicolumn{7}{c}{\bf (Abridged. Table with all H II regions will be published in the online edition of ApJ)} \\
   \enddata
\tablenotetext{a}{The first 93 entries are HeBCD H {\sc ii} regions with
names in alphabetical order. The remaining 271 entries are SDSS H {\sc ii}
regions. Their names are in the format xxxxx-yyyy-zzz, where xxxxx is the 
middle Julian date (MJD) of the
observation, yyyy is the plate number, and zzz is the fiber number.}
\tablenotetext{b}{Abundances are calculated adopting 
$T_e$=$T_e$(O {\sc iii}).}
\tablenotetext{c}{Abundance are calculated adopting 
$T_e$=$T_e$(He$^+$).}
\tablenotetext{d}{In units 10$^{-5}$.}
\tablenotetext{e}{In units 10$^{-7}$.}
\end{deluxetable}
   \begin{deluxetable}{lccccccc}
   \tablenum{4}
   \tablecolumns{8}
   \tablewidth{0pc}
   \tabletypesize{\scriptsize}
   \tablecaption{Parameters for the best solution for He
mass fraction in Table \ref{tab2} \label{tab3}}
   \tablehead{
   \colhead{Object\tablenotemark{a}} &
   \colhead{$T_e$(O {\sc iii})} &
   \colhead{$T_e$(He$^+$)} &
   \colhead{$N_e$} &
   \colhead{$\tau$($\lambda$3889)} &
   \colhead{$\Delta$H$\alpha$/H$\alpha$} &
   \colhead{$ICF$} &
   \colhead{$\chi^2_{min}$}
   }
   \startdata
CGCG 007-025 (No. 1) & 1.64$\pm$ 0.02& 1.64$\pm$ 0.04& 295$^{+ 102}_{-  51}$& 0.75$^{+ 0.16}_{- 0.23}$& 0.0144& 0.9957& 0.72E$+$00 \\ 
CGCG 007-025 (No. 2) & 1.65$\pm$ 0.03& 1.57$\pm$ 0.04&  13$^{+  96}_{-   2}$& 0.94$^{+ 0.26}_{- 0.59}$& 0.0001& 0.9963& 0.21E$+$01 \\ 
HS 0029+1748         & 1.28$\pm$ 0.02& 1.22$\pm$ 0.03&  12$^{+  47}_{-   1}$& 1.16$^{+ 0.24}_{- 0.36}$& 0.0000& 0.9959& 0.78E$+$01 \\ 
HS 0111+2115         & 1.11$\pm$ 0.06& 1.10$\pm$ 0.03& 459$^{+   1}_{- 268}$& 1.15$^{+ 1.54}_{- 0.59}$& 0.0427& 1.0000& 0.43E$+$00 \\ 
HS 0122+0743         & 1.76$\pm$ 0.02& 1.68$\pm$ 0.05&  11$^{+  66}_{-   0}$& 1.05$^{+ 0.12}_{- 0.50}$& 0.0003& 0.9957& 0.29E$+$01 \\ 
\multicolumn{8}{c}{\bf (Abridged. Table with all H II regions will be published in the online edition of ApJ)} \\
   \enddata
\tablenotetext{a}{The first 93 entries are HeBCD H {\sc ii} regions with
names in alphabetical order. The remaining 271 entries are SDSS H {\sc ii}
regions. Their names are in the format xxxxx-yyyy-zzz, where xxxxx is the 
middle Julian date (MJD) of the
observation, yyyy is the plate number, and zzz is the fiber number.}
   \end{deluxetable}

   \begin{deluxetable}{lc}
   \tablenum{5}
   \tablecolumns{2}
   \tablewidth{0pc}
   \tablecaption{Budget of different systematics effects in the
$Y_p$ determination \label{tab1}}
   \tablehead{
\colhead{Property}&\colhead{$\Delta$$Y_p$}
}
\startdata
He {\sc i} emissivity                                       &  $\la$ $+$1.7\% \\
$T_e$(He$^+$) = (0.95 -- 1.0)$\times$$T_e$(O {\sc iii}) &  $\la$ $-$1.0\% \\
Underlying He {\sc i} stellar absorption                    &  $\la$ $+$3.0\% \\
Collisional excitation of hydrogen emission lines           &  $\la$ $+$1.0\% \\
$ICF$(He$^+$ + He$^{2+}$)                                  &  $\la$ $-$1.0\% \\
\enddata
\end{deluxetable}


\begin{deluxetable}{llcrccrc}
\tabletypesize{\footnotesize}
\tablenum{6}
\tablecolumns{4}
\tablewidth{0pc}
\tablecaption{Maximum Likelihood Linear Regressions\label{tab4}}
\tablehead{
\colhead{Method} &\colhead{$N$}& \multicolumn{1}{c}{Oxygen} &
\multicolumn{1}{c}{Nitrogen} 
}
\startdata
\multicolumn{4}{c}{a) O/H=O/H[$T_e$(O {\sc iii})], N/H=N/H[$T_e$(O {\sc iii})]} \\
Old He {\sc i} emissivities & 93 &  0.2466$\pm$0.0012 + 53$\pm$9(O/H)  & 0.2485$\pm$0.0009 + \,~966$\pm$157(N/H) \\
Old He {\sc i} emissivities &364 &  0.2460$\pm$0.0011 + 66$\pm$7(O/H)  & 0.2473$\pm$0.0007 +   1121$\pm$104(N/H) \\
New He {\sc i} emissivities & 93 &  0.2509$\pm$0.0012 + 52$\pm$9(O/H)  & 0.2529$\pm$0.0009 + \,~914$\pm$158(N/H) \\
New He {\sc i} emissivities &364 &  0.2495$\pm$0.0010 + 65$\pm$7(O/H)  & 0.2518$\pm$0.0007 +   1095$\pm$106(N/H) \\
\multicolumn{4}{c}{b) O/H=O/H[$T_e$(He$^+$)], N/H=N/H[$T_e$(He$^+$)]} \\
Old He {\sc i} emissivities & 93 &  0.2472$\pm$0.0012 + 43$\pm$8(O/H)  & 0.2489$\pm$0.0009 + \,~824$\pm$140(N/H) \\
Old He {\sc i} emissivities &364 &  0.2457$\pm$0.0010 + 56$\pm$6(O/H)  & 0.2475$\pm$0.0007 +   1030$\pm$\,~96(N/H) \\
New He {\sc i} emissivities & 93 &  0.2516$\pm$0.0011 + 40$\pm$7(O/H)  & 0.2532$\pm$0.0009 + \,~756$\pm$135(N/H) \\
New He {\sc i} emissivities &364 &  0.2505$\pm$0.0010 + 50$\pm$6(O/H)  & 0.2522$\pm$0.0007 + \,~958$\pm$\,~94(N/H) \\
\enddata
\end{deluxetable}

\end{document}